\documentclass[12pt]{iopart}
\usepackage{epsfig, iopams, graphicx}

\newcommand\bb[1] {   \mbox{\boldmath{$#1$}}  }

\newcommand\del{\bb{\nabla}}

\def\gtsima{$\; \buildrel > \over \sim \;$}
\def\gtsim{\lower.5ex\hbox{\gtsima}}

\newcommand\bcdot{\bb{\cdot}}
\newcommand\oer{\over} 
\newcommand\btimes{\bb{\times}}
\newcommand\vv{\bb{v}}
\newcommand\BB{\bb{B}}
\newcommand\LL{{\cal L}}
\newcommand\boxi{\bb{\xi}}
\newcommand\BV{Brunt-V\"ais\"al\"a\ }

\newcommand\kva{ \bb{k\cdot v_A}  }

\newcommand\dd{\partial}

\newcommand\beq{ \begin{equation} }
\newcommand\eeq{ \end{equation} }

\begin{document}


\title[Surprises in astrophysical gasdynamics]{\bf \Large Surprises in astrophysical gasdynamics}

\author{ Steven A. Balbus$^{1,2,3}$ and William J. Potter$^{1}$}

\address{$^1$ Department of Physics, Astrophysics, University of Oxford, 
Denys Wilkinson Building, Keble Road, Oxford OX13RH}
  \ead{steven.balbus@physics.ox.ac.uk}

  \address{$^2$ Laboratoire de Radioastronomie, \'Ecole Normale
  Sup\'erieure, 24 rue Lhomond, 75231 Paris CEDEX 05, France}

  \address{$^3$ Institut universitaire de France,
  Maison des Universit\'es, 103 blvd.\ Saint-Michel, 75005
  Paris, France}

\begin{abstract}

Much of astrophysics consists of the study of ionised gas under the influence of gravitational and magnetic fields.  Thus, it is not possible to understand the astrophysical universe without a detailed knowledge of the dynamics of magnetised fluids.  Fluid dynamics is, however, a notoriously tricky subject, in which it is all too easy for one's {\it a priori} intuition to go astray.   In this review, we seek to guide the reader through a series of illuminating yet deceptive problems, all with an enlightening twist. We cover a broad range of topics including the instabilities acting in accretion discs, the hydrodynamics governing the convective zone of the Sun, the magnetic shielding of a cooling galaxy cluster, and the behaviour of thermal instabilities and evaporating clouds. The aim of this review is to surprise and intrigue even veteran astrophysical theorists with an idiosynchratic choice of problems and counterintuitive results.  At the same time, we endeavour to bring forth the fundamental ideas, to set out important assumptions, and to describe carefully whatever novel techniques may be appropriate to the problem at hand.   By beginning at the beginning, and analysing a wide variety of astrophysical settings, we seek not only to make this review suitable for fluid dynamic veterans, but to engage novice recruits as well with what we hope will be an unusual and instructive introduction to the subject. 

\end{abstract}
\submitto{\RPP}


\section{Introduction}


Fluid dynamics plays a defining role in shaping our understanding of the vast majority of
physical processes unfolding throughout the cosmos.   Indeed, the dominant composition 
of the current Universe in the form stars, accretion discs, galaxies, galaxy clusters 
and the all-pervasive gaseous medium between such overdensities, is at heart a vast ensemble
of inhomogeneous, magnetised, gravitating bodies of fluid.  
Astrophysical fluid dynamics provides the overarching structure needed to describe large-scale processes
at w ork in our Universe.  However, the physics of magnetised fluids is a messy business.
The theorist must contend with distressingly long mean-free-paths, turbulence, plasma
instabilities, magnetic stresses, radiation physics, and the list goes on.  Often
there is little choice but to resort to a highly phenomenological
formalism and, when, as often happens, the phenomenology takes on a
life of its own, the distinction between guesswork and
fundamentals is blurred.  At the end of complex simulation, what has actually
been explained?   The answer, more often that not, is unsatisfyingly ambiguous.

In this article, we will attempt to shine a bit of light through these murky depths, highlighting
a series of problems in a broad range of settings, all of which contain an enlightening twist.  We have tried to be careful to 
distinguish fundamental physics from any assumptions that need to be made.   We hope that the article will serve the needs both of seasoned researchers looking for something new, and of those who may be embarking on their own research programmes in astrophysical fluid dynamics, who would welcome a broader introductory review. 

The basic idea for this article stems from a series of monographs by Sir Rudolf Peierls on the theme 
of surprises in theoretical physics (Peierls 1979, 1991) which inspired the 2011 Spitzer Lectures given at Princeton Unversity given by one of us (SAB).  
Peierls discusses a series of problems, all of which contain some kind of
 enlightening twist: a contradiction that really isn't, something that looks hard but turns
out to be simple (or vice-versa), a technique that looks ideal for the problem
at hand and then crashes, a
technique that looks like it is being egregiously misapplied that works beautifully.
With welcome rare exceptions, standard texts do not
spend much time exploring seemingly promising approaches that actually fail. 
This is a pity, because it is all too easy to be fooled.  Just how easy will perhaps be made apparent in this article,  via several examples from a wide variety of different subject areas. 

More importantly, we hope that this article will serve to illustrate
how much remains to be understood of what would seem to be the most basic properties of
astrophysical gases.  The coupling between heating, cooling and buoyancy, the inertial quirks of
rotationally-dominated dynamics, and especially the
insidious effects of weak magnetic fields can all play havoc 
with one's physical intuition.  Theorists are still in the process
of learning first-hand just how counterintuitively magnetised fluids can behave.   

Thus motivated, the authors have selected a set of fundamental problems 
of general interest, each of which has a story to tell.  In a few cases, the problem
discussed has turned out to be important,
but not in all cases.  (At least not yet!)   We chose these
examples not for their immediate impact (though in a few cases the {\em ultimate} impact
has been considerable), but because
we have found them to be generally illuminating
of important physical principles, or because there was an 
intellectual novelty to the problem not widely appreciated, or just because they were interesting.  

This then is a review in the spirit of Peierls' books,
a collection of problems from the dynamics of astrophysical
fluids that a practising expert might be able to read with enjoyment, but which might also serve as
a technical introduction for a motivated beginner or outsider.  We make no apologies for the idiosyncratic 
choice of topics; the unusual blend of problems is deliberate.
We have tried to ensure that the astrophysical content of all problems is 
as clear and as self-contained as possible.

We begin by way of review with the fundamental set of governing equations relevant to
astrophysical gas flows.   This is followed by the first of our problems, that of
thermal evaporation of cool interstellar clouds by a hot ambient medium, 
a problem first treated in detail for isolated
clouds in the classic work of Cowie and McKee (1977).   After establishing an interesting 
analogy with ordinary electrostatics, we then move on to examine the problem of thermal
instability in a medium that is subject to bulk heating and cooling.   Another classic
problem of interstellar medium theory (Field 1965; Field, Goldsmith, \& Habing 1969), thermal
instability theory in a stratified medium turns out to be
rich in surprises with insights afforded by thermodynamic identities and
very complex dynamics.   

Continuing in this thermal vein, we revisit the thermal conduction
problem, this time with a weak magnetic field included.  For the very dilute plasmas of interest
in astrophysics, a magnetic field makes a huge difference to the way in which heat is conducted.   Because of the 
tiny electron Larmor radius, heat flows only along the magnetic lines of force, even for a very
weak magnetic field.  The surprises here are stunning: a host of new dynamical instabilities,
when there is more-or-less any temperature gradient in a stratified medium.   

For larger field
strengths in a stratified medium the Newcomb-Parker (Newcomb 1961, Parker 1966)
instability is an issue, and is the topic of the next section.  The surprise here is that 
the strong magnetic field makes no difference to the Schwarzschild instability criterion, 
when written as
a constraint on the vertical density gradient in a gravitational field.   
The twist is a very simple demonstration of this.  
Of course the classic calculation does not include
thermal conduction along field lines, so it would be amiss not to revisit Newcomb-Parker with this
in mind.   Surprise---or not, depending on your prior inclination---the Newcomb-Parker stability 
criterion is dramatically altered.   

Next is a sort of local ``theory-of-everything,'' at least in the linear adiabatic regime.   By this we mean a derivation of a very compact vector equation for small, three-dimensional Lagrangian displacements in an axisymmetric but otherwise arbitrary magnetised, differentially rotating and stratified background.   The equation is so powerful that many well-known problems---including the endlessly surprising magnetorotational instability---can be almost read off from it directly.   Our final two problems are (i) a discussion of the remarkable Papaloizou-Pringle instability, which keeps arising, often unrecognised, in different settings, and reminds us of the crucial role that boundary conditions may play in regulating stability behaviour; (ii) a very simple possible explanation of what seems to be a very complicated problem: the rotation pattern of the solar convection zone.   Here the realisation that convection tends to occur in cells of constant angular velocity and constant entropy (crucially, however, with the unstable driving radial gradient subtracted off!) allows a direct analytic solution for the shape of the isorotation contours.   A welcome surprise: turbulent flow need not be totally unfathomable.

\section {Fundamental equations}

For ease of reference, to establish notation, to highlight a few
points of interest, and with apologies to the well-initiated, let us remind ourselves of 
the fundamental equations of
astrophysical fluid dynamics (e.g.\ Shu 1991) and their immediate consequences.  Throughout this review,
we use the notation
$(x, y, z)$ for Cartesian coordinates, $(R, \phi, z)$ for (radial, azimuthal, axial) cylindrical
coordinates, and $(r, \theta, \phi)$ for (radial, colatitude, azimuthal) spherical coordinates.  

\subsection{Mass conservation}

If $V$ is a fixed volume in space and $S$ its bounding surface,
then the mass $M$ within $V$ changes with time only if there
is a net flux of mass integrated over $S$.  With $\rho$
equal to the mass density and $\vv$ the local velocity field,
we have
\beq
{dM\over dt} = \int_V {\dd\rho\over \dd t}\, dV  = -\int_S\rho\vv\bcdot\bb{dS}.
\eeq
Since $V$ is an arbitrary volume, the divergence theorem gives
immediately
\beq\label{mass}
{\dd\rho\over \dd t}+ \del\bcdot(\rho\vv) = 0,
\eeq
the equation of mass conservation.  It is also customary to use
Cartesian index notation, with $i$, $j$, $k$ used to represent
the $x$, $y$, and $z$ variables.  As usual, repeated indices are summed
over unless otherwise explicitly stated.  Thus, mass conservation may also be written
\beq\label{massi}
\dd_t\rho + \dd_i(\rho v_i) = 0,
\eeq
where the subscripted $\dd$ symbol denotes partial
differentiation with respect to cartesian coordinate $x_i$.  

It will prove very useful to introduce the Langrangian time derivative following
a fluid element, denoted $D/Dt$:
\beq\label{lagr}
{D\ \over Dt} = {\dd\ \over \dd t} + \vv\bcdot\del = \dd_t + v_i\dd_i.
\eeq
In terms of the Lagrangian derivative, the mass conservation equation becomes
\beq\label{masslagr}
{D\ln \rho \over Dt} = - \del\bcdot\vv
\eeq

\subsection{Entropy equation}

Let $S$ be the entropy per particle.  Up to an irrelevant additive constant,
for an ideal gas
\beq
S = {k_B\over \gamma - 1} \ln P\rho^{-\gamma}.
\eeq
Here, $k_B$ is Boltzmann's constant, $\gamma$ is the ratio of specific
heats at constant pressure to that of constant density, and
$P$ is the gas pressure.  Typically, $\gamma $ is $5/3$ for a
monatomic gas and $7/5$ for a diatomic molecular gas.   In the $\gamma\rightarrow 1$ isothermal limit, 
$$
S=-k_B\ln\rho.
$$

To change the entropy of a travelling fluid element, explicit heat
sources (or sinks) are required: for a given Lagrangian fluid element,
$TdS=dq$, where $dq$ is the change of heat per particle.    It
is more usual to work with the heat per unit volume.
For example, if $\bb{F}$ is the heat flux in units of energy
per area per time, then the volume specific heating rate is 
$-\del\bcdot\bb{F}$.  The volume specific change in entropy is
simply $ndS$, where $n$ is the number density of particles.
Thus, a typical astrophysical gasdynamic rendering of the entropy
equation will take the form
\beq\label{ent}
{P\over \gamma - 1} {D(\ln P\rho^{-\gamma})\over Dt} = 
-\del\bcdot\bb{F} +{\rm sources} - {\rm sinks.}
\eeq
where the ideal gas equation of state $P=nk_B T$ has been used.  
Possible sources might include ohmic or viscous heating, or high energy
external particles.  A common sink term is losses from radiation
(cf. \S2.6 below).

In contrast to the mass equation, whose form rarely changes, the entropy
equation must be constructed anew for each environment.  Happily,
in many applications, the adiabatic limit of zero heat exchange is a
sufficiently good approximation to use.

\subsection{Dynamical equation of motion}
Newton's law of motion ``${F}=ma$'' takes the form
\beq
\rho {D\bb{v}\over Dt} = -\del P -\rho\del\Phi
-\del\bcdot\bb{\Pi}+\bb{J}\btimes\bb{B},
\eeq
where $\bb{\Pi}$ is the viscous stress tensor whose precise form need
not concern us here (it is a complex subtopic unto itself), but
the term may very often be ignored.   Magnetic
fields make an appearance in the final $\bb{J\times B}$ Lorentz force term.  
Here $\bb{J}$ is the current density and $\bb{B}$ the magnetic
field vector.  In the problems that we will discuss, Maxwell's displacement current
may be ignored (since the velocities are non-relativistic), so that 
\beq\label{BS}
\mu_0 \bb{J} = \del\btimes\bb{B}
\eeq
where $\mu_0$ is the usual vacuum permeability, $4\pi \times 10^{-7}$
in SI units.   The Lorentz force becomes, using a standard vector identity,
\beq
{1\over\mu_0}\bb{(\del\times B)\times B} = - \del {B^2\over 2\mu_0} 
+{1\over \mu_0} \bb {(B\cdot\nabla)B}.
\eeq
This may be interpreted as magnetic pressure force arising from the first
term on the right, plus a restoring tension force from the second term.
The tension arises from the distortion of the magnetic field's
line of force.  Such distortions can propagate along a field line
in a manner precisely analogous to waves propagating along a taut
string under tension.  This mode of response is peculiar to a magnetized
gas, allowing it to host shear waves with no corresponding change in the density
or pressure.  These disturbances 
are called ``Alfv\'en waves,'' which play
a key role in the behaviour of astrophysical gases in many environments.
The equation of motion without the viscous term now reads:
\beq\label{eom}
\rho {D\bb{v}\over Dt} = -\del \left( P + {B^2\over 2\mu_0}\right) -\rho\del\Phi
+{1\over \mu_0} \bb {(B\cdot\nabla)B}.
\eeq
\subsection{Conservation of Vorticity}

In the absence of magnetic fields, the equation of motion simplifies to
\beq\label{Eull}
 {\dd \bb{v}\over \dd t} + ({\bb{v\cdot}}\del)\bb{v}
\equiv {D\bb{v}\over Dt}  = -{1\over\rho} \del P -\del\Phi
\eeq
Let us apply this equation to the following curious little problem.  What is
the Lagrangian derivative of a small line element $\bb{dl}$ as it moves
embedded in the velocity field?   Clearly, if one moves with
one end of the moving line element and watches what happens to the
other end, there is a change in $\bb{dl}$ only if this other end moves in this {\it relative} sense.
In that case, the change in $\bb{dl}$ after a time $\Delta t$ is given by
\beq\label{Eul0}
\Delta (\bb{dl}) = (\bb{dl}\bcdot\del)\bb{v} \Delta t,
\eeq
which gives
\beq\label{Eul1}
{D(\bb{dl})\over Dt} = (\bb{dl}\bcdot\del)\bb{v}.
\eeq
Now, if in equation (\ref{Eull}), 
we restrict ourselves to the case in which $P$ is a function only
of $\rho$ (a ``barotropic'' fluid), or if $\rho$ is a constant, then the right side of
equation (\ref{Eull}) is a pure gradient, say $\del\Upsilon$.
Then
\beq
{D(\bb{v\cdot dl})\over Dt} = {D\bb{v}\over Dt}\bcdot \bb{dl} + \bb{v}\bcdot{D(\bb{dl})\over Dt}
= \bb{dl}\bcdot\del(\Upsilon +v^2/2)
\eeq
where (\ref{Eull}) and (\ref{Eul1}) have been used.  If we integrate $\bb{v\cdot dl}$ around 
a finite loop moving with the fluid, then the rate of change of this integral vanishes:
\beq
{D\ \over Dt} \oint \bb{v\cdot dl} = \oint {D\ \over Dt} (\bb{v\cdot dl}) = 
\oint\bb{dl}\bcdot\del(\Upsilon +v^2/2) =0
\eeq
since the line integral of a pure potential vanishes around a closed loop.  
Notice that this holds {\it even} if the interior of the loop violates our potential
flow restriction, only the nature of the flow along the boundary curve matters.  The line integral of
the velocity $\bb{v}$ taken round a closed loop is known as the circulation, and we have just demonstrated the conditions under which it is conserved. 
 
Calculating the rate of change of area or volume elements embedded in the flow
is just a matter of repeated application of equation (\ref{Eul0}) and retaining first
order terms in $\Delta t$.   For example, applying (\ref{Eul0}) to each of 
$\bb{dx}$, $\bb{dy}$, and $\bb{dz}$ in the volume element expression
\beq
dV = (\bb{dx\times dy})\bcdot\bb{dz}
\eeq
and keeping terms linear in $\Delta t$ leads to 
\beq
{D(dV)\over Dt} = dV\, \del\bcdot\bb{v}
\eeq
The area element
\beq
\bb{dS} = \bb{dx\times dy}
\eeq
may also be treated exactly the same way, but the result is not 
quite as slick.
It is best to use index notation.  In this case, we find:
\beq
{D(dS_i)\over Dt} = dS_i\dd_jv_j - dS_j\dd_iv_j.
\eeq
Then, for an arbitrary vector $\bb{\omega}$,
\beq\label{omegda}
\bb{\omega\cdot}{D(\bb{dS})\over Dt} = (\bb{\omega\cdot dS})\del\bcdot\bb{v} -
[(\bb{\omega}\bcdot\del)\bb{v}]\bcdot\bb{dS}
\eeq
The utility of this result becomes
apparent when we consider the curl of equation
(\ref{Eull}).   We define the ${\it vorticity}$ as $\bb{\omega} = \bb{\nabla\times v}$.
Noting 
\beq
({\bb{v\cdot}}\del)\bb{v} = \del{v^2\over 2} +(\del\btimes\bb{v})\btimes\bb{v},
\eeq
the curl of equation (\ref{Eull}) yields the Helmholtz vorticity equation:
\beq\label{vorti}
{\dd \bb{\omega}\over \dd t} + \del\btimes(\bb{\omega\btimes\bb{v}}) = 
   {1\over\rho^2} \del\rho\btimes \del P.
\eeq
As before, we restrict ourselves to constant density or barotropic flow,
so that the right side of the equation vanishes.   Then, expanding the left
side of the equation produces 
\beq\label{omeguy}
{D\bb{\omega}\over Dt} = -\bb{\omega}  \del\bcdot\bb{v} +
(\bb{\omega}\bcdot\del)\bb{v},
\eeq
whence we find immediately
\beq
{D(\bb{\omega}\bcdot\bb{dS})\over Dt}=
\bb{dS}\bcdot {D\bb{\omega}\over Dt}+
\bb{\omega}\bcdot {D\bb{dS}\over Dt}  = 0
\eeq
The content of equation (\ref{vorti}) is that $\bb{\omega}\bcdot\bb{dS}$ is conserved
as it moves with the flow: any change in the vorticity contribution is compensated
by kinematic changes in the projected area element.  This also follows from our previous
result of the circulation $\oint \bb{v\cdot dl}$ remaining constant moving with the flow
and an application of Stokes' theorem.   Note as well that equation (\ref{omeguy}) with
$-D\ln\rho/Dt$ substituted for $\del\bcdot\bb{v}$ yields
\beq
{D\ \over Dt}{\bb{\omega}\over \rho} = \left({\bb{\omega}\over\rho}\bcdot\del\right)\bb{v},
\eeq
i.e. $\bb{\omega}/\rho$ satisfies the same equation as the embedded line element $\bb{dl}$, equation (\ref{Eul1}).
Thus vortex lines, normalised by the density, are in effect ``frozen'' into the fluid.

A revealing application of the Helmholtz vorticity equation is the case of
time-steady pure rotation,
$\bb{v} = R\Omega\bb{e_\phi}$.  Equation (\ref{vorti}) then has only a $\phi$ component,
and simplifies to
\beq\label{vorti2}
- R {\dd\Omega^2\over \dd z} =  {1\over\rho^2} (\del\rho\btimes \del P)\bcdot\bb{e_\phi}.
\eeq
A $z$-gradient of $\Omega$ is special.  If there are no other fluid motions,
such a gradient can be supported only if isobaric (constant pressure) and isochoric (constant density)
surfaces deviate from one
another.   If, by contrast, such surfaces coincide, then the external forces are derivable
from a gradient and $\Omega$ can only depend on $R$---it is constant on cylinders.   
In this case, one speaks of ``barotropic flow.''  Otherwise, if there is any $z$
dependence of $\Omega$ and isochores deviate from isobars, the flow is said to 
be ``baroclinic.''  In stars, a tiny offset between isochoric and isobaric surfaces can produce significant baroclinic differential rotation, as in the Sun (cf.\ \S9).

\subsection{The Induction Equation}

The addition of the magnetic field $\bb{B}$ into the problem requires an additional
equation governing {\em its} evolution.  This is provided by the Faraday induction
equation
\beq\label{far}
\del\btimes\bb{E}  = - {\dd\bb{B}\over \dd t} 
\eeq
where $\bb{E}$ is the local electric field in the gas.  Under many circumstances
of interest, the gas is an excellent conductor (a very small ionization fraction
will generally suffice), but we must not set $\bb{E}=0$ as a consequence.
Any particular fluid element will be in motion, and it is only in the local rest
frame {\em of the element}
that the electric field may be expected to vanish.\footnote{Yet more carefully, we should use the rest
frame of the electrons for this statement since they are much more mobile than the ions, but we will not need to make this nice distinction here.} This
means
\beq\label{f}
\bb{E} + \bb{v\times B} = 0,
\eeq
where $\bb{E}$ and $\bb{B}$ refer to the ``observer frame''
in which the fluid element has velocity $\vv$.  (It will not
have escaped the reader's notice that this statement is the same as
requiring that the Lorentz force on individual charges must vanish.)
Putting these last two equations together by eliminating $\bb{E}$
leads to the induction equation:
\beq\label{ind}
{\dd\bb{B}\over \dd t}=\bb{\nabla\times(v\times B)} .
\eeq
The induction equation allows us to calculate the evolution of the 
magnetic field.  As in our discussion of vorticity, this may be written in
the alternative useful form
\beq\label{indd}
{D\bb{B}\over Dt} = -\bb{B}\bb{\nabla\cdot v}+ (\bb{B\cdot\nabla})\bb{v}.
\eeq
The magnetic field follows the same equation as $\bb{\omega}$, so
that once again $\bb{B}/\rho$ are material lines, embedded in the moving fluid.

In the case of purely axisymmetric rotation, $\bb{v} = R\Omega\bb{e_\phi}$, and
the field evolves according to
\beq\label{ddni}
{DB_\phi\over Dt } = R(\bb{B\cdot\nabla})\Omega,
\eeq
whence
\beq\label{ddin}
B_\phi(t) = B_\phi(0) + Rt(\bb{B\cdot\nabla})\Omega,
\eeq
an example of shear drawing out an azimuthal field from a radial component.  
Finally, since $\bb{B}$ is derivable from a vector potential $\bb{B}=\del\btimes\bb{A}$,
we may ``uncurl'' (\ref{ind}) to obtain:
\beq
{\dd\bb{A}\over \dd t} = \bb{v\times(\nabla\times A)} +\del\Phi
\eeq
where $\Phi$ is an unspecified potential function, which may be chosen for covenience (e.g. to eliminate
$\del\cdot\bb{A}$).   Expanding the cross product
and switching to index notation
\beq
{D A_k\over Dt} = v_i \dd_k A_i + \dd_k\Phi
\eeq
Then, recalling (\ref{Eul1}),
\beq
{D (dl_k A_k)\over Dt} =  dl_k \dd_k(\Phi+A_i v_i)
\eeq
In other words, $D(\bb{A\cdot dl})/Dt$ integrated around a closed loop 
moving with the flow vanishes since it amounts to integrating a potential function
over a closed path.  This is precisely analogous to our treatment of the conservation of the circulation integral.  
By Stokes' theorem (a ``recurl'') 
$\bb{B\cdot dS}$ must also be conserved, where $\bb{dS}$ is the
area element bounded by the closed loop, in analogy to $\bb{\omega \cdot dS}$.  Magnetic flux is conserved in exactly the same manner as vorticity is conserved.      

Note the manner in which the Maxwell equations are satisfied.
The vanishing of the divergence of $\BB$ is assured
if $\del\bcdot\BB=0$ is an initial condition.  The Faraday equation
and Biot-Savart law are already explicitly invoked in the
fluid equation formulation.  
Finally, although the charge density computed from $\del\bcdot\bb{E}$ is
ignored in the equation of motion, this turns out to be a second
order relativistic term---the same order of smallness that is ignored
by dropping the displacement current term and using the Biot-Savart
magnetostatic limit.  Thus, to first order in the velocity
(normalized to the speed of light), the Maxwell equations are all satisfied.  

\subsection {Energy Conservation}

The equations (\ref{mass}), (\ref{ent}), (\ref{eom}), and 
(\ref{ind}) are a complete description of ideal magnetohydrodynamics
(MHD) in the sense that no additional field equations are required.  
Individual physical processes can be added as separate terms to these equations
as needed, but as long as we treat our system as a single fluid
no additional evolutionary equation is needed.  It is nevertheless
valuable to have an equation expressing the conservation of energy.
Such an equation is of course not independent from our earlier set, and can
in fact be derived from this group of four by taking the dot product
of equation (\ref{eom}) with ${\bf v}$ and simplifying. 
In conservation form, it reads:
\beq\label{energycons}
{\dd{\cal E}\over \dd t} +\del\bcdot\bb{{\cal F}_E} = 0 ,
\eeq
where the energy density ${\cal E}$ is
\beq
{\cal E} = {\rho v^2\over 2} + {P\over \gamma - 1} +\rho \phi
+{B^2\over 2\mu_0}
\eeq
and the energy flux $\bb{{\cal F}_E}$
\beq\label{energycons1}
\bb{{\cal F}_E} = \left(  {\rho v^2\over 2} +{\gamma P\over \gamma - 1}
+\rho\Phi \right)\bb{v} +{1\over \mu_0} \bb{ B\times (v\times B)}
+\bb{F}
\eeq
The magnetic field appears in $\bb{{\cal F}_E}$ as part of the electromagnetic
Poynting flux.  We have retained a thermal flux $\bb{F}$, but have
ignored viscous and resistive processes as well as external heating or cooling.  

In diffuse astrophysical gases, nonadiabatic heating and cooling
processes can operate on time scales of practical interest.  We
therefore introduce the net loss function ${\cal L}$, with dimensions
of energy per second per unit mass; $\rho{\cal L}$ is then the cooling
per unit volume.  ${\cal L}$ is generally
a function of two thermodynamic variables, $\rho$ and $T$ say.  
It is ordinarily {\em not} a function of spatial gradients, which
instead appear explicitly elsewhere in the equations (for example, \bb{F} above).  
On the other hand, ${\cal L}$
could include bulk heating, for example, in the form of impulsive
collisions of the thermal medium by high energy particles.  
The physics of the cooling process might be electron-ion thermal bremsstrahlung
or inelastic electron-ion collisons.  A typical form for ${\cal L}$
is given by 
\beq
\rho{\cal L} = n^2\Lambda(T) - n\Gamma
\eeq
where $\Lambda$ depends only on the temperature of gas and details of the
atomic processes involved, and $\Gamma$ is a heating rate.  In classical
applications, the heating is often due to cosmic rays, in which case it is not
sensitive to the properties of the thermal
gas.  The {\em number} density $n$ could be chosen to be the electron
density in an ionized gas, or the dominant species in a predominantly
neutral gas.  The fact that $\Lambda$ as written depends only upon $T$ is
a reflection of the assumption of thermodynamic equilibrium between the
ions and electrons, a very important restriction indeed.  It is often,
but not always, valid in cases of interest.  We will not treat the much
more complicated case of out-of-equilibrium gases here.

Amending the entropy and energy equations to include bulk losses,
we find
\beq\label{entbis}
{P\over \gamma - 1} {D(\ln P\rho^{-\gamma})\over Dt}  
+\del\bcdot\bb{F} =  - \rho {\cal L},
\eeq
and 
\beq
{\dd{\cal E}\over \dd t} +\del\bcdot\bb{{\cal F}_E} = -\rho{\cal L} 
\eeq

\subsection {The Boussinesq and anelastic approximations}

The Boussinesq limit, or Boussinesq approximation, is a powerful
simplification which allows progress to be made by filtering out
extraneous compressive modes, generally acoustic waves, from the system under study.
It often causes confusion, however, as the name is used in at
least two different ways.  A small digression is appropriate.

The limit in which the Boussinesq approximation applies is that of 
constant density everwhere, except in the buoyancy term of the equation
of motion.  In this case, the equation of mass conservation reads
\beq
\del\bcdot\bb{v} =0 \quad {\rm (Boussinesq)}
\eeq
The reason for confusion is that the Boussinesq approximation is used,
quite correctly, in situations where the density is {\em not} constant.

To set the stage, we write the general mass conservation equation
in the form
\beq\label{bouss}
\epsilon {D\ln \rho\over Dt} + \del\bcdot\bb{v}=0,
\eeq
where $\epsilon$ has been inserted as a small parameter to remind us
that the density changes are, in a sense we need to make more precise,
small.  The density changes in this equation might be small in the
relative sense that the spatial gradients of the velocity are 
very large, or in an absolute sense: the change in the relative
density of a liquid is likely to be ignorable even when the velocity
gradients involve the longest global length scales.  In either case,
the leading order Boussinesq approximation is that the velocity divergence
vanishes.  

In its classical implementation, the Boussinesq approximation is really a Lagrangian
statement: when displaced, a fluid element maintains its density.   For this statement
to have any content, the background density gradient cannot be exactly zero.
We can, however, impose the condition exactly that a fluid element's density is constant. 
A vertical salinity gradient in seawater, for example, will give rise to a very small density gradient, say $\rho'(z)$.
Then, a strictly vertical displacement $\xi$ of a fluid element will produce a density change relative to 
the element's new surroundings of $\delta\rho=-\xi \rho'$.   The displacement creates a buoyancy force
$$
{\delta\rho\over \rho^2}P'(z)
$$
where $P'$ is the background pressure gradient due to the presence of a gravitational accleration $g$,
 $P'=-\rho g$.   (Here and above $\rho$ is treated as an approximate constant.)    Thus, the equation of motion for $\xi$ is
\beq
{\ddot \xi} = {g \rho'\over \rho}\xi
\eeq
With $\rho'<0$, this gives rise to buoyant oscillations with an oscillation frequency $\omega^2 = -g\rho'/\rho$ 
(Lighthill 1978).
When heated from below, a fluid may locally invert the sign of $\rho'$, in which case $\omega^2<0$ and unstable thermal convection ensues.

There is an appealing internal consistency in this example: $D\rho/Dt$ is set to zero, which
is ``as it should be'' if $\del\bcdot\bb{v}=0$.  In astrophysical settings, gas dynamical considerations prevail,
not liquid water, 
and $D\rho/Dt$ is clearly not a vanishing quantity.  Instead it is the entropy $S$ that is modelled as strictly
conserved with moving fluid elements in the rigorous adiabatic limit $D\sigma/Dt=0$, where 
$\sigma=\ln P\rho^{-\gamma}$.  
How then, may we justify the constraint $\del\bcdot\bb{v}=0$?

The answer is more WKB (named for physicists Wentzel , Kramers, and Brillouin) than Boussinesq in spirit, though the vanishing velocity divergence condition is
often informally referred to in the astrophysical literature as ``Boussinesq.''  The often-made WKB approximatation
consists of assuming that over the scale of a perturbation, the change in the background quantities is
negligible.   This is also referred to as the {\em local approximation.}   The point is that the combination
of a very small perturbation length scale and a long time scale leads to a vanishing velocity divergence,
but only to leading order.   $D\ln\rho/Dt$ should then be interpreted as the error incurred. 

Moreover, displaced fluid elements remain very nearly in pressure equilibrium, differing only to the extent that
the (very short) sound crossing time over a wavelength is comparable to other times of interest.   
Relative pressure perturbations may be thus be
ignored compared with relative density perturbations, and then the adiabatic constraint becomes $\gamma\delta\rho=
\xi \rho \sigma'$.  The frequency of oscillation emerges as $\omega^2=-g\sigma'/\gamma$, a quantity known as the
\BV frequency.   We recover the classical Boussinesq limit by letting $\gamma\rightarrow\infty$.


What might be done to eliminate acoustic oscillations if
the background equilibrium density is very strongly stratified
in position, as is often the case in stars?  One approach, known as the {\em anelastic approximation}
is equivalent to the mass conservation asymptotic regime 
\beq\label{anel}
\epsilon {\dd\rho\over\dd t} + \del\bcdot(\rho\vv) = 0.
\eeq
Note the difference with equation (\ref{bouss}): there the Lagrangian
change was regarded as small, here it is the explicit Eulerian time
dependence that is small.  The anelastic approximation amounts to writing
\beq
\rho(\bb{x}, t) = \rho_0(\bb{x}) + \epsilon \rho_1(\bb{x}, t)
\eeq
and regarding the leading order mass conservation equation as
\beq
\del\bcdot(\rho_0\vv) = 0.
\eeq
The spatial dependence of $\rho_0$ is of course retained.  
The anelastic approximation is used extensively in simulations of the
convective zone in the Sun (Thompson et al. 2003) and other stars.  

The foundations of the anelastic approximation are less firm than those
of the Boussinesq limit, however.  When is the velocity large
enough to justify dropping the lead time derivative?  The velocity field
may itself be linear in $\rho_1/\rho_0$.  Moreover, it has recently
been pointed out that the standard anelastic approximation, lacking the
density time derivative, is not consistent with an energy conservation
formulation.  Efforts continue as of this writing to include modifications
to render the anelastic approximation compatible with this requirement
(e.g. Brown, Vasil, \& Zweibel 2012).

Throughout this paper we will work, when needed, in the Boussinesq $\del\bcdot\bb{v}=0$ limit,
appropriate to large wavenumber disturbances in a stratified background.

\section{Thermal conduction--electrostatic analogy}

\subsection{Preliminaries}

A common astrophysical environment involves high density gas concentrations
embedded within a more diffuse medium.
(Interstellar gas is a good example of this sort of ``cloudy
medium.'')  If the high density clouds are in approximate pressure equilibrium with
their environment, they must be cool relative to the warmer confining gas.
Heat will diffuse towards the cloud surface by thermal conduction from the hot surroundings, and if
the clouds are small, the concentrated flux will evaporate the surface layers.  The outer layer will
heat faster than it is able to cool.  It must then expand, and flow outward in a wind.   (See figure [1].)   An analogous process has been
studied in tokamaks: pellets containing a deuterium-tritium ice are ablated by laser heating, and the 
evaporative outflow in principle causes the pellet to implode (Mayer 1982).

We begin with the calculation of
the mass loss rate $\dot m$ from an isolated spherical cloud of radius $R$ 
immersed in a hot gas at a given ambient
temperature (Cowie \& McKee 1977).  
In its simplest form, the heat flux $\bb{F}$ is given by (Spitzer 1962)
\beq\label{minone}
\bb{F} = - \kappa \del T_e,
\eeq
where $T_e$ is the electron temperature and $\kappa$ is the thermal conductivity
coefficient, itself highly temperature dependent:
\beq\label{zer}
\kappa  \simeq 6\times 10^{-7} T_e^{5/2}
\eeq
The units of $\kappa$ are erg  cm$^{-1}$ s$^{-1}$ K$^{-1}$.   (Note, because of the specialised nature of the material, in this section only we will use cgs and esu units.)   

The evaporation rate from a spherical cloud in an ambient gas at temperature
$T_h$ was first worked out by Cowie and McKee (1977).  Gravity is negligible in this problem.   
If the heat flux is given by 
this simple diffusion approximation, the evaporation rate is
\beq\label{one}
{\dot m} = {16 \pi \kappa_h \mu R \over 25k_B }
\eeq
where $\kappa_h=\kappa(T_h)$ 
is the value of the thermal conductivity coefficient
in the ambient gas, $\mu$ is the
mean mass per particle, and $k_B$ is the Boltzmann constant.  
Note the interesting result that the rate is not proportional to the
surface area presented by the cloud, as one might naively expect, but to the radius.
There is a deep and interesting 
reason for this on which we shall have much to say in the next subsection.

The precise numerical prefactor in equation (\ref{one}) depends upon
the assumption of a spherical cloud (which is of course an idealization)
and the fact that the thermal conductivity 
due to Coulomb collisions is proportional to the temperature $T$ to
the 5/2 power.  The origin of this particular value is not difficult to understand.
The characteristic length at which the kinetic and potential energies of
two electrons are equal is $r_c\sim e^2/k_B T$, where $e$ is the electron
charge.  The Coulomb cross section ($\sim \pi r_{c}^{2}$) will then have a $1/T^2$
dependence, and the mean free path $\lambda_{ee}$ should then scale as
$T^2/n_e$, where $n_e$ is the number density of electrons.  The diffusion
of the heat flux in the small mean free path limit is proportional to
\beq
(m_e n_e) c_e \lambda_{ee}\del c_e^2 \propto {T^{5/2}\over \sqrt{m_e}}
\del T
\eeq
where $c_e^2=k_BT/m_e$ is the square of the electron thermal velocity.
This shows the origin of the $5/2$ power and justifies the assumption
that the electrons are the heat carrying population---at least when the
electron and ion temperatures are equal.

The precise calculation of the thermal conductivity coefficient $\kappa$ in an
ionized gas is by no means simple, involving a detailed kinetic treatment
of the electron distribution function and delicate handling of
a logarithmically-divergent integral (the {\lq}Coulomb logarithm{\rq}) in the course of calculating the scattering cross section.
The problem was worked out in complete detail in the 1950s by Lyman Spitzer
and his colleagues (Spitzer 1962).  For our purposes, it will suffice to use
equation (\ref{zer}).
(This number actually has a very weak temperature and density dependence
stemming from a formal ``Coulomb-logarithm'' term.  The value selected in equation (\ref{zer})
is appropriate for most dilute astrophysical plasmas.)  
\begin{figure}
          \centering
          \includegraphics[width=10cm]{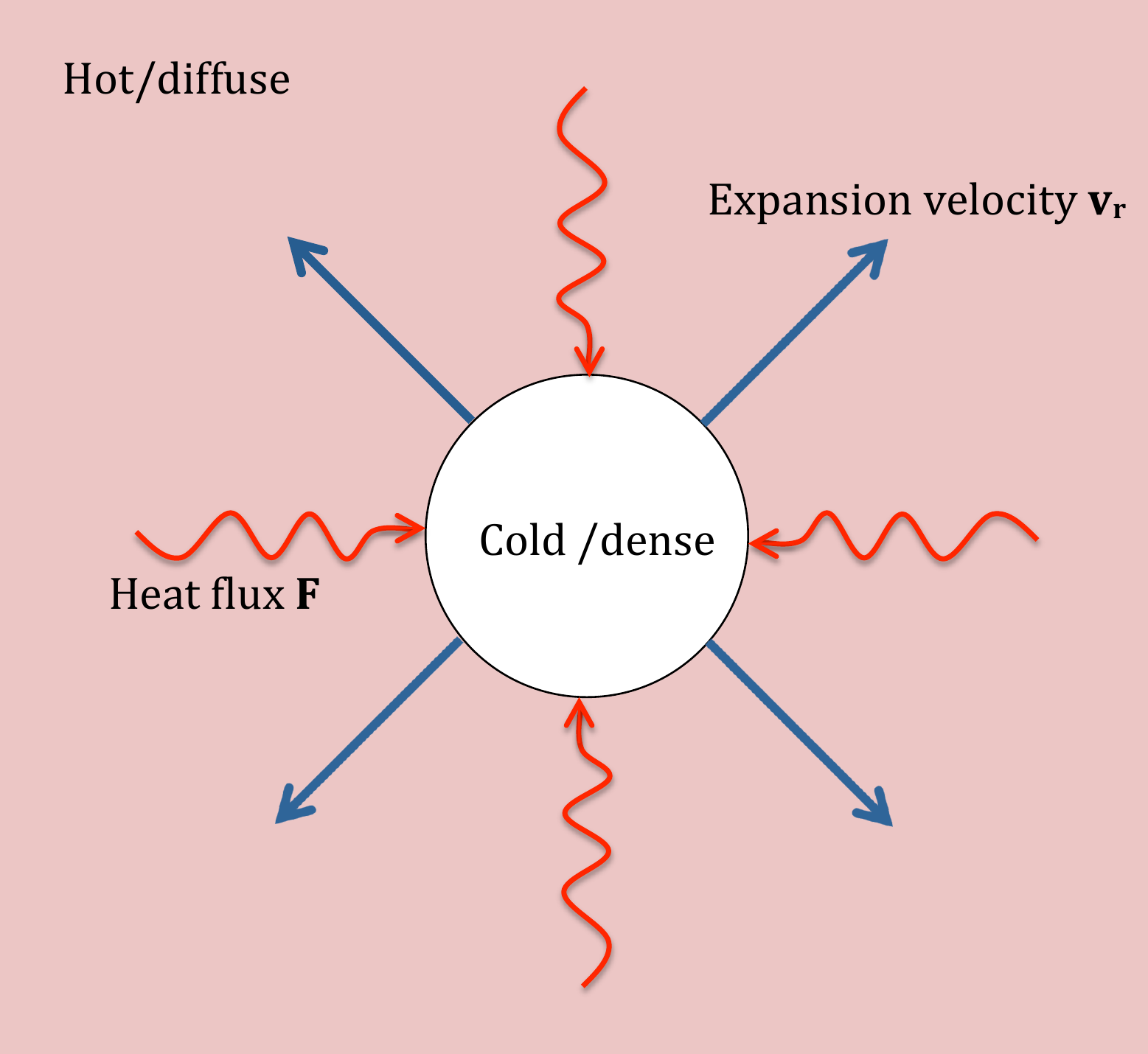}
          \caption{A schematic diagram of an evaporating cold cloud, embedded in a hot diffuse medium. Heat flows from the medium into the cloud, causing the outer cloud layers to heat up, expand, and flow outwards in an evaporative wind.}
          \label{evaporation}
\end{figure}

\subsection {Mass loss as capacitance}

Even with a nonlinear temperature dependence, the thermal evaporation problem
can be cast in a form in which it is entirely analogous to classical
electrostatics.   This is our first surprise, and it is a pleasant one,
for we may then bring to bear on evaporation problems
a powerful mathematical formalism and an intuition honed from electrostatic potential theory.
In particular, it will be shown that
$T^{5/2}$ satisfies the Laplace equation, and the mass loss rate of
equation (\ref{one}) is equal to known constants times the formal {\em
geometrical capacitance} of the evaporating cloud (e.g., the radius $R$
for a sphere, $2R/\pi$ for a thin disc).  Let us see how this arises.

Consider a two-phase medium, with cool high density clouds  
embedded in a hot substrate.  The hotter diffuse gas is in pressure
equilibrium with the clouds.  Heat is diffused by thermal conduction
into the clouds, which ``evaporate'' as a consequence.  The evaporation 
is described by a total mass loss rate from the initial 
spherical cloud surface, 
\beq
\dot{m}=4\pi R^{2}\rho(R) v_{r}(R),
\eeq
a quantity that in steady-state remains constant throughout the flow.  We 
assume that the flow is highly subsonic so that the dominant balance is
one between the thermal energy flux and thermal conduction flux.
The energy equation (\ref{energycons}) under these circumstances reduces to 
\beq \label{Ecoolsph}
\del\bcdot\left( {5P\over2}\vv -\chi T^{5/2}\del T \right) = 0
\eeq
where $\chi =\kappa/T^{5/2}$ and the pressure $P$ may both be regarded as constant.  This is 
equivalent to 
\beq
\vv = {2\chi\over 5P} T^{5/2}\del T +\del\btimes\bb{\alpha}
\eeq
where $\alpha$ is an arbitrary vector field.  

For irrotational flow $\bb{\omega} = 0$, $\bb{\alpha}$ may be ignored, and the velocity follows
the temperature gradient.  This is a well-posed mathematical
problem.  Imposing the constraint of a vanishing mass flux divergence in (\ref{Ecoolsph}) and assuming
an ideal gas equation of state,
the temperature $T$ then satisfies a remarkably simple equation (Balbus 1985):
\beq\label{lap}
\nabla^2 T^{5/2} = 0
\eeq
A unique solution is obtained by specifying the temperature
in the ambient gas (i.e., at ``infinity'') and on each cloud. 
These are classical Dirichlet boundary conditions.  
Clearly, $T^{5/2}$ must be proportional to the electrostatic potential
$\Phi$ in this analogy.  The constant of proportionality is
set by demanding that the electric field analogue $-\del \Phi$
be our mass flux:
\beq
\rho\vv =   {4\mu\chi\over 25 k_B} \del T^{5/2}\equiv -\del\Phi
\eeq
where $k_B$ is the Boltzmann constant and $m$ the mean mass per particle.
Hence,
\beq
\Phi = - {4\mu\chi\over 25 k_B}T^{5/2}
\eeq
If we now use the analogue of Gauss' Law (``the integral of $\bb{E\cdot dS}$
over a closed surface is $4\pi$ times the enclosed change''), then
\beq\label{mdoteq}
{\dot m} = \int\rho\vv\bcdot\bb{dS} = 4\pi C \Delta\Phi
\eeq
where we have represented the enclosed charge as the product of
the capacitance $C$ and the potential difference $\Delta \Phi$
between the common ground of the cloud surfaces and infinity.
(The clouds are all supposed to be 
cold, or at zero common potential, relative to the hot surroundings.)
Thus, with $T_h$ the temperature of the hot gas far from the clouds,
\beq\label{dphi}
\Delta\Phi= {4\mu\chi\over 25 k_B}T_h^{5/2},
\eeq
hence (Balbus 1985),
\beq\label{tada}
{\dot m} = {16\pi \mu\chi C \over 25 k_B}T_h^{5/2}
\eeq
As for $C$, this must be the electrostatic capacitance of $...$ exactly what? 

By way of an answer, note that in contrast to our potential and electric
field, $C$ is not some mathematical analogue quantity, it really is
the {\em actual} capacitance in esu units, a length.  We are, in effect, solving the
standard Dirichlet problem of a system of conductors with specified surface potentials
(all zero in this case) and given potential at infinity.   If we were calculating the
evaporation of a lone spherical cloud, for example, then $C=R$, the
cloud radius (Cowie \& McKee 1977).  Then the result (\ref{tada}) for ${\dot m}$ is exactly
the expression (\ref{one}).   But why should we limit
ourselves to a single evaporating spherical cloud?  What about a hemispherical shell
of radius R? Then,  $C=R(1/2 +1/\pi)$.   Two spherical
clouds of radius $R$ in contact?  
$C=2R\ln 2$.  

While one can look up the esu capacitance for any
particular shape in a compiled table and thereby determine how rapidly the corresponding gas cloud would
evaporate (e.g. Smythe \& Yeh 1972), the real value of equation (\ref{tada}) is not this handy 
convenience.  It is the more general insight that $T^{5/2}$ and
${\dot m}$ behave respectively as a potential and a
capacitance in complex systems percolating with clouds.

\subsection {Evaporation in a Cloudy Medium}

One consequence of the electrostatic analogy is the remarkable (and generally
unappreciated) Faraday cage behaviour of a cloudy medium.  Just as a
Faraday cage can shield its interior from external fields when the cage
wires cover only a tiny fraction of the effective surface, a very low
filling factor of clouds in a two-phase medium will profoundly affect
the system's thermal behaviour.   The easiest way to compute this effect
is to imagine the mathematically equivalent but reverse problem of hot
spherical clouds (radius $R$, temperature $T_h$) embedded in a cold $T=0$ medium.  
An isolated cloud would exhibit the ``monopole'' temperature profile
\beq
T^{5/2} (r) = {T_h^{5/2} R\over r}
\eeq
When does the assumption of isolation break down?  It breaks down when, at a typical
cloud's surface, the superposed
contribution from all other clouds results in a net augmentation of $T^{5/2}$
that is comparable to $T_h^{5/2}$. 

Consider a spherical medium of radius $X$ which is actually comprised of individual 
embedded spherical clouds, each of radius $R$.  
The volume fraction of the medium filled by
clouds, hereafter the ``filling factor,'' is $f$.  Then, the number of clouds
per unit volume is $3f/4\pi R^3$.  The condition for a cloud near the center
of the system to be marginally isolated from the other clouds may then be defined as
\beq
\int_0^X 4\pi r^2\ (3f/4\pi R^3)\ (R/r)\, dr = 1
\eeq
or
\beq
f={2R^2/3X^2} \quad {\rm (sphere)}
\eeq
The same calculation for a gross slab of vertical 
thickness $h$ and radius $X$ gives
\beq
f={2R^2/3Xh}  \quad {\rm (slab)}   
\eeq
These are much more stringent criteria than the naive guess $f\ll 1$!
Indeed, in most astrophysical applications they are far from satisfied.
We are facing a sort of ``Olbers paradox'', in which the superposed small
effects from numerous distant sources overwhelms the local contribution.  

What are the consequences of this?   The primary physical consequence
is that if cool clouds are evaporating into a hot medium at temperature
$T_h$, the process unfolds in two stages.  The first is evaporation from
the clouds into a local intercloud medium with a temperature at some
value intermediate between the cool cloud temperature (effectively zero)
and $T_h$.  Let us call this temperature $T_I$.  The potential function
$T_I^{5/2}$ in our electrostatic analogy cannot obtain an extremal value
in the interior of its domain since it satisfies the Laplace equation.
Accordingly, we expect little variation in $T_I$ within this intercloud
region.  The second stage of the evaporation is that this intercloud medium
at temperature $T_I$ is heated by 
an extended
hot phase at temperature $T_h$, driving a large scale evaporative outflow.
(See figure \ref{evaporation}.) 

\begin{figure}

     \centering
     \includegraphics[width=12cm]{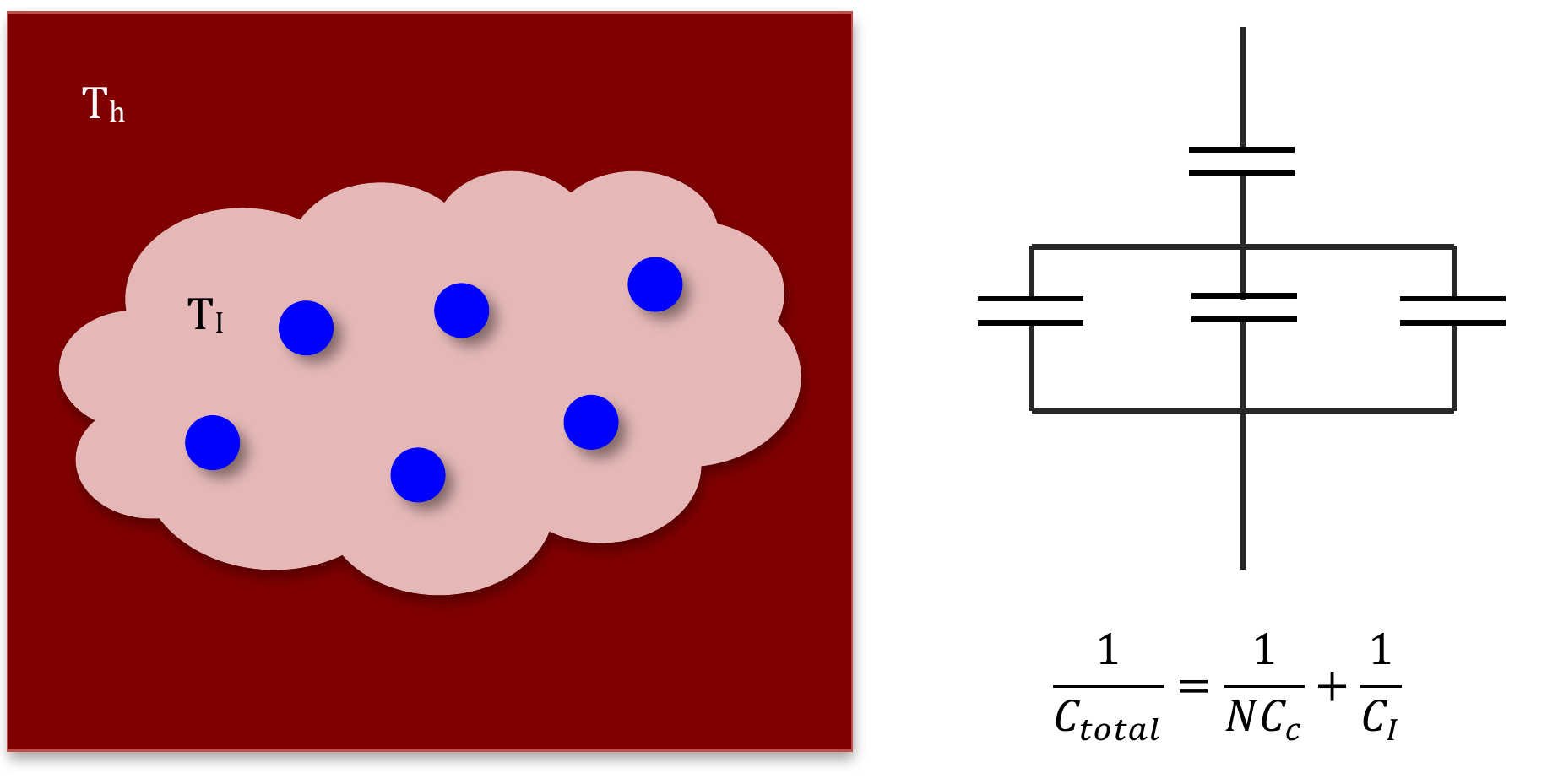}

      \caption{A schematic diagram of an ensemble of cold clouds
      surrounded by an intermediate temperature gas and a hot ambient
      medium. This is represented by an electric circuit component on the right
(for 3 ``clouds'').  The total geometrical capacitance of the system is given
      by the sum of the capacitances $C_C$ of the cold clouds as though they
were in parallel ($NC_{C}$ for $N$ clouds), added to the
      geometrical capacitance of the intermediate temperature region $C_I$ of the ensemble as though
it were in series  ($1/NC_{C} +1/C_{I}$).}

      \label{evaporation}
\end{figure}

The total change in the temperature potential $T^{5/2}$ is the change
going from the clouds $T^{5/2}=0$ to $T^{5/2}=T_I^{5/2}$ in the immediate
intercloud medium, and thence to $T_h^{5/2}$.  In terms of the system capacitance
$C_s$ of the ensemble of clouds and the gross capacitance $C_g$ of the large
scale shape assumed by the ensemble,
\beq
{{\dot m}\over C_s} = {  {\dot m}_c\over R} +{{\dot m}\over C_g}
\eeq
where ${\dot m}_c$ is the evaporation rate of an individual cloud.  
But if $N$ is the total number of clouds, then $N{\dot m}_c = {\dot m}$
since mass is conserved.  Hence
\beq
{1\over C_s} = {1\over N R}  +{1\over C_g}
\eeq
As though they were components in a laboratory circuit, the individual
cloud capacitances first add in parallel, then in series with a capacitance
$C_g$.  (Note: this is unlike resistors!)  Expressed in terms of the filling factor $f$ and gross system
volume $V$, $C_s$ is thus given by
\beq
C_s = {C_g\over 1 + 4\pi R^2 C_g/(3fV)}
\eeq
For the specific case of the gross slab geometry we introduced above,
\beq
C_s = {2\pi/X \over 1 + 8 R^2 /(3\pi fXh )}
\eeq
As a practical matter, an astrophysical cloudy medium in (say) the disc
of a galaxy, would evaporate in a very hot gas as though there were no
clouds at all, as if the entire disc were a continuum of gas ($C_s\simeq C_g$).
This is the first surprise.  The second surprise is that radiative losses,
which are often unimportant when considering the thermal evaporation of
a single cloud, make a great impact on the cloudy medium problem---one
that is still not well understood.

The intermediate temperature $T_I$ may be determined by mass conservation: 
the total mass evaporation rate from all the clouds into the gas at $T=T_I$
must equal the evaporation rate from $T_I$ into the gas at $T=T_h$.  If there
are total of $N$ spherical clouds in our system, then
\beq
NRT_I^{5/2} = C_s T_h^{5/2}.
\eeq
Once again, we consider a slab galaxy with a cloudy medium.
Using $C_s=2X/\pi$, the intermediate temperature gas has a temperature
of 
\beq
T_I= \left[ 8R^2\over 3\pi fXh\right]^{2/5} T_h
\eeq
Typical values might be 1 pc for $R$, $2\times 10^4$ pc for $X$,
100 pc for $h$, and $10^{-2}$ for $f$.  With a hot gas temperature
of $T_h=10^8$ K expected for the intracluster medium of a rich X-ray
cluster, $T_I$ is a few million degrees.  While this value fits nicely
into the intermediate asymptotic regime which is both very large
compared with the cloud temperature and very small compared with $T_h$, it
is not a temperature regime that can maintain itself for long: radiative losses
generally cannot be offset by conductive heating.     

We return once again to our canonical evaporation problem of a slab of
clouds immersed in a very hot gas at $T_h = 10^8$ K, a disc galaxy in a
hot intracluster medium.  The clouds start to evaporate as per above,
rapidly driving the hot gas out of immediate contact with the clouds,
replacing it with gas at $T=T_I \sim 10^6$ K.  Thus far, all is well.

But now the $10^6$ degree gas starts to cool (by thermal bremsstrahlung
and atomic collisions) more rapidly than it evaporates.  Our solution is
not quite self-consistent:  the $T_I$ gas loses pressure support as it
cools, and the ambient $T_h$ gas re-enters the galaxy!  This now creates fresh
$T_I$ gas, and our problem simply repeats.

The problem seems to be intrinsically time-dependent, and it is yet
to be solved in any quantitative sense.
The most likely scenario is one in which after an initial transient
(assuming the clouds do not all evaporate), the gas settles into an
evolving outwardly-moving, mass-loaded (because of ongoing evaporation),
cooling conduction front.  How such a system would manifest itself observationally
is an interesting astrophysical question.  

The surprises here are several.  A seemingly complicated problem (the
thermal evaporation of many clouds at once) turns out at first to
be simple (ordinary electrostatics) but in the end complicated again
(because of radiation effects).  But the key insight that flows from clouds have
important long range interactions even at low filling factors is one that
is very likely to survive beyond the idealizations we have adopted here.

\section {Thermal Instability}

\subsection{Preliminaries}

A fascinating and important property of a gas subject to bulk heating and cooling
is the tendency for its thermodynamic behaviour to be unstable.  
The underlying cause is easy to understand.  Imagine a slightly overdense region
embedded in pressure equilibrium in a surrounding gas.  The increased density leads to
a higher collision rate among the constituent gas particles, which in turn leads to
a higher rate of thermal energy loss.  If this enhanced loss prevails over any
corresponding enhanced heating (as it generally does: heating depends less 
sensitively on the density), the overdense region becomes yet more dense as it is 
compressed by the surrounding medium.  The losses become yet more rapid and the process runs
away.  

This notion was first put on a quantitative footing by Field (1965).  
In terms of the radiative loss function ${\cal L}$, Field showed that
the thermal instability in the manner descibed above occurs when 
\beq\label{Field}
\left(\dd {\cal L}\over \dd T \right)_P < 0.
\eeq
In retrospect, this is a rather obvious mathematical rephrasing of the
physical description that preceeds the equation.  Note the crucial point that
the temperature derivative must be taken with the pressure $P$ held
constant; the loss function is more naturally given as a function of density
$\rho$ and temperature $T$.  Fixing $P$ in the temperature derivative reflects the 
constant pressure conditions imposed by the ambient medium.  

But are matters really so simple? After all, astrophysical gases do not,
in general,
pervade the universe homogeneously.  Rather,
they are trapped and held in
gravitational potential wells.  Under such conditions,
an overdensity will not
stay put as it grows, it will fall down, like all
heavy objects, in the direction the
gravitational field is pointing!  As it falls, the element of fluid
does not keep its pressure fixed, the pressure grows to match that
of the surrounding fluid at lower depths.  Indeed, the growing pressure
can squeeze the element and {\em adiabatically} heat it to the point
at which the element becomes warmer and thus {\em less} dense than the
nonadiabatic surroundings.
We now no longer have an overdensity, we have an underdensity that will
rise back up in the opposite direction of
the gravitational field.  In other words,
the thermal instability is trying to play out in a fluid element
that is actually undergoing buoyant oscillations in the gas.  These oscillations 
will generally be more rapid than the timescale associated with the 
cooling, so to leading order it is the entropy $S$ of the fluid element
that remains constant as the fluid element evolves, not the pressure $P$!
Should our criterion be
\beq
\left(\dd{\cal L}\over \dd T\right)_S < 0 
\eeq
for instability?  This is a very different, and in practice
much more difficult, inequality to satisfy.  The surprises begin.  

Suppose, for example that the cooling function had the canonical form
\beq
\LL = A\,   \rho T^d - B,
\eeq
where $A$ and $B$ are constants and $d$ is a positive real number less than but
of order unity.   Then
\beq
\left( {\dd\LL\over \dd T}\right)_P = (d-1) A\rho T^{d-1} <0.
\eeq
But with the entropy $S$ held constant,
\beq
\left( {\dd\LL\over \dd T}\right)_S = 
\left(d+{1\over \gamma - 1}\right) A\rho T^{d-1} >0.
\eeq
The two results are clearly inconsistent with one another.  

To confuse the issue thoroughly, we offer the following plausible argument
(Defouw 1970), which suggests that we had the criterion correct the first time,
in equation (\ref{Field}).  Consider a convectively stable oscillating fluid element at the peak of
its upward (against the sense of the gravitational field)
displacement.  As before, it is in pressure equilibrium with its
surroundings.  Assume that the buoyant oscillations are growing in amplitude
with time, due to thermal instability.  Then, cooling at the maximum upward displacement will be
enhanced relative to the equilibrium ${\cal L} = 0$ ambient surroundings.
But the element's temperature $T$ must be less than the surroundings,
consistent with the sense of the downward buoyant force on this phase of the
oscillation.  Imagine now crossing 
from the ambient medium into the fluid element along an isobar (constant $P$).
The change in ${\cal L}$ will be positive, the change in $T$ negative.
In other words, equation (\ref{Field}) must be satisfied for thermal
instability, even when the instability is actually an oscillating overstability.

So what is going on here?  The problem is the tacit assumption of the
{\it existence} of convectively stable buoyant oscillations, independently of whether Field-style thermal
instability is present or not.  The surprise is that this assumption is
incorrect.  

\subsection {Eulerian and Lagrangian Perturbations}

\subsubsection {A useful digression.}\ To probe more deeply, we need to sharpen the distinction between perturbations
affecting a particular fluid element, and perturbations referring to a particular
spatial location.  The former are {\em Lagrangian} perturbations, the latter are
{\em Eulerian} perturbations (Lynden-Bell \& Ostriker 1967, Shapiro \& Teukolsky 1983).  
Let $\boxi (\bb{r}, t)$ be the vector field
representing the displacement length of a fluid element located at $\bb{r}$
in equilibrium, i.e. the displaced fluid element is located at $\bb{r} + \bb{\xi}$.
The displacement is considered to be small compared with all
scale heights of interest.  To be definite, we work here with 
the cooling function $\LL$, but the results apply to any flow quantity of interest, including
vector field components.  
The equilibrium cooling function will be denoted simply as $\LL$, the perturbed
cooling function as $\LL'$.   The Eulerian change in $\LL$, denoted $\delta \LL$, is 
\beq\label{Euler}
\LL'(\bb{r}, t)- \LL(\bb{r}) \equiv \delta  \LL
\eeq
The Lagrangian change in the $\LL$, denoted $\Delta \LL$, is then,
to leading order in $\boxi$:
\beq \label{Lagr}
\LL'(\bb{r}+\boxi, t)- \LL(\bb{r}) \equiv \Delta \LL = \LL'(\bb{r}, t) -\LL(\bb{r})
+(\boxi\bcdot\del )\LL =\delta \LL +(\boxi\bcdot\del)\LL
\eeq
where in the final term we have used the equilibrium $\LL$, the distinction with
$\LL'$ introducing only higher order corrections.   Abstracting 
the key equality from the above,
\beq\label{LEq}
\Delta \LL = \delta \LL +(\boxi\bcdot\del)\LL
\eeq
our fundamental relation between Lagrangian and Eulerian perturbations.  
We will use this definition for both scalar and vector functions alike, taking
care to include any unit vectors in the $\bb{\xi\cdot\nabla}$ derivative.

What is the relationship between between the Eulerian radial velocity disturbance
$\delta\bb{v} $ and the Lagrangian change $\Delta\bb{v}$?
Here some care is needed.
The Lagrangian change of the velocity field $\delta\bb{v}$
is by definition
the difference between the time derivative of the displacement of a perturbed
element and its unperturbed counterpart, i.e. $\Delta\bb{v} = D\boxi/Dt$.  
Hence,
\beq\label{LEq1}
{D\bb{\xi}\over Dt}\equiv {\dd\bb{\xi}\over \dd t}+(\vv\bcdot\del)\boxi
 = \bb{\delta v} + (\bb{\xi\cdot\nabla})\bb{v},
\eeq
the latter equality following from (\ref{LEq}).  In other words,
\beq
\bb{\delta v} = {\dd\bb{\xi}\over \dd t}+(\vv\bcdot\del)\boxi -
 (\bb{\xi\cdot\nabla})\bb{v},
\eeq
an important result.

To illustrate the power of the Lagrangian approach as well as 
to derive some powerful equations that we shall use throughout this work,
consider the equations of
mass conservation, (adiabatic) entropy conservation, and the MHD induction equation
of a perfect conductor.  These are all of the form
\beq\label{Qq}
{\dd Q\over \dd t} = L(\bb{v})
\eeq
where $Q$ is a non-velocity flow attribute, and the operator $L$ is {\em linear}
in the velocity $\bb{v}$, and independent of time derivatives.
In equilibrium, which of course need {\it not} be static, $L(\bb{v}) = 0$.
Both $Q$ and $L$ may be either scalar or vector quantities. 

Imagine continuously deforming an equilibrium solution, altering
the structural variables (i.e., pressure, density, magnetic field) by adding
a finite velocity field $\bb{w}$ (not necessarily small compared with $\bb{v}$)
to the unperturbed velocity $\bb{v}$, then
tabulating the brief evolution over an infinitesimal time interval $\delta t$.  This
will normally cause a small change in $Q$, as per the governing equation (\ref{Qq}).  
Since the partial derivative $\dd Q/\dd t$ is Eulerian, taken at
a fixed spatial location, it may equally well be regarded as the Eulerian
change $\delta Q$ induced by the deformation, divided by the interval $\delta t$.  
Using the fact that $L(\bb{v})=0$, and that $L$ is linear in the velocity and free of time derivatives,
\beq
\delta Q =  [L(\bb{v+w })]\delta t = [L(\bb{v})+ L(\bb{w})]\delta t = 0+ L(\bb{w})\, \delta t=
L(\bb{w\,} \delta t)=L(\bb{\xi})
\eeq
where $\bb{\xi}\equiv \bb{w}\delta t$ is the infinitesimal
displacement of a fluid element associated with
the continuous deformation.  From this general result follows three equations of 
{\em constraint,}
in the sense that they display no explicit time evolution, but instead 
embody the Lagrangian conservation of mass, entropy, and magnetic flux.  The relations 
allow a direct calculation of the Eulerian changes associated with fluid displacements:
\beq \label{consmpert}
\delta\rho = -\del\bcdot(\rho\bb{\xi})
\eeq
\beq
\delta\ln P\rho^{-\gamma} \equiv \left({\delta P\over P} - \gamma {\delta\rho\over \rho}
\right) = -\bb{\xi}\bcdot\del\ln P\rho^{-\gamma},
\eeq
\beq
\delta\bb{B} = \del\btimes( \bb{\xi}\btimes\bb{B}).
\eeq
These may also be derived directly from the fundamental equations and the definitions of Eulerian and Lagrangian perturbations, but with considerably more effort.

In this work, we will be working in the WKB limit of local
plane wave behaviour $\exp(i\bb{k\cdot r})$, with $|\bb{k}|$
large compared with reciprocal equilibrium scale heights.  
We will also exclude acoustic disturbances by working in
the Boussinesq limit, meaning here that the terms comprising the $\bb{\del\cdot\xi}$ divergence are
large compared with $\bb{\xi\cdot\nabla}\rho$ in ($\ref{consmpert}$).  
This also means that disturbances are very close to pressure equilibrium
with their surroundings, so that $\delta\ln  P$ may be neglected in comparison
with $\delta \ln \rho$.\footnote{Formal verification of this claim requires consideration
of the dynamical equation of motion.}  The pressure balance argument should include a contribution 
from the magnetic field as well, but our interest will be in those cases in which the magnetic field
is sufficiently weak that its pressure contrbuition may be ignored---but not its tension.  
(An explicit exception to this restriction is the Newcomb-Parker problem, \S 6.)  
To leading order, our three equations then simplify to
\beq\label{con1}
\bb{\del\cdot\xi}=0,
\eeq
\beq\label{con2}
{\delta\rho\over \rho}=  {1\over \gamma} \bb{\xi}\bcdot\del\ln P\rho^{-\gamma}
\eeq
\beq\label{con3}
\delta\bb{B} = i(\bb{k\cdot B})\bb{\xi}
\eeq
These three equations of constraint are extremely useful across a wide variety of fluid problems.

\subsection {When does thermal instability mean convective instability?}

We return to the problem at hand.
For a medium whose equilibrium cooling is described by
$\LL=0$,  the distinction between $\delta {\cal L}$ and $\Delta L$ is lost, and it is 
an interesting fact that the earlier considerations presented above
may now be distilled to a single apparently trivial equation:
\beq\label{magic}
\left(\Delta\LL\right)_S = \left(\delta \LL\right)_P.
\eeq
What this says is the following.  Follow a local fluid element (a ``blob''), 
initially in an $\LL=0$ state, remaining in pressure equilibrium with its surroundings,
on its adiabatic upward displacement.    
(As before, ``upward'' means against gravity.   Downward displacements work too.)  Once displaced, the blob's
local thermodynamic variables no longer satisfy $\LL=0$; there is some $\Delta \LL$.   
But $\Delta\LL$ due to this adiabatic displacement must be the
same change in $\LL$ that would be found by crossing from the immediate surrounding
undisturbed $\LL=0$ medium (at the same pressure) into the displaced
fluid element.   By either the Lagrangian adiabatic or Eulerian isobaric route,
we start at $\LL =0$ and end up in the same blob with the same new value of
$\LL$.  From this obvious little equation follows everything.

Equation (\ref{magic}) may be written
\beq\label{mbis}
\Delta T \left(\dd\LL\over \dd T\right)_S = \delta T 
 \left(\dd\LL\over \dd T\right)_P 
 \eeq
For an adiabatic displacement,
\beq
{\Delta T\over T} = {1\over \gamma}(\gamma-1) {\Delta P\over P}
\equiv 
{1\over \gamma}(\gamma-1)\left( {\delta P\over P} + \xi {\dd\ln P\over \dd r}
\right),
\eeq
where $\xi$ is the radial component of the displacement.
(We are assuming that
the background is spherically symmetric in radius $r$.)
Thus, since $\delta P = 0$,
\beq\label{magic2}
{\Delta T\over T} = {\xi\over \gamma}(\gamma-1) {\dd\ln P\over \dd r}.
\eeq

On the other hand,
\beq
{\delta T \over T} \equiv  {\Delta T\over T} - \xi{\dd\ln T\over \dd r}
\eeq
Putting the last two equations together with $d\ln T = d\ln P-d\ln\rho$,
\beq\label{magic3}
{\delta T \over T} = \xi\left( - {1\over \gamma} {\dd\ln P\over \dd r}
+{\dd\ln \rho\over \dd r} \right)
\eeq
When used in (\ref{mbis}), equations (\ref{magic2}) and (\ref{magic3})
lead to 
\beq
-(\gamma-1){\dd\ln P\over \dd r} \left(\dd\LL\over \dd T\right)_S=
{\dd\ln P\rho^{-\gamma}\over \dd r}\left(\dd\LL\over \dd T\right)_P
\eeq
Finally, assuming the medium is in hydrostatic equilibrium with a gravitational
field $g$ satisfying $-\rho g = {\dd P/\dd r}$, 
our equation becomes
\beq\label{magic4}
(\gamma-1) {g\over c_S^2} \left(\dd\LL\over \dd T\right)_S= 
{\dd\ln P\rho^{-\gamma}\over \dd r}\left(\dd\LL\over \dd T\right)_P
\eeq
where $c_S^2=P/\rho$ is the square of the isothermal sound speed.  This
remarkable equation states that if 
the adiabatic temperature gradient of $\LL$
with respect to $T$ is positive,
as is almost always the case for any
astrophysical loss function,
the corresponding {\em thermodynamic} isobaric temperature
gradient must have the same sign as the {\em spatial gradient} of
the entropy!  In particular the Field criterion, which requires a negative
isobaric temperature gradient for thermal instability, will always be
accompanied by a negative spatial gradient of the specific entropy.
The latter is a prescription for convective instability, normally a much
more rapidly growing disruption than a simple thermal stability.

Now here is a nice surprise: were we to try to construct an atmosphere
in both hydrostatic and thermal equilibrium subject to bulk heating
and cooling, then for realistic thermal loss functions, classic thermal
instability would not occur without a simultaneous, much more disruptive,
convective instability.   In this case, thermal runaway is an afterthought
to adiabatic heating and cooling.  Rising convective plumes are warm
relative to their surroundings and are simply made warmer by unbalanced
external heating; the same holds in reverse for descending cool parcels.

This resolves the immediate puzzle: we should not have tried to analyze
a simple thermal instability in a gas in both hydrostatic and thermal
equilibrium under the assumption that there were simple buoyant oscillations
present.  But the surprises, as we shall presently see, are only just beginning.

\subsection {Cooling in a cooling medium}

Although the problem we analyzed in the previous section was well-posed
and physically sensible, in realistic astrophysical settings there are
seemingly slight differences from our idealization that in fact profoundly
affect the thermal behaviour of the gas.  What is the fate, for example,
of a gas that is cooling without a significant heat source present?
This is an interesting question because a typical astrophysical environment
will be characterized by a thermal cooling time much longer than a
dynamical sound crossing times: as it cools, the gas passes from one
state of hydrostatic equilibrium to another.  (This is similar to the
notion of individual fluid elements passing from one state of collisional
thermal equilibrium to another in flow.)  At one time (e.g. Cowie \& Binney 1977),
this was a simple standard model for the very hot X-ray emitting gas one finds
cradled within the potential of a megaparsec scale cluster of galaxies, though
current models of the cluster tend to be far more complex.  (This almost certainly
means there is something important still missing in our understanding of the dynamics.)
The question then arises of the stability of the bulk cooling process.
Even using what we've just learned, this is a {\em surprisingly} tricky problem, as the
cooling instability time scale and the background flow time scale are comparable.
This is analogous to the classical cosmology problem of linear perturbations growing by gravitational instabilty
in a critical Friedmann universe.  

\subsubsection {Equilibrium cooling}

The equations governing the unperturbed flow, taken to be spherical,
are (i) mass conservation:
\beq
\dot m = 4\pi\rho r^2 v = {\rm constant}
\eeq
where the equilibrium velocity is $\bb{v} = v \bb{e_r}$;
note that both $\dot m$ and $v$ will be negative for inflow.
Next, we impose (ii) hydrostatic equilibrium:
\beq
{dP\over dr} = -\rho g
\eeq
where $-g$ is the radial gravitational field.  Finally, we require (iii) the entropy
loss equation without a heating term in ${\cal L}$:
\beq
{Pv\over \gamma - 1}{d\ln P\rho^{-\gamma}\over dr} = - \rho {\cal L}
\eeq
In fact, these equations do not represent a rigorous global equilibrium
(a finite amount of material eventually collapses to the origin $r=0$), but rather an extended quasi-static phase of a slowly cooling atmosphere.  

\subsubsection {Cooling disturbances.}

To understand the growth of {\it local} disturbances in this inflowing gas, we will
follow the cosmologists and work in a comoving, Lagrangian coordinate 
frame.    The transformation between a small physical radial separation $\delta r$,
and the separation represented in local comoving coordinates, $\delta r'$, in which radial stretching caused by the inflow does not appear, is given by:
\beq
\delta r=a \delta r'. 
\eeq
where we take $a$ to be a function of $r'$ and $t$.
This is equivalent to defining $a$ by
\beq\label{aaa}
a = \left(\dd r\over \dd r'\right)_t,
\eeq
The radial scale of the perturbations is taken to be much smaller
than the global scales of the problem, so that while $a$ in general would depend on $t$ and $r'$,   
in a small Lagrangian 
neighbourhood $r'+\epsilon$ around our fluid element, with no restrictions on $t$, 
$a$ may sensibly be written as $a(t)$.  
Equation (\ref{Eul1}) then tells us
\beq
 {D(\delta r)\over Dt} = \delta r {dv_r\over dr}
\eeq
or using (\ref{aaa})
\beq\label{dvr/dr}
{dv_r\over dr} = {1\over a}{Da\over Dt}
\eeq

The idea is that while the original equations are written in terms of the 
usual $r$ spatial coordinates, it is only in small neighbourhoods of these embedded comoving $r'$
coordinates that the displacement perturbations exhibit a simple WKB plane
wave form.  A ``local'' calculation is truly local only in these coordinates.
If (say) the radial displacement $\xi_r$ has a spatial dependence
$\sim \exp(ik'r')$, with $k'$ independent
of time in these Lagrangian coordinates, and $k'r' \gg 1$, then
\beq
{\dd\ \over \dd r}  = {\dd r'\over \dd r}{\dd \over \dd r'}= {ik'\over a}
\eeq
The wavenumber is, in effect, frozen into the flow expansion.   Note as well that (\ref{dvr/dr}) implies
\beq
{Dv_r\over Dt} = v_r {dv_r\over dr} = {v_r\over a} {D a\over Dt},
\eeq
so that $a$ is linearly proportional to the the velocity $v_r$.  

\subsubsection {Linear dynamical response}

The perturbed dynamical equation of motion is
\beq
{D\bb{\delta v}\over Dt} + (\bb{\delta v}\bcdot\del)\vv = 
-{1\over \rho}\del \delta P +{\delta \rho\over \rho^2}{\del P}
\eeq
Now, the $r$ component of the left side of this equation is
\beq\label{radir}
{D{\delta v_r}\over Dt} + {\delta v_r}{dv_{r}\over dr} = 
{1\over a} {D{a \delta v_r}\over Dt}, 
\eeq
where we have used (\ref{dvr/dr}). A similar expression emerges for the angular component of $\bb{\delta v}$,
denoted $\bb{\delta v_{ang}}$, but with $r$ replacing $a$.  Now the left side of the angular part of
the equation of motion becomes
\beq
{1\over r} {D(r \bb{ \delta v_{ang}})\over Dt}
={1\over r} {D\oer Dt}\left[ {r^2 {D(\bb{\xi_{ang}}/r)\over Dt}}\right]
\eeq
where the final equality is aided by the angular component of
equation (\ref{LEq1}).  We note that only if there is a change in the ratio of the angular
displacement relative to $r$ is there a true change in the angular velocity at
a particular location, a sensible result.  

To make further progress, we need to specify our perturbations more precisely.
We will use {\em spheroidal perturbations} of the form
\beq
\bb{\delta v} = \delta v Y_{lm} \bb{e_r} + {\cal R}(r)\bb{\nabla_{ang}}Y_{lm}
\eeq
where $Y_{lm}$ is the usual spherical harmonic function of colatitude $\theta$
and azimuth $\phi$, and 
$\bb{\nabla_{ang}}$ is a vector operator whose components are the angular
components of the standard $\del$ gradient.   The quantities $\delta v$ and ${\cal R}$
are $r$-dependent amplitudes of the $Y_{lm}$ and their angular gradients.
These should not be confused with $\delta v_r$ and $\bb{\delta v_{ang}}$, the full
vector components depending on $r$, $\theta$, and $\phi$.  
The use of spherical harmonics here is useful to display the $1/r$ dependence 
explicitly in the angular structure, because fluid elements may traverse an extended distance
in radius over the course of the calculation.   

Note that
\beq
\del\bcdot\bb{\delta v} =
\left[{1\over r^2}{\dd r^2\delta v \over \dd r} - {l(l+1)\over r^2}{\cal R}\right]Y_{lm}
\eeq
Upon integration over angles, the angular equation of motion takes the compact form
\beq
 {D\over Dt}\left(Y_{lm} {\cal R}\right) = - {\delta P\over \rho}.
\eeq
Finally, we use the standard Boussinesq approximation $\del\bcdot\bb{\delta v}\simeq0$
to eliminate sound waves.  The previous two equations lead immediately to 
\beq\label{delro}
{D\over Dt}\left({Y_{lm} \over l(l+1)}
 {\dd r^2\delta v \over \dd r}   \right)
= - {\delta P\over \rho}.
\eeq
Using this in the radial equation of motion (\ref{radir}) (to eliminate $\delta P$),
together with the large wavenumber approximation, we find that upon grouping terms,
\beq\label{deltarho}
{1\oer a} {D\over Dt}a\delta v_r \left( 1 + {k^2r^2\over a^2 l(l+1)} \right) 
\equiv
{1\oer a} {D\over Dt}\left( a\delta v_r \over \beta^2 \right) 
 = -{\delta\rho\over \rho}g
\eeq
which defines the geometrical parameter $\beta\le 1$.

\subsubsection{Linear thermal response}

Our final step begins with the linearly perturbed entropy equation,
\beq
{D\over Dt}\left( {\delta P\over P} -\gamma{\delta\rho\over \rho} \right)
	+\delta v_r {d\ln P\rho^{-\gamma}\over dr} = - (\gamma -1)\left(
	{\delta T\over T}\ T\Theta_{T|P} +{\delta P\over P}\ P\Theta_{P|T} \right)
\eeq
where $T$ is the gas temperature.  We have introduced the following
notational scheme:
\beq
\left[\dd(\rho{\cal L}/P) \over \dd X\right]_Y
\equiv
\left(\dd\Theta \over \dd X\right)_Y\equiv \Theta_{X|Y}
\eeq
Notice that the stability discriminant is no longer the gradients of ${\cal L}$, as in 
the static problem of \S 4.3, but the gradients of $\Theta =\rho{\cal L}/P$.  
This is a destabilising influence, increasing the effect of the cooling as the temperature drops.  

It will be noted that we have chosen to regard the cooling parameter $\Theta$
as a function of $P$ and $T$.
The disturbances of interest are
very nearly isobaric so that the relative change in pressure is much
smaller than that of either the density or temperature.  (The quasi-isobaricity
stems from $\delta P\sim 1/kr$ scaling.)  With the $\delta P$ term dropped,
the entropy equation becomes
\beq
\left[ {D\over Dt} + \left(\gamma-1\over\gamma\right)T\Theta_{T|P}\right]
{\delta\rho\over \rho} -{\delta v_r \over \gamma} {d\ln P\rho^{-\gamma}\over dr} =0
\eeq
Finally, substituting from $\delta\rho/\rho$ from equation (\ref{deltarho}), we arrive
at the governing equation for $\delta v_r$:
\beq\label{delvr}
\left[
{D\ \over Dt} +\left(\gamma-1\over \gamma\right) T\Theta_{T|P} 
\right]
{1\over ag} {D\ \over Dt} \left(a\delta v_r\over\beta^2\right) +
 {N^2\over \epsilon^2 g} \delta v_r =0,
\eeq
where 
\beq
N^2 \equiv {g\over \gamma}{\dd\ln P\rho^{-\gamma}\over \dd r} 
\eeq
is the \BV (buoyancy) frequency we first encountered in \S 2.  Free oscillations
in an adiabatic gas respond at this characteristic frequency and
more generally propagate as internal gravity waves.  Note that we
have inserted a small parameter ``tag'' $\epsilon$ in order to set
the scale for the final term as being large, and thus to formalize a 
perturbation treatment.  This will allow for
a (surprisingly) revealing WKB solution of this differential equation.

We seek a solution to equation (\ref{delvr}) of the form (e.g., Bender \& Orszag 1978):
\beq\label{wkb1}
y\equiv \left(a\delta v_r\over\beta^2\right) = A\exp\left[ {S_0\over\epsilon} +S_1\right]
\eeq
where $A$ is a fiducial constant whose value is immaterial and the $S_i$ are
functions of time to be determined.  The differential equation to be solved is
\beq\label{reveqn}
\left[
{D\ \over Dt} +\left(\gamma-1\over \gamma\right) T\Theta_{T|P}
\right]
{1\over ag} {D y\over Dt} + {\beta^2 N^2 y \over \epsilon^2 ga} =0,
\eeq
Substituting (\ref{wkb1}) into (\ref{reveqn}), and sorting out terms of
order $1/\epsilon^2$ and $1/\epsilon$ leads to the two equations:
\beq\label{S0}
\left(S_0'\right)^2 +{\beta^2 N^2} = 0,
\eeq
\beq\label{S1}
2S_1' + {S_0''\over S_0'} -{(ag)'\over ag} + 
\left(\gamma-1\over \gamma\right) T\Theta_{T|P}  = 0.
\eeq
where we have used the primed $'$ notation to represent $D/Dt$.
This pair of first order differential equations decouples, 
is solved by elementary methods, and 
$\delta v_r$ may then be obtained via (\ref{wkb1}):
\beq
\delta v_r = \beta^2 \left(g\over a\beta N\right)^{1/2} 
\exp  \int^t \left( \pm  i \beta N - {\gamma -1\over 2\gamma}
\, T\Theta_{T|P} \right) dt
\eeq

\subsection {Eliciting the instability}

At first glance, it might seem that
the stability of the Eulerian velocity perturbation
is regulated by the thermodynamic derivative $\Theta_{T|P}$: if this
is negative, there is exponential growth.  This is just the classical
Field (1965) result.  But of course this assessment is too crude.
Despite appearances, the amplitude modulation terms are on the same 
footing as the exponential cooling.  More surprises. 

It is a bit more revealing to work with the Lagrangian displacement $\xi_r$, which
to leading WKB order is just $\delta v_r/(\beta N)$.  This implies
\beq
{\xi_r\over \beta} = \left(1\over \beta N\right)^{1/2} \left(
g\over a N^2\right)^{1/2}
\exp  \int^t \left( \pm  i \beta N - {\gamma -1\over 2\gamma}
\, T\Theta_{T|P} \right) dt
\eeq
Now,
\beq
\left(a N^2\over g\right) \propto v\ {d\ln P\rho^{-\gamma}\over dr}   \propto \Theta.
\eeq
where the last relation follows from the entropy equation.  Continuing our
equation juggling,
\beq
{1\over\Theta^{1/2}} = \exp\left( - {1\over 2}\int{d\Theta\over\Theta}\right)
\exp\left( - {1\over2}\int{(d\Theta/dr)\over\Theta/v}\, dt\right)
= \exp\left( {\gamma-1\over 2}  \int{(d\Theta/dr)\, dt\over d\ln P\rho^{-\gamma}
/dr}\right),
\eeq
where in the final equality we have once again used
the entropy equation.  We have thus found
\beq\label{xiP}
{\xi_r\over \beta} = \left(1\over \beta N\right)^{1/2} 
\exp  \int^t  dt\, \left( \pm  i \beta N - {\gamma -1\over 2\gamma}\left[
\, T\Theta_{T|P} -   {\gamma(d\Theta/dr)\over d\ln P\rho^{-\gamma}
/dr}\right]\right),
\eeq
Our last trick is to bring a thermodynamic
identity into play (Balbus \& Soker 1989):
\beq
\gamma {d\Theta\over dr} = {T\Theta_{T|P} } {d\ln P\rho^{-\gamma}\over dr}
+(\gamma - 1 ) T\Theta_{T|S} {d\ln P\over dr},
\eeq
which, it will be noted, reduces to the remarkable equation (\ref{magic4})
in the limit that $\Theta=0$.  The $\Theta$ terms in our expression
for $\xi/\beta$ collapse, and upon using hydrostatic equilibrium to 
substitute for $d\ln P/dr$, we are left with
\beq\label{xiS}
{\xi_r\over \beta} = \left(1\over \beta N\right)^{1/2}
\exp  \int^t dt\, \left( \pm  i \beta N - {(\gamma -1)^2\over 2\gamma}
{g\over c_S^2} \left[
\, T\Theta_{T|S} \over d\ln P\rho^{-\gamma} /dr\right]\right), 
\eeq
where, as before, $c_S^2$ is the square of the isothermal sound speed, $P/\rho$.
Thus, the behaviour of $\xi_r/\beta$ ultimately consists of a buoyant 
oscillation, modified by a thermal loss term whose stability is regulated by an
{\em adiabatic} thermal gradient.  (Knowledgable readers will recognise the action-conserving
amplitude, $[\beta N]^{-1/2}$.)  Whether there is growth or not depends
not just on the sign of this gradient, but on how the thermal behaviour competes with
the evolution of $\beta$.  The classical
1965 Field stability criterion $\Theta_{T|P}>0$ is nowhere to be found.   

This in itself is surprising and very often misunderstood;
perhaps more surprising still is the sheer complexity of the
problem.  The reader who has navigated through this somewhat harrowing
section from
beginning to end will have developed a healthy respect for the
subtlety of understanding thermal instability in moving backgrounds,
with the full interplay between the dynamics of the background and
the developing perturbation.  

\section {Magnetothermal and heat-flux buoyancy instabilities}

\subsection{Magnetised heat flux}

Consider an ionized plasma whose thermal physics is dominated by heat
conduction.   In astrophysical environments this will typically mean
a hot diffuse plasma such as a galactic halo or an intracluster medium.   
The kinetics of such a gas will be dictated
by any magnetic field that might be present, even a very weak one.  
This is because in what we shall call a ``dilute'' gas,
both the ion and electron gyroradii are much smaller than
the respective particle's mean free path.  Under these circumstances,
the standard form of the collisional heat flux (equation [\ref{minone}])
is no longer valid.  Instead, it must be modified to take into
account (i) that only the component of the gradient parallel to the
local magnetic field lines contributes significantly to the flow of heat, and (ii)
that the resultant heat flux flows parallel to the same field lines.
In other words, we must replace the scalar thermal conductivity
$\kappa$ with a 
tensor conductivity $\kappa b_i b_j$ where $b_i$ and $b_j$ 
are components of the unit vector parallel to the magnetic
field $\bb{B}$.  (For the scalar conductivity case, $b_ib_j$ in effect reverts to
the Kronecker delta $\delta_{ij}$.)
The magnetic heat flux then takes the form (Braginskii 1965):
\beq\label{fie}
F_i = - \kappa b_ib_j\dd_j T_e
\eeq
where $\dd_j$ is the partial derivative with respect to the Cartesian
variable $x_j$, other $x_i$ held constant.    What makes this interesting is that the $b_i$ are more than just labels for a field-based coordinate system, they are dynamical variables in their own right.

The change of the form of the conductivity has particularly important 
consequences for the stability of hot plasmas.  Given a time steady equilibrium,
the Eulerian change in the heat flux $\delta F_i$ in the WKB limit of rapidly
spatially varying perturbations is given by (Balbus 2000, 2001)
\beq\label{delf}
\delta F_i = -\kappa (\delta b_i\, b_j\dd_j T_e+ b_i\, \delta b_j\dd_j T_e+b_i b_j\dd_j \delta T_e)
\eeq
The first term redirects a pre-exisiting heat flux, the second alters the directional temperature
gradient being tapped as a heat source, and the final takes into account the new temperature
gradient along a pre-existing field line.   The complexity of 
equation (\ref{delf}) should be compared with $-\kappa\dd_i\delta T_e$, the sole term that 
would be present in
$\delta F_i$ for a scalar conductivity.

\subsection {Magnetothermal Instability}

To see this process in its clearest manifestation, consider a slab of hot gas, hot on
the bottom and cooler on top.  The temperature stratification is entirely
along the $z$ axis.  There is a very weak magnetic field present, of no dynamical
signficance whatsoever.  The field's kinetic significance is that it limits the electron 
gyroradius to be much less than a Coulomb mean free path, so that equation (\ref{fie})
describes the conductive heat flux.  In equilibrium, the field is taken to be uniform along the $x$ axis,
$\bb{B}= B\bb{e_x}$, which ensures that there is no heat flux unless the system is disturbed.
The gravitational field points in the $-z$ direction.  We assume equal electron
and ion temperatures, and that the specific entropy $S$ satisfies $dS/dz>0$,
so that in the absence of heat conduction, any disturbances would correspond to stable buoyant oscillations.

This static equilibrium is disturbed by displacements of the form 
$$
\bb{\xi} = \xi\exp(ikx+\sigma t)\bb{e_z}.
$$
Equation (\ref{con1}) is then trivially
satisfied, equation (\ref{con2}) does not apply because the presence of heat flow
renders the dynamics nonadiabatic, and equation (\ref{con3}) has only one component:
\beq
\delta B_z = ikB\xi.
\eeq
The change in the unit magnetic field vector $\bb{b}$ is simply
\beq
\delta b_z = {\delta B_z\over B} = i k\xi,
\eeq
implying a perturbed heat flux of 
\beq\label{131hf}
\delta F_x = -\kappa\left[ {\delta B_z\over B}\dd_z T + ik\delta T\right] = -ik \kappa\left[ \xi\dd_z T +\delta T
\right] \equiv -ik\kappa\Delta T
\eeq
The heat flux is proportional to the {\em Lagrangian} change in the temperature $\Delta T$.
This should be contrasted with the zero field result, $\delta F_x = -ik\kappa \delta T$, where only
the {\em Eulerian} change is important.  The difference is profound: Eulerian and
Lagrangian temperature changes may well have opposite signs.    
\begin{figure*}
          \centering
          \includegraphics[width=17cm]{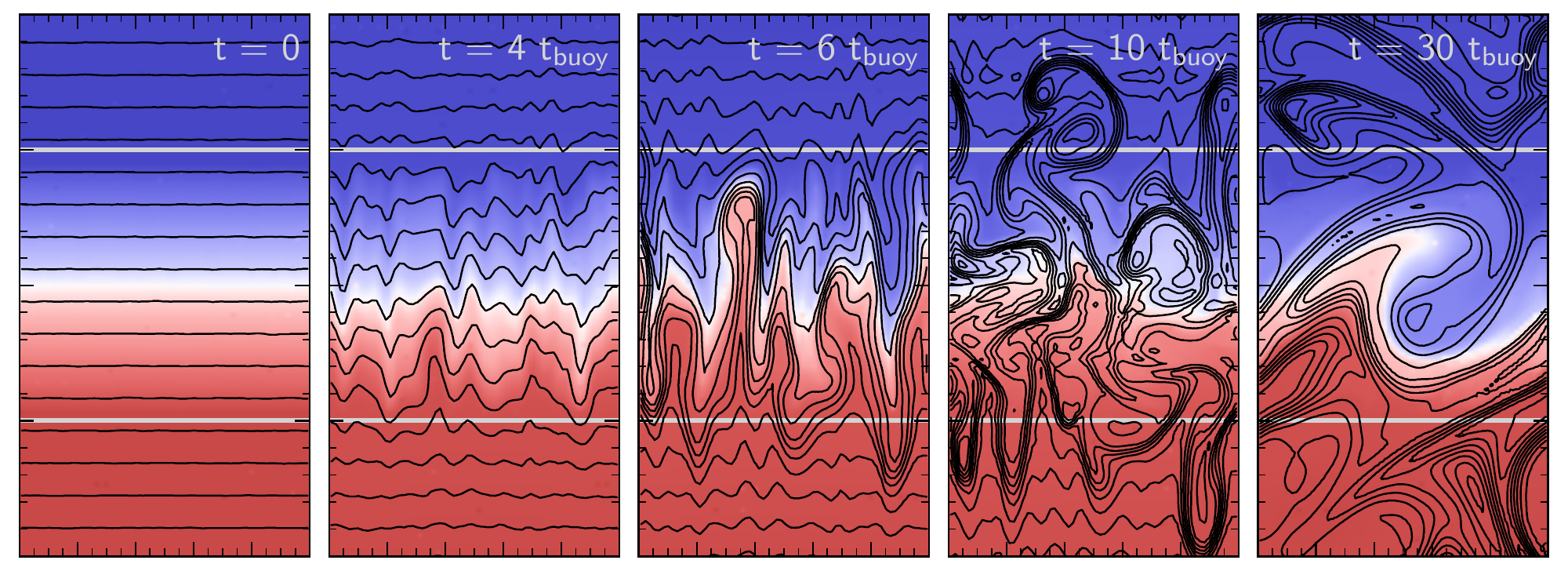}
          \caption{Numerical MHD simulation of the magnetothermal instability (McCourt et al. 2011). Field lines are shown in black, and colour indicates temperature (with red hotter, blue colder).  Small perturbations in the initial horizontal field are convectively unstable. This instability eventually leads to the complete disruption of the initial configuration and the onset of sustained non-linear turbulence. }
          \label{MTI}
\end{figure*}

The perturbed dynamical equation of motion in the $x$ direction reduces to $ik\, \delta P =0$,
so that the pressure perturbation is identically zero, implying $\delta\ln\rho=-\delta\ln T$.  
In the $z$ direction, with $\delta v \equiv\sigma\xi$, 
\beq\label{xig}
\sigma^2\xi = {\delta\rho\over \rho^2} {dP\over dz} = -{\delta\rho\over \rho} g
\eeq
where $-\rho g =dP/dz$ defines the gravitational field strength $g$.  Finally, the perturbed
entropy equation reads
\beq\label{133en}
-\sigma\gamma{\delta\rho\over \rho}  +\sigma\xi {d\ln P\rho^{-\gamma}\over dz} = -{\gamma-1\over P}
\del\bcdot\bb{\delta F}.
\eeq
Replacing $\delta\ln\rho$ with $-\delta\ln T$ and combining (\ref{131hf}) and (\ref{133en}) we obtain
\beq\label{139}
\left[ \sigma\gamma + {k^2\kappa T(\gamma-1)\over P}\right]{\delta\rho\over \rho} = 
\left[\sigma {d\ln P\rho^{-\gamma}\over dz} + {k^2\kappa T(\gamma-1)\over P}{d\ln T\over dz}
\right]\xi.
\eeq
Finally, substituting for $\delta\rho$ from equation (\ref{xig}), the dispersion relation emerges:
\beq\label{dispmti}
\sigma^3 +\sigma^2 {\gamma-1\over \gamma}{k^2\kappa T\over P} +\sigma N^2 +
{\gamma-1\over \gamma}{k^2\kappa T\over P} g{d\ln T\over dz} = 0,
\eeq
where we have used a standard notation for the \BV oscillation frequency,
\beq
N^2 \equiv = {g\over \gamma}{d\ln P\rho^{-\gamma}\over dz},
\eeq
which is a positive quantity for stable adiabatic perturbations.
Equation (\ref{dispmti}) may be compared with the dispersion relation for 
nonmagnetised thermal conduction
\beq\label{dispdbv}
\sigma^2 +\sigma {\gamma-1\over \gamma}{k^2\kappa T\over P} + N^2 =0;
\eeq
here the thermal term serves only to damp the adiabatic \BV oscillations.  
On the other hand, when even a tiny magnetic field is included, one branch of long
wavelength disturbances are characterised by a balance between the third
and fourth terms in equation (\ref{dispmti}) and the growth/damping rate is
\beq\label{123}
\sigma = - (\gamma-1){k^2\kappa T\over P} { {d\ln T/ dz}\over
d\ln P\rho^{-\gamma}/ dz}  \quad
{\rm (long\ wavelengths)}, 
\eeq
the other two solutions correspond to buoyant oscillations, a dominant balance between the
first and third terms.   For short wavelengths, the interesting solution is a dominant balance between
the second and fourth terms:
\beq
\sigma^2  = - g{d\ln T\over dz} \quad {\rm (short\ wavelengths)}.
\eeq
Evidently, if $dT/dz <0$, the equilibrium is unstable.   This is quite surprising:
in the presence of dissipative thermal conduction, short 
wavelength disturbances might normally be thought to be prone to strong damping, not a 
robust instability growing on a free-fall time!   Figure (3) shows a simulation of the full nonlinear development of the MTI.

Perhaps the greatest surprise here is that the magnetised dispersion relation makes no reference
at all to a magnetic field, yet produces results that are
completely different from a nonmagnetised gas. 
The key point is that the magnetic field affects the dynamics when the Lorentz
force becomes comparable to either the pressure or rotational forces.   
On the other hand, the magnetic field affects the {\em kinetics} when the 
electron or ion gyroradius (depending upon whether thermal conduction or viscosity
is involved) is small compared with a Coulomb mean free path, 
a very different requirement.  For the dilute astrophysical plasma in which 
Coulomb thermal conduction is important, an astonishingly minute magnetic field
will suffice---$10^{-21}$ T is enough to make an electron gyroradius smaller than the
Coulomb mean free path for a $10^6$K plasma with an electron density of $10^4$m$^{-3}$.
Interstellar magnetic fields are more than ten orders of magnitude in excess of this.
The effect of the magnetic field is masked in our use of equation (\ref{fie}) for the 
heat flux, serving only to set the cross-field diffusion equal to zero.   For a sufficiently
weak magnetic field cross-field diffusion would have to be included, in which case
the transition from magnetic to nonmagnetic plasma would be quite smooth.  
Even when our interest is purely at the level of dynamics, it is often not a
straightforward matter to decide when a seemingly weak magnetic field might
be important.  The question always is,"weak" compared to what?  We will have surprises
in store for us along these lines in sections (7) and (8).  

To summarise: the most rapidly growing modes are at short wavelengths, with buoyantly unstable behaviour 
reminiscent of violating the Schwarzschild criterion for the onset of convection---except
that it is not the entropy gradient that enters, it is the {\em temperature} gradient.

\subsection {Physics of the magnetothermal instability}

Recall the reason for classical convective instability in a gas: an adiabatically
displaced fluid element will cool at constant entropy on an upward displacement,
in pressure balance at all times with its surroundings.
If this {\em constant} entropy element is nevertheless still warmer than the equilibrium background at the
displaced location then (i) the background entropy must have been {\em decreasing} upward, and
(ii) the element will continue to rise by buoyant forces.  Thus, the criterion for instability 
is that the equilibrium entropy profile must be decreasing upwards.
The presence of thermal conduction would only lessen the growth rate, by lessening the temperature difference (and thus the density difference) between the element and its immediate surroundings.  

When a weak magnetic field alters the thermal conduction, heat flows most efficiently
along the field line, and at large wavenumbers the flux is so efficient that nearly isothermal
conditions are maintained along a field line (i.e. $\Delta T \simeq 0$) and it is the component
of $\del T$ across the field line and parallel to gravity that is important.   
Therefore, if the background temperature decreases with height,
an upwardly displaced element is always warm relative to the surroundings,
and vice-versa for a downwardly displaced element.  This must be convectively unstable.  At
small wavenumbers (long wavelengths), the growth rate is small.   To first order in $\sigma$,
equation (\ref{xig}) suggests ignoring changes in $\delta\rho$, hence also in
$\delta T$ since $\delta P=0$.   The instability is only in
the perturbed velocity.   Indeed, at first sight it appears not to be a buoyant instability at all:  
the growth rate (\ref{123}) is independent of the gravitational field.
The dominant balance is thermal, between the heat deposition along the perturbed
field line and the corresponding rise in entropy of an upwardly moving fluid element.
But the role of gravity is hidden: in this regime, $g$ is in effect a large parameter. 
Indeed, taking $g\rightarrow \infty$ in equation [\ref{dispmti}] produces the same growth rate.
This ensures that an element moves ``instantly'' to the proper $\delta\rho=0$ position as the gas is
heated by conduction along the displaced field line.   Equation (\ref{xig}) shows that the precise
$\delta\rho$ is small but negative, i.e. the element really {\em is} driven by buoyant forces.  

The final surprise is the astrophysical significance of the instability: a weak to moderate
magnetic field cannot thermally isolate a hot dilute plasma in a gravitational field.  
In fact, such configurations appear to be candidates for a dynamo amplification of the
magnetic field (Parrish \& Stone 2007, McCourt et al. 2011).  

A problem of widespread interest is one in which cool gas is stratified at the
bottom of a gravitational potential well surrounded by hotter exterior plasma.
This is the disposition of the intracluster medium in rich clusters of galaxies.
The question is whether thermal conduction from the hot gas will evaporate
the cool gas or whether the cool gas reservoir will grow by radiative losses.
It came as a great surprise when it emerged that the act of 
heating cool gas from above by thermal conduction is actually unstable. When a weak magnetic field is present, that is.

\subsection {A buoyant heat flux instability}

Remarkably, as first shown by Quataert (2008), the magnetothermal instability
is only half the story.  
Let us reconsider the effects of a heat flux in the background equilibrium.
As before, gravity points in the downward $z$ direction.  The simplest
manifestation of the heat flux instability involves a purely vertical magnetic
field, parallel to the gravitational field.  We must also allow
for {\em both} vertical and horizontal displacements $(\xi_x, \xi_y)$, and the same for
the wavenumber components $(k_x, k_z)$. 
As before, the disturbances
have the leading order WKB form $\exp(\sigma t -i\bb{k\cdot r})$; the background field is presumed nearly
constant on the scale of the perturbation.  
Mass conservation (\ref{con1}) then implies
\beq
k_x\xi_x + k_z\xi_z = 0.
\eeq

There is now an equilibrium heat flux present, assumed to be divergence free:
\beq
\bb{F} = -\kappa\bb{b} b_z{\dd T\over \dd z}= - \bb{e_z} \kappa {\dd T\over \dd z}.
\eeq
The perturbed heat flux is given once again by equation (\ref{delf}):
\beq
\delta F_i = -\kappa (\delta b_i\, b_j\dd_j T_e+ b_i\, \delta b_j\dd_j T_e+b_i b_j\dd_j \delta T_e),
\eeq
but now the
first term in this expression, which was not present in our earlier
magnetothermal calculation, serves to
redirect the equilibrium flux.  The perturbed magnetic field unit vector,
\beq
\bb{\delta b} = \delta\left(\bb{B}\over B\right)=
\bb{ b\times} \left( {\bb{\delta B}\over B} \btimes \bb{b}\right),
\eeq
must always be orthogonal to the unperturbed $\bb{b}$, and has only an $x$ component:
\beq
\delta b_x = i (\bb{k\cdot b})\xi_x,
\eeq
where (\ref{con3}) has been used.

The equations of motion are
\beq
\sigma^2\xi_x =  -ik_x {\delta P\over \rho} 
\eeq
\beq
\sigma^2\xi_z = - {\delta \rho\over\rho} g -ik_z {\delta P\over \rho} 
\eeq
and with $\bb{k\cdot\xi} = 0$, we may eliminate $\delta P$ from the above to find
\beq\label{129}
{\delta\rho\over \rho} =    - {\sigma^2\over g} {k^2\over k_x^2} \xi_z
\eeq
where $k^2 = k_x^2 + k_z^2$.  The perturbed heat flux is
\beq
\bb{\delta F} = -\kappa\left(\delta b_x {\dd T\over \dd z}\bb{e_x} +i(\bb{k\cdot b})
\delta T \bb{e_z}\right)
\eeq 
and
\beq
\del\bcdot\bb{\delta F} = - \kappa\left(  ik_x\delta b_x {\dd T\over \dd z} -
(\bb{k\cdot b})^2 \delta T\right)= -k_z^2 \kappa(\xi_z\dd_zT -\delta T )
\eeq
With the zero pressure condition $\delta\ln T = -\delta\ln\rho$ and 
equation (\ref{129}), this becomes
\beq\label{therm}
\del\bcdot\bb{\delta F}= \kappa T {k_z^2}\xi_z\left( {\sigma^2\over g} {k^2\over k_x^2}- 
\dd_z\ln T\right)
\eeq
Even before we arrive at a dispersion relation, it is clear that efficient 
heat transport will lead to a buoyant instability in this problem.  As
$k_z\rightarrow \infty$,  there must be a near cancellation of the two terms in
the flux divergence of the right side of equation (\ref{therm}).   The first 
is the ordinary thermal diffusion term arising from temperature fluctuations, the second
represents heat flow along redirected field lines.  We will presently see that the 
precise growth rate is slightly {\em less} than that given by a balance by the
terms on the right, so $\xi_z> 0$ corresponds to $\del\bcdot\bb{\delta F} <0$.
In other words, the redirected field lines converge when $\xi_z>0$, and diverge when
$\xi_z <0$.  This is a prescription for buoyant instability, since there is 
conductive heating (cooling)
for an upward (downward) displacement.  The dispersion relation becomes:
\beq
\sigma^3 +{\gamma -1 \over \gamma}{\kappa T\over P}{k_z^2} \sigma^2
+{k_x^2\over k^2} {g\over \gamma}\dd_z\ln P\rho^{-\gamma} \sigma
-{\gamma -1 \over \gamma}{\kappa T\over P} g{k_z^2k_x^2\over k^2}\dd_z\ln T = 0
\eeq
Just as suggested by equation (\ref{therm}), there is now instability if 
$\dd T/\dd z >0$, exactly the opposite of the magnetothermal instability configuration
of the previous section.   In other words, even if the cold gas is on the bottom,
the configuration is unstable!   Now that really {\em is} a surprise.   

The problem is not so much the cold gas on the bottom,
it is the heat flux flowing into this gas, trying to disrupt this happy configuration, that is the root of the difficulties.
This {\em heat-flux buoyancy instability}, or HBI as it is known, achieves its most
important astrophysical application in X-ray clusters of galaxies.  The space between
the galaxies in such a cluster is filled with a hot, diffuse plasma, cooler toward
the central region at the base of the cluster well.   We have just seen that any flux
into this cool region will be unstable, a fact not known until Quataert's important
2008 paper.  The nature of the response of the gas to the HBI is to try to draw out
a tangential field, thermally shielding the interior (see the simulation of figure 4), but in real clusters conditions may
be too disturbed. 

As an epilogue to this story, there is a Braginskii viscosity as well as a Braginskii
conductivity (Braginskii 1965, Islam \& Balbus 2005), which channels the momentum flux
in an ionised plasma along field lines in a manner similar to the heat flux.  A recent study 
(Latter \& Kunz 2012) points to the importance of Braginskii viscosity in limiting the thermal
insulation properties of the HBI in the nonlinear regime.  (The magnetothermal instability is much
less affected.)  The most basic properties of the
thermal behaviour of the gas in 
X-ray clusters is still not at all well-understood.  How the various Braginskii diffusion coefficients
and their associated instabilities play themselves out promises to be full of surprises, 
ripe for future elucidation.   

\begin{figure*}
          \centering
          \includegraphics[width=17cm]{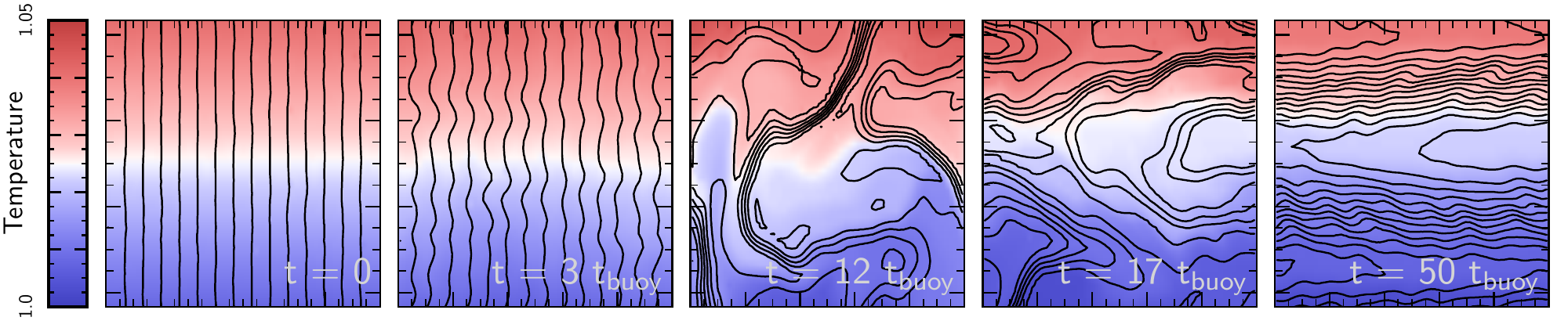}
          \caption{A numerical MHD simulation by McCourt et al. 2011 of the heating buoyancy instability. Colour indicates temperature (with red hotter, blue colder) and the magnetic field lines are drawn in black. Small perturbations to the initial vertical field are unstable due to effective conductive heating (cooling) along the magnetic field lines associated with upward (downward) displacements.  This induces convective instability. The instability eventually saturates, rearranging the field lines into a stable horizontal configuration.  The horizontal field contains components of both senses to conserve
flux.}
          \label{HBI}
\end{figure*}

\section {The Newcomb-Parker Problem}

\subsection{Introduction}

Often just called the ``Parker Instability'' in the astrophysical literature, 
the Newcomb-Parker problem addresses the behaviour of a gas in a gravitational
field with substantial magnetic support.  Before the astrophysicists took it up
(Parker 1966), the problem had been thoroughly studied in the 1950s and early
1960s by the magnetic confinement
community of plasma physicists.   This is perhaps not too surprising: if one's 
goal is to confine a thermonuclear plasma in the laboratory, 
it behooves one to understand 
any instabilities that might be present.    

The astrophysical side of the Parker Instability arose from attempts
to understand hydrostatic equilibrium in a vertical gravitational
field, in practice the disc of the Milky Way Galaxy.   (More recent
applications include accretion disc theory.) In the Galactic problem,
in addition to the presence of significantly magnetised interstellar
gas, cosmic rays are also thought to comprise an important dynamic
component.  In some discussions of this problem, the cosmic rays
are central to an explanation of the instability itself, with these
energetic particles moving upwards along magnetic field lines
anchored in place by sinking cool interstellar gas.   In the 1970s,
the Parker Instability was studied as a means for initiating star
formation (Mouschovias 1974, Blitz \& Shu 1980).

Discussion of the Parker Instability in the astrophysical literature often tends
to be somewhat confusing; Parker's 1966 paper is itself 
far from an easy read.  The surprise here is that the essence of the 
instability is really simplicity itself.   The inclusion of cosmic
rays is somewhat of a red herring:
this is just Schwarzschild convection in a more general guise.  Make an upward
displacement.  Is $\delta\rho<0$?  If so, the system is unstable.   In the plasma
literature this point is understood, but the variational
techniques used are a bit too powerful, hiding some of the
interesting dynamics.   Here, we are motivated to set up
an ``asymptotic matching zone'' that both plasma- and astrophysicists
may appreciate.   

\subsection {Equilibrium state}

Consider a slab of interstellar gas lying in the $xy$ plane with vertical
coordinate $z$.   The gravitational field $\bb{g} = -g\bb{e_z}$ points
downward in the $-z$ direction.  In equilibrium, the gas contains a magnetic field pointing
in the horizontal $\bb{e_x}$ direction, depending only upon $z$:
\beq
\bb{B} = B(z)\bb{e_x}
\eeq
The gas also contains cosmic rays, with pressure $P_{cr}(z)$.  The cosmic
ray pressure is taken to remain constant along field lines; i.e., any gradient
is immediately eliminated by rapid particle streaming: 
\beq
\bb{B}\bcdot\del P_{cr} = 0
\eeq
This is our effective equation of state for the cosmic rays.  
The equation for hydrostatic equilibrium is
\beq\label{hse}
{d\ \over dz}\left[ P + P_{cr} + {B^2\over 2\mu_0} \right] = - \rho g
\eeq
The field $g$ may be any suitable function of $z$.  

\subsection {Departures from equilibrium}

Next, consider displacement perturbations in the vertical direction,
$\bb{\xi}= \xi\bb{e_z}$, the most
unstable modes.  We are free to assume an $x$ dependence of $\exp(ikx)$
since there is no $x$ dependence in the equilibrium state.  The $z$ 
dependence is left unspecified.  We will work at the point of marginal
stability, which will allow an exact treatment of the problem.  Accordingly,
there is no time dependence in either the equilibrium or perturbed states.  

The linearly perturbed equation of motion (here, hydrostatic balance) is:
\beq
0 = - {1\over\rho}\del\left(\delta P+\delta P_{cr}+{ \bb{B\cdot\delta B}\over \mu_0}\right)
+ {\delta\rho\over \rho} \bb{g} + {B \dd_x\bb{\delta B}\over \rho\mu_0}+
\bb{e_x} {\bb{\delta B\cdot}\del B\over \rho\mu_0}.
\eeq
(The subscripted notation $\dd_i$ denotes partial differentiation with respect to
coordinate $i$.)
The induction equation for the magnetic field is
\beq
\bb{\delta B} = \del\btimes(\bb{\xi}\btimes\bb{B}) = \del\btimes(\xi B\bb{e_y}),
\eeq
or
\beq
\delta B_z = B\dd_x\xi, \quad \delta B_x = -\xi\dd_zB
\eeq
Notice that we retain background gradient terms that would be dropped in a WKB treatment.  
Using the above relations,
\beq
\delta B_z \, \dd_z B = (B\dd_x\xi)\dd_z B = B\dd_x(\xi\dd_z B) = - B\dd_x \delta B_x
\eeq
so that the final two terms in the $x$ equation of motion exactly cancel.  
This leaves 
\beq
0 = - {1\over\rho}\dd_x \left( \delta P  +\delta P_{cr}+{ \bb{B\cdot\delta B}\over \mu_0}\right)
\equiv - {1\over\rho}\dd_x(\delta P_{tot})
\eeq
Since the operator $\dd_x$ amounts to multiplication by $ik$, the total perturbed pressure 
$\delta P_{tot}$ vanishes for these most unstable modes.   The $z$ force balance is then 
very simple:
\beq
0= {\delta\rho\over\rho} g +(kv_A)^2\xi
\eeq
where the Alfv\'en velocity is given by
\beq
\bb{v_A} = {\bb{B}\over (\rho\mu_0)^{1/2}},
\eeq 
with $\mu_0$ being the vacuum permeability.

\subsection{Stability Criterion}

The most unstable modes are clearly those with $k\rightarrow 0$, so the marginal
stability condition is just neutral buoyancy:
\beq
{\delta\rho\over \rho} = 0.
\eeq
No magnetic field in sight.  
Neither Karl Schwarzschild nor Baron Rayleigh would have been surprised at this
simple and intuitive result.   Astrophysical complexity
should not be allowed to obscure this basic physical point.  

\subsubsection {Adiabatic disturbances}

For adiabatic conditions, equation (\ref{con2}) gives
\beq\label{con2bis}
{\delta\rho\over \rho} = {1\over \gamma}\left( {\delta P\over P} + \xi 
{\dd\ln P \rho^{-\gamma}\over \dd z}\right)
\eeq
In the problem without the magnetic field or cosmic rays, 
the $\delta P$ term vanishes, and we recover the classical Schwarzschild 
criterion that the entropy should increase upward:
\beq
{\dd\ln P \rho^{-\gamma}\over \dd z} > 0 \quad{\rm (Schwarzschild\ Stability\ Criterion)},
\eeq
for stability.  (The sign is determined by requiring $\delta\rho > 0$
for $\xi>0$).   Since $\dd P/\dd z = - \rho g$, this may also be written
\beq\label{xgener}
{\dd\ln \rho\over \dd z} + {g\over a^2} < 0 \quad {\rm (Stability)},
\eeq
where $a^2=\gamma P/\rho$ is the adiabatic sound speed.   In {\em this} form,
the result is in fact completely general, even with cosmic rays and magnetic fields,
for an adiabatic gas.

To see this, note the following cosmic ray manipulations:
\begin{eqnarray}
0 & = & \delta(\bb{B\cdot\nabla}P_{cr}) = \bb{\delta B}\bcdot\del P_{cr} + \bb{B\cdot}\del
\delta P_{cr}\\
 & = & \delta B_z\dd_z P_{cr} + i k B\,  \delta P_{cr}\\
 & = & (B_x\dd_x\xi)\,\dd_z P_{cr} + ikB\, \delta P_{cr}\\
 & = & ikB(\delta P_{cr} + \bb{\xi\cdot\nabla}P_{cr} ) = ikB\, \Delta P_{cr}
\end{eqnarray}
In other words, the Lagrangian pressure disturbance of the cosmic rays vanishes,
$\delta P_{cr} = - \xi\dd_zP_{cr}$.  

Similarly, for the magnetic pressure,
\beq
\mu_0\delta P_{mag} = 
\bb{B\cdot\delta B} = B\, \delta B_x = B(-\xi\dd_z B) = -\xi \dd_z(B^2/2)            
\eeq
and since 
\beq
0 = \delta P_{tot} = \delta P + \delta P_{cr} + \delta P_{mag},
\eeq
we find that
\beq\label{delpeq}
\delta P = \xi {\dd\ \over \dd z} \left( P_{cr} + {B^2\over 2\mu_0} \right).
\eeq
Inserting this result into (\ref{con2bis}), using equation (\ref{hse}),
and demanding $\delta\rho>0$ for
stability, leads us directly to
the condition (\ref{xgener}) once again.

The destabilising role of the cosmic rays and magnetic field becomes more apparent if the stability
diagnostic is the vertical temperature gradient.  Then, combining equations (\ref{xgener}) and (\ref{hse})
yields the stability criterion
\beq
{\dd\ln a^2\over \dd z} > -\left[{g\over a^2} + {1\over P}{\dd\ \over\dd z}\left( {B^2\over 2\mu_0} +P_{cr}\right)\right]
\quad {\rm (Stability)}
\eeq
In other words, a less negatively steep temperature gradient will destabilise.

There are three surprises here.  The first is that once the gravitational field
is specified, the presence of magnetic fields and cosmic rays
makes {\em no} difference to the upper limit of the inverse density scale height for buoyant
stability.  It is always $g/a^2$.   (The critical temperature scale is, however, affected.)
The second surprise is that the derivation has been
simple and general, at both the conceptual and technical levels.   We are dealing
with elementary buoyancy forces and nothing more.  Students of astrophysical
gasdynamics would do well to examine the more extravagent claims on behalf of the
Parker Instability with a discerning eye.  

The third surprise is that the criterion is entirely incorrect for a dilute plasma.

\subsubsection {Parker--Newcomb--Magnetothermal Instability}

To include the effects of thermal conduction along the field lines, for the calculation of
$\delta\rho/\rho$ from the entropy equation we
return to equation (\ref{139}).  This goes through just as before, only now we must retain the
$\delta P/P$ term because of cosmic ray pressure and a dynamically important magnetic field:
\beq\label{139b}
\left[ \sigma\gamma + {k^2\kappa T(\gamma-1)\over P}\right]{\delta\rho\over \rho} = 
\left[\sigma {d\ln P\rho^{-\gamma}\over dz} + {k^2\kappa T(\gamma-1)\over P}{d\ln T\over dz}
\right]\xi +\sigma{\delta P\over P}
\eeq
Equation (\ref{delpeq}) is unchanged by the inclusion of thermal conduction, so (\ref{139b})
becomes
\beq\label{139c}
\left[ \sigma\gamma + {k^2\kappa T(\gamma-1)\over P}\right]{\delta\rho\over \rho} = 
\left[-\sigma\gamma\left({g\over a^2} + {d\ln\rho\over dz} \right) + 
{k^2\kappa T(\gamma-1)\over P}{d\ln T\over dz} \right]\xi 
\eeq
In the absence of thermal conduction ($\kappa = 0$), we recover the classic result of the 
previous section.  But when thermal conduction is present, the $\sigma\rightarrow 0$
limit is a bit more delicate.  
In the limit of small temperature gradients, the two terms in square brackets on the right
side of the equation form the dominant balance. Thus, we obtain not only a stability criterion,
but a leading order growth rate!  When we are stable by the classic Newcomb-Parker criterion,
there is instability when the temperature gradient decreases upwards.   This is precisely
the magnetothermal instability criterion (and leading order growth rate), recovered just as
the classic Schwarzschild criterion is recovered in the conduction free case: without reference
to the magnetic field at all.   

The classic Newcomb-Parker stability criterion is simply 
not applicable to dilute astrophysical plasmas.   Surprise.

\section{Local, 3D, weakly-magnetized, adiabatic perturbations in 2D rotating,
 stratified backgrounds}

We begin rather formally, deriving and examining a
kind of master equation for the evolution of three-dimensional Lagrangian displacements
in axisymmetric, two-dimensional, magnetized, shearing, stratified backgrounds.
The magnetic field is considered to be weak in that it does not affect the 
equilibrium state, only the perturbations.  We are then in the regime of weak fields and large
wavenumbers.

This is a very general equation.  Moreover, it has a sort of elegant
transparancy:  the emergent force balance is pleasingly intuitive.  We will
use various limits of the equation to (i) derive directly a very
general form of the magnetorotational instability 
(normally a tedious affair); and (ii)  derive some
new results on {\em linear} convection theory in a background shear
similar to that of the Sun.  

The surprises here are many, and they will be highlighted in the development
to come.   One of the most important will be obvious in the early stages of
the development, and that is the deceptive ease with which magnetism can
be included in the analysis, yet its effects are often profound.  We have already
seen evidence of this in our discussions of various thermal conduction driven instabilities
due to kinetic heat flow along field lines---a tiny field goes a long way.
That was all kinetic transport.
We will see here that the direct {\em dynamical} influence of a supposedly 
weak magnetic field can be no less subtle.   

\subsection{Governing equation}

\subsubsection{Equilbrium.}

Our background equilibrium state might be a star, disc, or model galaxy.  
The gas is axisymmetric about a rotation axis, denoted by $z$.   It will
be convenient to use either spherical $(r, \theta, \phi)$ or cylindrical
$(R, \phi, z)$ coordinates depending upon the problem at hand.  The density
$\rho$, pressure $P$, and angular velocity $\Omega$ are then regarded as functions either
of $r$ and $\theta$, or $R$ and $z$.    The
same holds for any other variables constructed from these quantities.
As usual, the velocity vector is denoted by $\bb{v}$.  

The partial derivatives of entropy
and angular momentum cannot be chosen entirely arbitrarily,
even if our interest is a simple local calculation in which these quantities are viewed
as given background-defined constants.   Rather, because the magnetic
field is negligible in the equilibrium state, the derivatives
are linked by the $\phi$ component of the vorticity equation (\ref{vorti2}),
which reads:
\beq\label{164bis}
R{\dd\Omega^2\over \dd z} = {1\over \rho^2} \left( {\dd\rho\over \dd R}{\dd P\over \dd z} -
{\dd\rho\over \dd z}{\dd P \over \dd R}\right).
\eeq
For future reference, it is also convenient to have an alternative form of this equation in
terms of the entropy-like variable $\sigma=\ln P\rho^{-\gamma}$ rather than the density $\rho$,
\beq\label{165bis}
R{\dd\Omega^2\over \dd z} = {1\over \gamma \rho} \left( {\dd\sigma\over \dd z}{\dd P\over \dd R} -
{\dd\sigma \over \dd R}{\dd P \over \dd z}\right) .
\eeq

\subsubsection{Inertial terms from Lagrangian derivatives.}

We are interested in linear departures from the fundamental equation given by (\ref{eom}).  
We shall work in the local WKB limit, so that for the weak magnetic field $\bb{B}$ only
terms involving the gradients of $\bb{\delta B}$ are retained, as these are boosted
by the assumed large wavenumber. 
If we divide equation (\ref{eom}) by $\rho$ and
take Eulerian perturbations, there results
\beq\label{master1}
{D\ \over Dt} \bb{\delta v} + (\bb{\delta v}\bcdot\del) \bb{v}
={\delta\rho\over \rho^2}\del P - {1\over \rho}\del\left(
\delta P +{\bb{B\cdot\delta B}\over \mu_0}\right) +
{1\over\rho  \mu_0} (\bb{B}\bcdot\del)\bb{\delta B}
\eeq
where now 
\beq
{D\ \over Dt} = {\dd\ \over \dd t} + \Omega{\dd\ \over \dd\phi}.
\eeq

It is easiest to begin with a representation in cylindrical coordinates.
The projected components of the left-side operator of equation (\ref{master1}) are:
\beq\label{158}
\bb{e_R}\bcdot \left[ {D\ \over Dt} \bb{\delta v} + (\bb{\delta v}\bcdot\del) \bb{v}\right]
= {D\delta v_R\over Dt} - 2\Omega\delta v_\phi
\eeq
\beq\label{159}
\bb{e_\phi}\bcdot \left[ {D\ \over Dt} \bb{\delta v} + (\bb{\delta v}\bcdot\del) \bb{v}\right]
= {D\delta v_\phi\over Dt} + 2\Omega\delta v_R + R(\bb{\delta v\cdot\nabla})\Omega
\eeq
\beq\label{160}
\bb{e_z}\bcdot \left[ {D\ \over Dt} \bb{\delta v}+(\bb{\delta v}\bcdot\del)\bb{v}\right]
={D\delta v_z\over Dt}.
\eeq
Let $\bb{\xi}$ be the Lagrangian displacement field.  From (\ref{Lagr}),
\beq\label{161}
{D\bb{\xi}\over Dt} \equiv \Delta\bb{v} = \bb{\delta v} + \bb{\xi\cdot\nabla}(R\Omega
\bb{e_\phi}) =  \bb{\delta v} + \bb{e_\phi}  \bb{\xi\cdot\nabla}(R\Omega)  - 
\bb{e_R}\xi_\phi\Omega,
\eeq
whence
\beq\label{162}
\delta v_R = {D\xi_R\over Dt}, \quad \delta v_\phi = {D\xi_\phi\over Dt} - R \bb{\xi\cdot
\nabla}\Omega, \quad \delta v_z = {D\xi_z\over Dt}.
\eeq
Using (\ref{162}) in (\ref{158})-(\ref{160}), we arrive at
\beq\label{163}
{D\ \over Dt} \bb{\delta v} + (\bb{\delta v}\bcdot\del) \bb{v} = 
\bb{\ddot\xi} +2\bb{\Omega\times\dot\xi} +\bb{e_R}(R \bb{\xi\cdot\nabla})\Omega^2
\eeq
where $\bb{\Omega} = \Omega\bb{e_z}$. 
The overhead ``dot'' notation indicates a time derivative of the vector components but not the
unit vectors.  In other words, the coordinates are now treated as locally Cartesian,
with $\bb{e_x}$ replacing $\bb{e_R}$, $\bb{e_y}$ replacing $\bb{e_\phi}$, and 
$\bb{e_z}$ remaining as such.  The physical meaning of the additional two
inertial forces on the right side of (\ref{163}) is
readily grasped: the first is obviously the Coriolis force, and the second is difference between centrifugal
and rotational forces, a tidal force that is in balance only at $\bb{\xi} =0$.   

Since the time derivatives are taken following fluid elements, the appropriate 
fixed spatial coodinates
when a time derivative is taken are those following the unperturbed fluid elements.
Denoting these coordinates by primed notation $R'$, $\phi'$, and $z'$, we have
\beq\label{164}
R' = R, \quad \phi' = \phi -t\Omega(R,z), \quad z'=z
\eeq
The usual chain rule then gives
\beq\label{165}
{\dd\ \over \dd R} = {\dd \over \dd R'} - t{\dd\Omega\over \dd R} {\dd\ \over\dd\phi'}
\eeq
and more generally
\beq\label{166}
\bb{\nabla} = \bb{\nabla'} - t(\bb{\nabla}\Omega){\dd\ \over \dd\phi'}
\eeq
In the WKB limit, the embedded perturbations of interest will take the plane
wave form $\exp i(R'k'_R +z'k'_z+m\phi')$.  (As $m$ is the same in both frames,
there is no primed superscript.)  Equation (\ref{166}) then implies
\beq\label{167}
\bb{\nabla} = i\bb{k'} - im t(\bb{\nabla}\Omega)\equiv i \bb{k}(t),
\eeq
when operating upon perturbed $\delta$-variables.  The poloidal components of $\bb{k}$ 
thus tend to align with $-\bb{\nabla}$ as time goes on.  A particularly useful
result follows with some help from equation [\ref{ddin}]:
\beq
\bb{k}(t)\bcdot \bb{B}(t) = \bb{k'}\bcdot\bb{B}(0).
\eeq
That is, even though the (Eulerian) wavenumber $\bb{k}$ and toroidal field
component $B_\phi$ are both time dependent, the dot product $\bb{k}\bcdot \bb{B}$
is a local constant.   

\subsubsection{External forces.}

There remains the right side of (\ref{master1}).  The constraint equations 
(\ref{con1}) and (\ref{con2}) now yield
\beq\label{172}
{\delta\rho\over \rho^2} \del P = {\del P \over\gamma\rho} \bb{\xi\cdot\nabla}\ln P\rho^{-\gamma}
\eeq
\beq\label{173}
{1\over\rho\mu_0}(\bb{B\cdot\nabla})\bb{\delta B} = -{(\bb{k\cdot B})^2 \over\rho\mu_0}\bb{\xi} 
\equiv -(\bb{k\cdot v_A})^2
\bb{\xi}
\eeq
where as earlier we use the Alfv\'en velocity,
\beq\label{174}
\bb{v_A} = {\bb{B}\over \sqrt{\rho\mu_0}}.
\eeq
Equation (\ref{173}) implies a very
simple ``Hooke's Law'' behaviour for magnetic tension: the force is restorative
and proportional to the displacement.  Surprisingly, ``restorative'' forces in 
rotating systems can have precisely the opposite effect, as we shall soon see.

Putting together equations (\ref{master1}), (\ref{163}), (\ref{166}),
(\ref{172}), (\ref{173}) yields the set of equations
$$
\bb{\ddot\xi}
+ (\bb{k\cdot v_A})^2\bb{\xi} +2\bb{\Omega\times\dot\xi} +\bb{e_R}(R\bb{\xi\cdot\nabla})\Omega^2
- 
{\del P \over\gamma\rho} \bb{\xi\cdot\nabla}\ln P\rho^{-\gamma}
$$
\beq\label{175}
\qquad \qquad\qquad\qquad\qquad\qquad\qquad\qquad\qquad +{i\bb{k}(t)\, \delta P_{tot}\over\rho}=0
\eeq
\beq\label{175bis}
\delta P_{tot} = \delta P +{\bb{B\cdot\delta B}\over \mu_0}
\eeq
\beq\label{176}
\bb{k}(t)\bcdot\bb{\xi} = 0,
\eeq
\beq\label{177}
\bb{k}(t) = \bb{k'} - m t(\bb{\nabla}\Omega)
\eeq
It is useful to eliminate the final total pressure term $\delta P_{tot}$, 
a process effected by taking the dot product of (\ref{175}) with $\bb{k}$ to 
isolate the final term of equation (\ref{175}).  Then,
equation (\ref{175}) is replaced with
\beq\label{178}
\left( \bb {\cal I} - {\bb{k k}\over k^2}\right)\bcdot\bb{{\cal L}(\xi)}=0
\eeq
where $\bb {\cal I}$ is the identity matrix with entries $\delta_{ij}$ (Kronecker delta),
and
$\bb{{\cal L}(\xi)}$ is the linear operator 
\beq\label{L196}
\bb{{\cal L}(\xi)}=
 \bb{\ddot\xi}
+ (\bb{k\cdot v_A})^2\bb{\xi} +2\bb{\Omega\times\dot\xi} +\bb{e_R}(R\bb{\xi\cdot\nabla})\Omega^2
-
{\del P \over\gamma\rho} \bb{\xi\cdot\nabla}\ln P\rho^{-\gamma}
\eeq 
Each of the individual terms in $\bb{{\cal L}(\xi)}$ is readily identifiable.  From left
to right we have acceleration, magnetic tension, Coriolis, tides, and finally buoyancy.  Note as well the similar $\bb{\xi\cdot\nabla}$ couplings that the tidal and
buoyant forces both invoke.  We shall argue in \S9 below that this entropy-angular momentum
dynamical pairing is important to an understanding of the Sun's rotation pattern.  

The lead factor in (\ref{178}) is a projection operator, ensuring that only $\bb{{\cal L}(\xi)}$
forces perpendicular to the wavevector ${\bf k}$ enter into dynamical consideration.
Finally, given the generality of our assumptions, the magnetic force is surprisingly simple, 
appearing only as the spring-like term $(\bb{k\cdot v_A})^2\bb{\xi}$.  

\subsection{The magnetorotational instability}

\subsubsection{Reduced system.}
To begin our exploration of the remarkable equation (\ref{178}), consider first
a disc in which the local equilibrium pressure gradient is negligible.   (Such 
conditions typically prevail in the midplane of a rotationally supported disc.)

For axisymmetric ($m=0$) disturbances, the wavenumber is independent of time,
and we will begin here.  With
disturbances proportional to $\exp(ik_z z)$, only displacements in the
plane of the disc enter, and the reduced equations are:  
\beq\label{179}
\ddot {\xi}_R - 2\Omega \dot{\xi}_\phi
 +\left( {\dd\Omega^2\over \dd\ln R} + (\bb{k\cdot v_A})^2\right)
\xi_R
=0,
\eeq
\beq\label{180}
\ddot {\xi}_\phi + 2\Omega \dot{\xi}_R + (\bb{k\cdot v_A})^2\xi_\phi
=0.
\eeq

\subsubsection {Hydrodynamic limit}

When the magnetic field vanishes, the two equations are yet more simple,
\beq\label{181}
\ddot {\xi_R} - 2\Omega \xi_\phi + {\dd\Omega^2\over \dd\ln R} \xi_R
=0,
\eeq
\beq\label{182}
\ddot {\xi_\phi} + 2\Omega \xi_R =0.
\eeq
The local normal modes with time dependence $\exp(-i\omega t)$ satisfy the dispersion relation
\beq\label{183}
\omega^2 = 4\Omega^2 +{\dd\Omega^2\over \dd\ln R}  = {1\over R^3}
{\dd(R^4\Omega^2)\over \dd R}\equiv \kappa^2,
\eeq
a quantity known as the epicyclic frequency.  Displaced fluid elements, viewed from
the point-of-view of an observer at the undisturbed circular orbit location, execute 
a retrograde ellipse.   It is retrograde in the sense that the element moves round the 
ellipse clockwise (counterclockwise) 
when the main circular orbit is counterclockwise (clockwise). 
The center of the ellipse is the undisturbed location of the element's circular orbit.

This is a simple consequence of angular momentum conservation.  The
element is moving on the ellipse---the ``epicycle''---around the
central point while maintaining the angular momentum of the undisturbed
circular orbit.  Thus, when the element is on its outward excursion at a greater
radial distance, it
must rotate a little more slowly.   It lags behind relative to
a point on the undisturbed circular orbit.  When the element
is on its inward excursion, it must rotate a little more rapidly.
This results in retrograde circulation.  In astrophysical discs, generally
$\kappa<2\Omega$.   From equation (\ref{182}), the epicycle is elongated along
the circular orbit, with major-to-minor axis ratio of $2\Omega/\kappa$.   

Equation (\ref{183}) indicates that if the specific angular momentum $R^2\Omega$ 
increases outward, then $\kappa^2 >0$ and our description of elliptical epicycles
is entirely self-consistent.  But if the specific angular momentum decreases with
increasing $R$, $\kappa^2<0$ and the displacements do not form a bounded ellipse,
but instead grow exponentially.  In this case, the flow is unstable to infinitesimal
axisymmetric disturbances.  The constraint that the specific angular momentum increase
outward for stable flow is known as the Rayleigh criterion, after its discoverer Lord
Rayleigh (1916).  The Rayleigh criterion is satisfied for Keplerian flow $(R^2\Omega
\propto R^{1/2})$, galactic flat rotation curves $(\propto R)$, and essentially
all other astrophysical environments.

The Rayleigh criterion clearly is not a guarantee of stability to nonaxisymmetric
disturbances.   A sharp outwardly increasing discontinuity in $\Omega$ in a disc
would be liable to Kelvin-Helmholtz instabilities;
less dramatically, a simple inflection point in the flow (where $d^2\Omega/dR^2 =0$)
may be enough for destabilization (Acheson 1990).   But these are nonlocal flows 
in the sense that the rotational profile exhibits a marked deviation from
simple shear on a length scale is that is comparable to or smaller than the scale
of the perturbation.  The Rayleigh criterion is a local criterion, and as 
such is relevant to the important question of whether turbulence could 
develop locally in 
astrophysical discs, using the free energy of the shear itself.
Plane parallel shear flow certainly does break down into turbulence at high Reynolds numbers,
and this has undoubtedly encouraged the belief that discs will behave in a similar
manner.   The problem is that, unlike plane-parallel flow,
discs are characterised by a strongly stabilising,
outwardly increasing angular momentum gradient.
The question is whether this gradient is enough to prevent {\it nonlinear} stability
from occurring.

The onset of turbulence is a deeply contentious subject, one that is difficult to study analytically,
numerically, or in the laboratory.   Laboratory experiments designed to measure the 
rotation velocity field by Doppler techniques find no nonlinear shear instabilities
in Keplerian-like flows (Ji et al.  2006, Schartman et al. 2012).  
This is in accord with the findings of
early numerical work (Balbus, Hawley, \& Stone 1996) and a very detailed follow-up
(Lesur \& Longaretti 2005). 
A recent claim for turbulence and enhanced angular momentum transport
in Keplerian shear flow (Paoletti \& Lathrop 2011) is now understood
to be a boundary layer effect stemming from presence of
the axial endcaps (Avila 2012, Nordsiek et al. 2015).  It was precisely to mitigate these effects
that the Princeton experiment, described in Schartman et al. 2012, split the endcaps of their Couette
apparatus into differentially rotating rings.  Even with such precautions, end effects can still
induce turbulent flow.   When viscous effects are sufficiently small, however,
the turbulence is spatially confined, and as of this writing experiments and simulations seem 
to be in good accord (Avila, private communication).    Whatever turbulence is present in laboratory
Couette flow is due to viscous boundary layers, not local shear flow.

\subsubsection{Magnetic fields.}
The inclusion of the apparently restorative magnetic acceleration term
$(\kva)^2\boxi$ in equations (equations [\ref{179}]-[\ref{180}]) produces
a surprisingly expanded dispersion relation.   If we seek solutions of the 
form $\exp(ikz-i\omega t)$, the resulting equation is
\beq\label{184}
\omega^ 4 - \omega^2[\kappa^2 +2 (\kva)^2] + (\kva)^2\left[(\kva)^2 + {\dd\Omega^2\over \dd\ln R}
\right] = 0.
\eeq
This is a quadratic equation in $\omega^2$, and it is easily seen that if the final 
constant term is negative, solutions exist with $\omega^2<0$, i.e. local instabilities.
But it is easy for this constant term to be negative because $\dd\Omega^2/\dd R <0$ quite 
generally in astrophysical discs.  Thus, provided the magnetic field is not too strong,
$\kva$ can always be chosen sufficiently small at long wavelengths so that instability
is present.  This is the magnetorotational instability, or MRI, and it is thought
to be the underlying cause of turbulence in even moderately ionised astrophysical discs
(Balbus \& Hawley 1991, 1998).  Figure (5) shows an early two-dimensional nonlinear simulation of the MRI, 
starting with a tube of magnetic flux.

The MRI is endlessly surprising.  The reader may enjoy the algebraic exercise of showing that the
maximum growth rate of the MRI is
\beq\label{185}
|\omega_{max}| = {1\over 2} \left| d\Omega\over d\ln R \right|
\eeq
which is achieved at wavenumbers corresponding to 
\beq\label{186}
(\kva)^2_{max} = \Omega^2\left[1- \left(\kappa\over 2\Omega\right)^4\right],
\eeq
for a displacement eigenvector corresponding to 
\beq\label{187}
\xi_R = - \xi_\phi,
\eeq
Though we have not shown it here, this
applies to the case of any magnetic field geometry, including a purely toroidal field
(Balbus \& Hawley 1998).  

\noindent
In other words:\hfill

\bigskip

\noindent
{\em Instability occurs only in the presence of a magnetic field, but neither the
instability criterion, nor the maximum growth rate, nor the most unstable
displacement eigenvector depend upon any
properties of the magnetic field, including its geometry.} 

\bigskip

\begin{figure*}
          \centering
          \includegraphics[width=12cm]{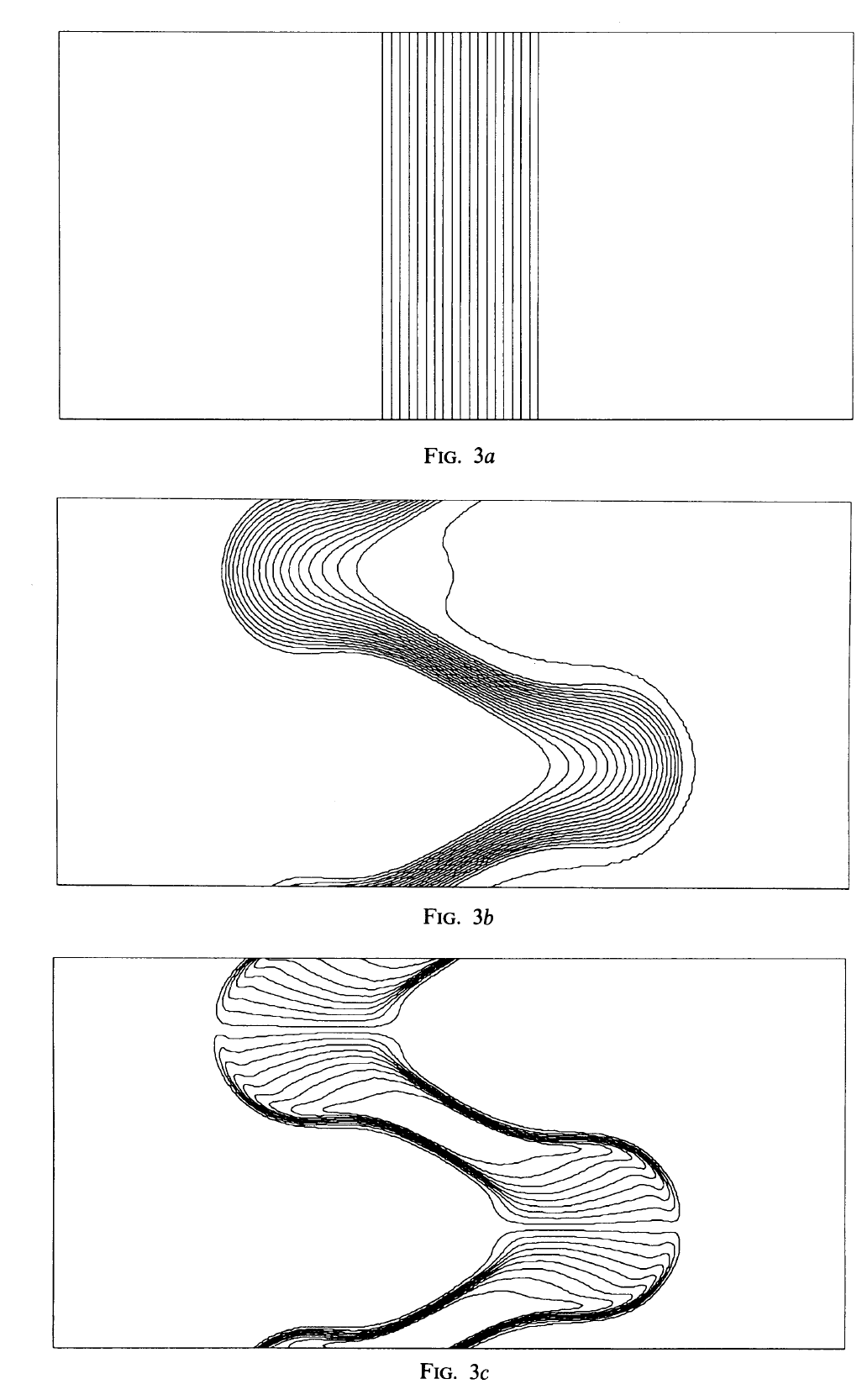}
          \caption{The original 2D numerical MHD simulation by Hawley \& Balbus (1991) showing the fastest growing mode of the magnetorotational instability with an initial net magnetic field in the $z$-direction. The perturbations grow until they form channeling solutions which disrupt the initial configuration. In 3D, the non-linear saturation of the MRI leads to sustained turbulence and angular momentum transport.  }
          \label{MRI}
\end{figure*}

Now all of this is one whopping surprise, and the fact that 
none of this was even remotely anticipated undoubtedly
contributed to preliminary MRI work languishing uninvestigated for more than thirty years, despite some interesting leads on global instabilities of axial magnetic fields in Couette experiments (Velikhov 1959, Chandrasekhar 1961).  Chandrasekhar's text, for example,
noted in passing that the Rayleigh criterion was not recovered in the limit 
$B\rightarrow 0$, speculated that field-freezing ought to be 
involved somehow, and left it at that.

These seemingly remarkable properties of the MRI can be understood with the help of a
simple physical model.  We have noted that the acceleration brought on by the magnetic 
field line tension is identical in form to a simple spring-like coupling in which the
force is proportional to (minus) the displacement.  Imagine then two nearby point masses,
connected by a weak spring, which are in orbit about a central mass.  The relative 
displacement of the two masses would satisfy a system of equations mathematically identical
to (\ref{179}) and (\ref{180}).  We will refer to the spring constant (per unit mass) 
as $K$.  It corresponds to $(\kva)^2$ in the MHD system.


\bigskip
\noindent 
\begin{figure*}
          \centering
          \includegraphics[width=12cm]{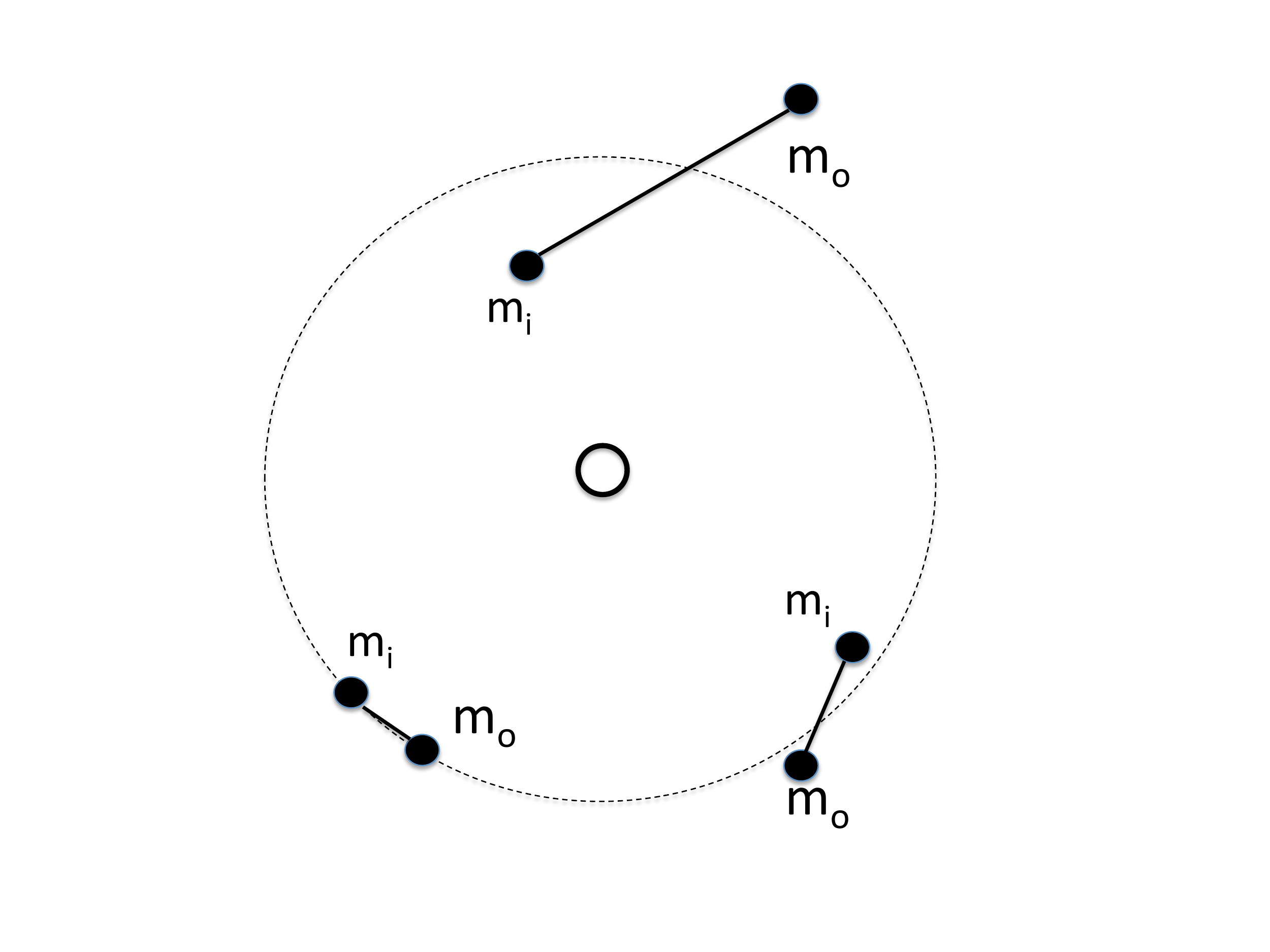}
          \caption{The magnetic force between two orbiting
fluid elements may be represented as a
spring-like force connecting two masses.  
Here, the spring is a simple blue line, the masses
are $m_i$  and $m_0$, and the figure shows the tethered configuration at three subsequent 
times, proceeding anticlockwise round the orbit.  See text for an explanation.}  
          \label{MRI}
\end{figure*}

Let $m_i$ refer to the mass that is orbiting slightly closer to the centre and
$m_o$ to the mass slightly farther out.  The spring pulls back on $m_i$ because it is
tethered to the more slowly orbiting mass $m_o$, and forward on $m_o$ for the oppositely analogous
reason.  But the backward-pulling torque on $m_i$ causes it to lose angular momentum,
so the mass to drop to an orbit closer in, where it actually {\it speeds up.}   Conversely,
the forward-pulling torque on $m_o$ causes it to acquire angular momentum, so the mass
moves outward to a more slowly rotating orbit.  The net torque between the two masses increases,
$m_i$ spirals inward more rapidly, $m_o$ spirals out more rapidly, and the process runs away (see figure 6).  The
very act of transporting angular momentum from one mass to another by a reactive, spring-like
force is intrinsically unstable.  

This description tacitly assumes that the spring constant $K$ is not so large that there
are many oscillations over a time scale rapid compared with an orbital period.   Were this
the case, the orbital dynamics would become irrelevant, and only the spring-induced
oscillations would
matter.  Indeed, when $\kva$ is large, the MHD dispersion relation reduces to 
$\omega^2 =(\kva)^2$, i.e. only Alfv\'en waves are present
\footnote{Technically, the waves include an Alfv\'enic and slow mode branch.}.

Moreover, in the limit of very small $K$, the instability is
still present, but the rate of growth becomes arbitrarily long.
Therefore, there must be a well-tuned value of $K$ that maximises the angular 
momentum transfer rate between the masses without causing a coupling so strong that
it isolates the mass pair from the orbital dyanamics, thereby stabilising the interaction.  
This value of $K$, call it $K_{max}$, is in fact the right side of equation (\ref{186}).  
Since $K_{max}$ depends only on the properties of $\Omega$, so will the maximum growth
rate $|\omega_{max}|$.  Thus, in the MHD problem, the magnetic field strength sets the absolute scale
for the fastest growing wavenumber $k_{max}$, but plays no role 
in determing $|\omega_{max}|$.

For the astrophysically important case of Keplerian flow, $\kappa=\Omega$ and
\beq\label{188}
(\kva)^2 = 15\Omega^2/16, \qquad |\omega_{max}| = 3\Omega/4\qquad {(\rm Kepler)}
\eeq
The most rapidly growing wavenumber increases a factor in amplitude of 111 each orbit!

There are a host of surprises associated with the detailed behaviour of the MRI,
but perhaps the biggest of all is the fact that the calculation is worth doing at all:
the MRI is formally present and vigorous in a rotating disc when the magnetic field is,
it would seem, ``negligibly small.''   The perceived complexities of treating a significantly
magnetised disc in which an embedded magnetic field was generating yet more field by 
radial shear was the principal reason for avoiding a detailed MHD analysis of accretion
discs.   Throughout the 1970s and 80s, the
topic remained anathema to most disc theorists.  The
understanding that the most salient features of MHD disc theory remain accessible
in the limit $B\rightarrow 0$ with $\kva$ finite, a regime that 
marginalises the importance of the {\it equilibrium} magnetic field behaviour, 
was the key realization that started our modern
understanding of accretion disc turbulence.    

\subsubsection{General axisymmetric disturbances.}
Using equation (\ref{178}), the full axisymmetric problem presents no particular
difficulties.  Equation (\ref{180}) for $\xi_\phi$ remains valid.
The normal modes have the plane wave form
\beq\label{189}
\boxi \propto \exp(ik_R R +ik_z z -i\omega t)
\eeq
with $k_R\xi_R + k_z\xi_z = 0$. 
The dispersion relation resulting from the $z$ component of (\ref{178}) is:
\beq\label{190}
{k^2\over k_z^2}\,  \varpi^4 +\varpi^2\left[ {1\over \gamma\rho}({\cal D}P)\, {\cal D\sigma} +
{1\over R^3}{\cal D}l^2\right] - 4\Omega^2 (\kva)^2 = 0,
\eeq
where
\beq\label{191}
\varpi^2 = \omega^2 - (\kva)^2, \quad {\cal D}\equiv \left( {k_R\over k_z}
{\dd\ \over\dd z} - {\dd\ \over \dd R} \right),
\quad l^2 = R^4\Omega^2,  \quad \sigma = \ln P\rho^{-\gamma}.
\eeq
(The $R$ component produces only a longer route to the same equation.) 
It is telling to compare the marginal stability condition of equation (\ref{190}) with and
without the magnetic field.  In the absence of any magnetic field, the marginal stability
condition is simply
\beq\label{192}
{1\over \gamma\rho}({\cal D}P)\, {\cal D\sigma} +
{1\over R^3}{\cal D}l^2 = 0.
\eeq
With $x\equiv k_R/k_z$, this may be written as a quadratic equation in $x$:
\beq\label{200}
x^2 N_z^2 +{x\over \gamma\rho}\left( {\dd P\over \dd R}{\dd\sigma\over \dd z} +{\dd P\over \dd z}{\dd P
\over \dd R} -{\gamma\rho\over R^3}{\dd l^2\over \dd z}\right) +N_R^2 +
{1 \over R^3}{\dd l^2\over \dd R} = 0 .
\eeq
where we have introduced the \BV frequencies
\beq\label{193}
N_z^2 = - {1\over \gamma\rho}{\dd P\over \dd z} {\dd \sigma\over \dd z} , \quad
N_R^2 = - {1\over \gamma\rho}{\dd P\over \dd R} {\dd \sigma\over \dd R}
\eeq
To ensure stability, the polynomial on the left of (\ref{200}) should be positive
somewhere and have no zeros for real $x$.  The most economical way of doing this is
to insist that the sum of the $x^2$ coefficient and constant term be positive (then
at least one of the terms must itself be positive), and that the quadratic discriminant
(``$b^2-4ac$'') is everywhere negative.  The first of these requirements is easily written
down:
\beq\label{202}
N^2 + {1 \over R^3}{\dd l^2\over \dd R} >0, \quad N^2 = -{1\over \gamma\rho}(\del P)\bcdot\del\sigma.
\eeq
The second is more of an algebraic challenge and will not be repeated here.   
The result of the calculation is the condition (Tassoul 1978, Balbus 1995) 
\beq\label{203}
-{\dd P\over \dd z}\left( {\dd l^2\over \dd R}{\dd\sigma\over \dd z} - {\dd l^2\over \dd z}
{\dd \sigma\over \dd R} \right) > 0 .
\eeq
The motivated reader who may wish to verify this directly should note that equation (\ref{164bis}) is 
needed on more than one ocassion in the course of the derivation.  

Normally, $-\dd P/\dd z>0$, so that equation (\ref{203}) states that the $\phi$ component of 
$\del\sigma\btimes\del l^2$ should be positive for stability.  An example of such a stable configuration
is $l^2$ increasing from the rotation axis, stratified on constant $R$ cylinders, with a spherically symmetric
entropy profile increasing outward.   Should the $\sigma$ gradient acquire a negative $z$ component with a positive
$R$ component, the same angular momentum distribution would be unstable.  

The surprise comes when we add a weak but otherwise arbitrary magnetic field, with at least some 
poloidal component.  Repeating our calculation in exactly the same way,  but
now including the $(\kva)^2$ terms in equation (\ref{190}), leads to the following stability criteria (Balbus 1995;
see also Fricke 1969, Acheson \& Hide 1973, Papaloizou \& Szuszkiewicz 1992):
\beq\label{204}
N^2 + {\dd\Omega^2 \over \dd\ln R} >0, 
\eeq
\beq\label{205}
-{\dd P\over \dd z}\left( {\dd \Omega^2 \over \dd R}{\dd\sigma\over \dd z} - {\dd \Omega^2 \over \dd z}
{\dd \sigma\over \dd R} \right) > 0 .
\eeq
The sole difference between (\ref{202})-(\ref{203}) and (\ref{204})-(\ref{205})
is that the $l$ gradients have been replaced
by $\Omega$ gradients!  This is remarkable considering the general disposition of the
magnetic field, but there is a certain logical consistency to all this:  by tethering
fluid elements, the presence of the magnetic field
creates a pathway tapping into the true free energy source in this problem, which is the shear itself.
We encountered a similar phenomenon in our study of magnetothermal behaviour, in which the stability
went from being regulated by a conserved quantity gradient (the entropy) to a free energy gradient 
(the temperature).

Free energy gradients are special.
Not only do they make life possible, they are palpable in a way that entropy or 
angular momentum gradients are not: they
hurt.  The extreme unpleasantness of a punch in the nose is due to the free energy 
source provided by the relative shear present between one person's head and another's
fist.  A burn from a hot stove is a thermal counterpart.  It is not the large entropy difference that gives us pain, it is $dT$.  We don't particularly care about how many microstates gives rise to the 
stove's macrostate, the salient feature is that it is hot.  
It is interesting, therefore, that while free energy gradients normally make their presence known in fluids only
through rather ghostly diffusive effects (thermal conduction and viscosity), a magnetic field
is a catalyst for turning them into agents of active dynamical destabilisation.  

Equations (\ref{202}) an (\ref{203}) are known as the H\o iland Criteria for the stability
(Tassoul 1978);
so (\ref{204}) and (\ref{205}) may be thought of as magnetic H\o iland Criteria.  They 
pertain to axisymmetric perturbations, and this is one case where we must closely heed to 
this restriction.   Our next surprising example will show why.

\subsection{Convection and rotation}

We return once again to the master equation (\ref{175} {\it et seq.}) to consider 
nonaxisymmetric
perturbations in rotating but nonshearing systems.   The space-time dependence of
our displacements is now $\exp i(k_R R +m\phi +k_z z -\omega t)$.  Choosing any two 
components of this equation together with the mass conservation condition (\ref{176}) leads
to a dispersion relation quite analogous to the general {\em axisymmetric} condition
(\ref{190}), but with the addition of one more remarkable term proportional to $m^2$:
\beq\label{206}
\varpi^4 +\varpi^2\left[ {k_z^2\over k^2} \left( {1\over \gamma\rho}({\cal D}P)\, {\cal D\sigma} -
4\Omega^2\right) +{m^2\over\gamma\rho R^2k^2}\del P\bcdot\del\sigma\right]- 4{k_z^2
\over k^2}\Omega^2 (\kva)^2 = 0.
\eeq
where now $k^2= k_R^2 +k_z^2 +m^2/R^2$.  
This is the magnetised version (Balbus \& Schaan 2012) of a classical result due to
Cowling (1951), but this is one classic whose lessons have not been fully absorbed.  
What we should learn from (\ref{206}) is that moderate rotation
cannot stabilise an unfavourable alignment of the pressure and entropy gradients
(entropy and pressure increasing in the same direction).  This is because
purely azimuthal wavenumbers with $k_z=0$ are apparently immune to Coriolis forces.
This is all the more surprising because when $k_R=0$ as well, the growth rate is the maximum 
possible.   In the case of a slowly rotating star like the Sun, in which the background 
pressure gradient is
(spherically) radial to a part in $10^5$, the fluid displacements associated with this
maximum growth rate are also radial to the same high level of accuracy.   Coriolis
deflections surely ought to produce distortions much larger than this!   So what is happening?  

What is happening is that a condition well-known to geophysicists is being set-up:
geostrophic balance.  Latitudinal and longitudinal motions on the Earth's
surface are subject to the Coriolis force due to the planet's rotation.  A time steady
balance can be achieved between this velocity dependent force and the pressure
gradient.   This gives rise, in the northern hemisphere, to clockwise circulating high
pressure regions and counterclockwise circulating low pressure regions.   (The pressure
gradient force 
is balanced by a velocity flow along isobars.)  Here, ``heliostrophic
balance'' is achieved by balancing the azimuthal Coriolis term $2\Omega\dot{\xi}_r$ with
$(1/\rho r)(\dd P /\dd \phi)$, and there is no $\phi$ deflection.   
The radial equation of motion therefore has no Coriolis term, and is instead a simple balance
between the radial acceleration and the unstable buoyancy force.
As far as fluid element displacements are concerned, it is as though the 
Coriolis force were absent.  

That ought to be a surprise, and more than a bit unsettling.  
As we shall see in the next section, there is evidence
that the Sun is a few degrees warmer on average at the poles compared with the equator.
This is often said to be a consequence of the Coriolis force: after all, it must be
easier for
a hot, outwardly rising gas parcel to move parallel to the rotation axis at high
latitudes toward the pole
then for it to move orthogonal to the rotation axis at lower latitudes toward the equator.
If so, it is a more subtle process, for 
we have just seen that there is nothing in the simplest version of buoyant 
instability to suggest that Coriolis forces have any affect on the most rapidly moving
convective blobs.   

Remember, these considerations apply only to {\it uniform} rotation.  
Does background shear at a level present in the Sun
make a difference?   Indeed it does.   That however brings its own plate of surprises,
which we will discuss in section 9.   

\section{The Papaloizou-Pringle Instability}

The Coriolis force exerts a powerful stabilising influence on the linear response of Keplerian
or near-Keplerian discs.  A constant density disc, for example, responds to perturbations 
with propagating
inertial waves driven by the angular momentum gradient, and there are no linear instabilities.  

Matters are no longer so straightforward when a boundary is present, which will bring in 
its own dynamical behaviour.  One such remarkable instability was discovered
by Papaloizou \& Pringle in 1984.  It caused a great stir at the time, because it demonstrated
that a then-popular (non-Keplerian) disc model of launching jets was in fact quite unstable.
The Papaloizou-Pringle instability is surprisingly
subtle, yet, as shown by Goldreich, Goodman, \& Narayan (1986), occurs in
very simple systems.   Our inventory of gas dynamical surprises would be incomplete without
at least a cursory visit.   

\subsection{Setting the stage}

Start with a standard disc.  For simplicity, we will take the density to be constant.
We ignore vertical structure, so the disc is really an axial cylinder.  The radial 
extent of the disc is assumed to be very narrow.   The mid-radius will be denoted
as $R_0$, and at a radius $R$ in the disc we define a local Cartesian coordinate
by $R= R_0+x$ with $x \ll R_0$.   The central rotation rate $\Omega_0$ is given by
\beq\label{207}
\Omega_0^2 = {GM\over R_0^3},
\eeq
where $M$ is the central mass.   Expanding to linear order around $R_0$, 
\beq\label{208}
\Omega = \Omega_0 (1-qx/R_0).
\eeq
We introduce the enthalpy function, $d{\cal H}=dP/\rho$, where as usual $P$ is pressure and
$\rho$ density.   Then the equation of motion for the unperturbed flow may be written
\beq\label{209}
R\Omega^2 = {d{\cal H}\over dR} +{GM\over R^3},
\eeq
or, expanding to leading order in $x$:
\beq\label{210}
{d{\cal H}\over dx} = (3-2q)\Omega_0^2 x.
\eeq
This is our formal background equilibrium condition.

Consider Eulerian velocity perturbations $\bb{\delta u}$ with radial ($x$) and
azimuthal ($y$) components.   The perturbed enthalpy is denoted $\delta {\cal H}$.
A perturbed variable $\delta Q$, is assumed to have a space-time dependence of
\beq\label{210bis}
\delta Q (x, y, t) =\delta Q(x) \exp[ik(y-R\Omega_0 t)-i\omega t]
=\delta Q(x) \exp[iky-i(\omega +kR\Omega_0) t]
\eeq
We work in an inertial frame, so the effective frequency   
has a rotational kinematic boost 
$kR\Omega_0$ included.   Note the distinction between $\Omega_0$
and $\Omega$, and the use of $R$, as opposed to $R_0$.     

The equation of mass conservation is
\beq\label{211}
{d(\delta u_x)\over dx} + i k \delta u_y = 0.
\eeq
Introducing the notation
\beq
\tilde\omega = \omega + q\Omega kx,
\eeq
the equations of motion may be taken from (\ref{158}) and (\ref{159}):
\beq\label{212}
-i\tilde{\omega}\delta u_x - 2 \Omega \delta u_y = - {d(\delta {\cal H})\over dx},
\eeq
\beq\label{213}
-i\tilde{\omega}\delta u_y +{\kappa^2\over 2\Omega}\delta u_x = -ik\delta {\cal H}, 
\eeq
where $\kappa^2$ is given by (\ref{183}), and we have dropped the 
``0'' subscript on $\Omega$, now a constant.  

Eliminating $\delta {\cal H}$ and $\delta u_y$ from (\ref{211})--(\ref{213}) yields a Laplace
equation for $\delta u_x$:
\beq\label{214}
{d^2(\delta u_x)\over dx^2} - k^2 \delta u_x = 0.
\eeq
Surprisingly, since the equations depend upon $x$ through $\tilde\omega$, this
is independent of everything except the constant wavenumber $k$.
The general solution is a superposition of $\sinh$ and $\cosh$
functions:
\beq\label{215}
\delta u_x = A\cosh(kx) + B\sinh(kx)
\eeq
A similar result is found for
surface water waves (Lighthill 1978),
a consequence of divergence- and curl-free flow. 
As with water waves, 
the crucial dynamics here lies in the
free surface boundary condition: the Lagrangian
change in the pressure (or enthalpy here) must vanish.
In other words, at the free surface $x=\pm s$, 
\beq\label{216}
\delta {\cal H} = - \xi{d{\cal H}\over dx},
\eeq
where $\xi$ is the radial displacement,
\beq\label{217}
\xi = {i\delta u_x\over {\tilde \omega}}.
\eeq
and ${d{\cal H}/dx}$ is the equilibrium enthalpy profile (\ref{210}).
Expressing (\ref{216}) in terms of $\delta u_x$ yields
\beq
{\tilde\omega}^2 {d(\delta u_x)\over dx} + \left(k{\tilde\omega} {\kappa^2 \over 2 \Omega} + k^2 {d{\cal H}\over dx}
\right)\delta u_x = 0.
\eeq
applied at $x=\pm s$, the thin disc boundaries.  The emergent
dispersion relation, in the limit $ks\rightarrow 0$, is
\beq
\omega^4 - \Omega^2\omega^2 + 3 (3-q^2)k^2s^2\Omega^4 = 0.
\eeq
The condition for unstable modes to be present is simply $q>\sqrt{3}=1.732$ (Papaloizou \& Pringle 1985).  
This contrasts with the {\em local} Rayleigh instability criterion of $q>2$.
It is easier to destabilise the flow if the disc boundaries are free.  (An example of Papaloizou-Pringle
instabilities in protostellar discs
is discussed by Lyra \& Mac Low [2012], in which the instability is mediated by Rossby-like modes.)

\begin{figure*}
          \centering
          \includegraphics[width=17cm]{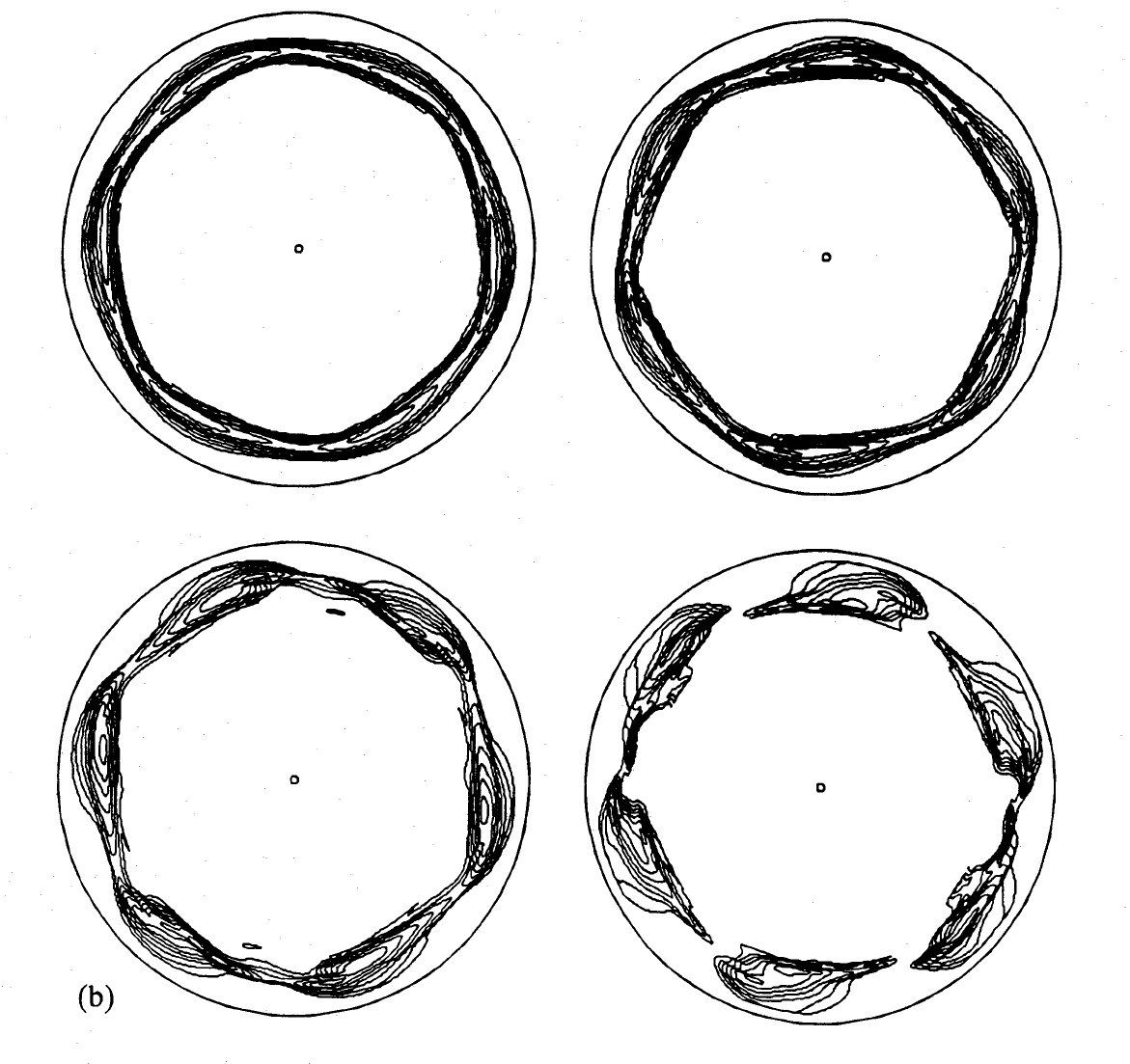}
          \caption{An early numerical MHD simulation by Hawley (1987) showing the growth of non-axisymmetric density perturbations in a disc. The non-axisymmetric perturbations grow and create asymmetric structure due to the Papaloizou-Pringle instability. }
          \label{MRI}
\end{figure*}

Goldreich et al. (1986) provided a detailed physical explanation for how this arises.  It depends
on the fact that a wave propagating through a moving fluid can either increase the
local energy density on average, in which case it is a positive energy wave, or
it can {\em decrease} the energy density, in which case it is a negative energy wave.
The onset of instability for $q>\sqrt{3}$ corresponds to the appearance of 
regions of both positive and negative perturbation energy densities, lying on
either side of a so-called corotation radius.  Energy flows from the region
of negative energy, and the loss causes an {\em increase} in amplitude, i.e.,
it is yet more negative.  The energy flows into the region of positive energy
density, also increasing its amplitude.   In other words, the flow of energy from
the negative to the positive energy region is intrinsically unstable!  

The enlightening surprise here is that although one's intuition can be shaped
by local flow behaviour, under conditions in which the edge of the system
is well-defined, entirely new dynamics can appear.   In the particular example
we have analysed, it might be said that conditions were artificial and certainly
not directly applicable to any known astrophysical environment.  But the 
physical content of the Goldreich et al. explanation suggests that this casual
dismissal misses the point.   The key notion is one of trapped waves, the 
existence of finite regions in waves of positive or negative energy
density are confined.    In the problem we chose this was set up in a simple
way by the use of edge dynamics; more complex examples might involve forming
such regions by appropriate background gradients in the equilibrium flow.
The elegant physics responsible for the destabilisation is just the same.

\subsection {Another example}

There is more to learn and more interesting surprises to be found with
our simple example.  Consider the same problem, this time with a hard wall
in place at $x=0$, so that the local relevant boundary condition is $\delta
u_x=0$.  Then $\delta u_x$ is proportional to $\sinh(kx)$.
At $x=s$, the free surface condition becomes
\beq\label{disrel}
{\tilde\omega}^2 +\left[ {\tilde\omega} {\kappa^2\over 2\Omega}
+ks\Omega^2 (3-2q)\right]\tanh ks = 0
\eeq
where $\tilde\omega$ is understood to be evaluated at $x=s$.
This is analogous to the dispersion relation that emerges for 
Rayleigh's surface water
waves in a sea of depth $s$ (Lighthill 1978):
\beq
\omega^2 = gk \tanh(ks)
\eeq
where $g$ is the (uniform) gravitational acceleration.  Indeed, if
we set $q=2$ so that $\kappa^2=0$ and epicylic oscillations are
eliminated, our dispersion relation becomes
\beq
{\tilde\omega^2} = (s\Omega^2)\, k\tanh(ks),
\eeq
the equivalent of water waves (Doppler boosted in frequency
by $qks\Omega$) with $g=\Omega^2 s$.

The surprise comes when we put back the epicyclic waves, creating an interplay
between the two types of response.  With $\kappa^2= 2\Omega^2(2-q)$,
the boundary condition at $x=s$ becomes the dispersion relation
\beq
{\tilde\omega}^2 + {\tilde\omega}\Omega (2-q)\tanh(ks) + ks\, \tanh(ks)\Omega^2 
(3-2q) = 0
\eeq
It suits our present purposes to leave this in terms of $\tilde\omega$ instead
of $\omega$.  

Instability corresponds to $\omega$, and thus $\tilde\omega$, acquiring a positive
imaginary component.  This, in turn, necessitates the condition
\beq
\Omega^2(2-q)^2 - 4ks\, \Omega^2 \tanh(ks) <0
\eeq
or
\beq
{\tanh ks\over ks} < {(3-2q)\over (2-q)^2}
\eeq
The left side function has a maximum of 1 at small $ks$, falling to a minimum
of zero at large $ks$.  There is instability (in fact, overstability),
if $q<1.5$. Keplerian flow is once again stable, but uniform rotation ($q=0$) is
not!  The key is the sign of the pressure gradient: in the presence of
a hard wall at the ``bottom'', all unstable 
modes are characterised by an {\em increasing} outward pressure.  The edge modes reinforce
the epicyclic oscillations under these conditions---an increasing outward pressure
makes the restoring gravity effectively more powerful---whereas no such reinforcement 
is present when the pressure decreases outward.  

Even in this very simple system, we have discovered two very different types
of instability depending upon which boundary condition is used.   One depends
upon extracting the free energy of differential rotation by the 
joint presence of positive and negative energy waves,
the other taps into an adverse pressure gradient to mutually reinforce surface
gravity and epicyclic oscillations.   The surprise here is that all this occurs
quite apart from local shear instability, the classic focus of the
rotational destabilisation.  Boundary conditions, which can be difficult to
pin down for accretion discs, cause qualitative changes in behaviour.

\section {Convection and rotation in the Sun}

\subsection {Helioseismology results}

The elucidation of the dynamical state of the Sun's interior, including a detailed rotation profile,
is one of the most impressive achievements of 20th century astronomy.  Not only
do we have more information for the velocity field of the Sun then 
for any other astrophysical fluid, we have far more
information about how the interior of the 
Sun is rotating than we have for a typical laboratory fluid 
experiment!  The physics behind this remarkable observational feat is the ability to
extract thousands of global eigenmode frequencies in the observed acoustic oscillation 
spectrum of the Sun (Thompson et al.\ 2003).
Differences between the frequencies corresponding to different azimuthal wavenumbers
then allow the precise angular rotation rate $\Omega$ to be determined as a function of
spherical radius $r$ and colatitude $\theta$.
By analogy with terrestrial earthquakes, which allow the Earth's interior to be
probed, the techniques that allow the precision determination of the Sun's interior
state is known as {\em helioseismology}.

\begin{figure} [t]
\includegraphics[width=12cm] {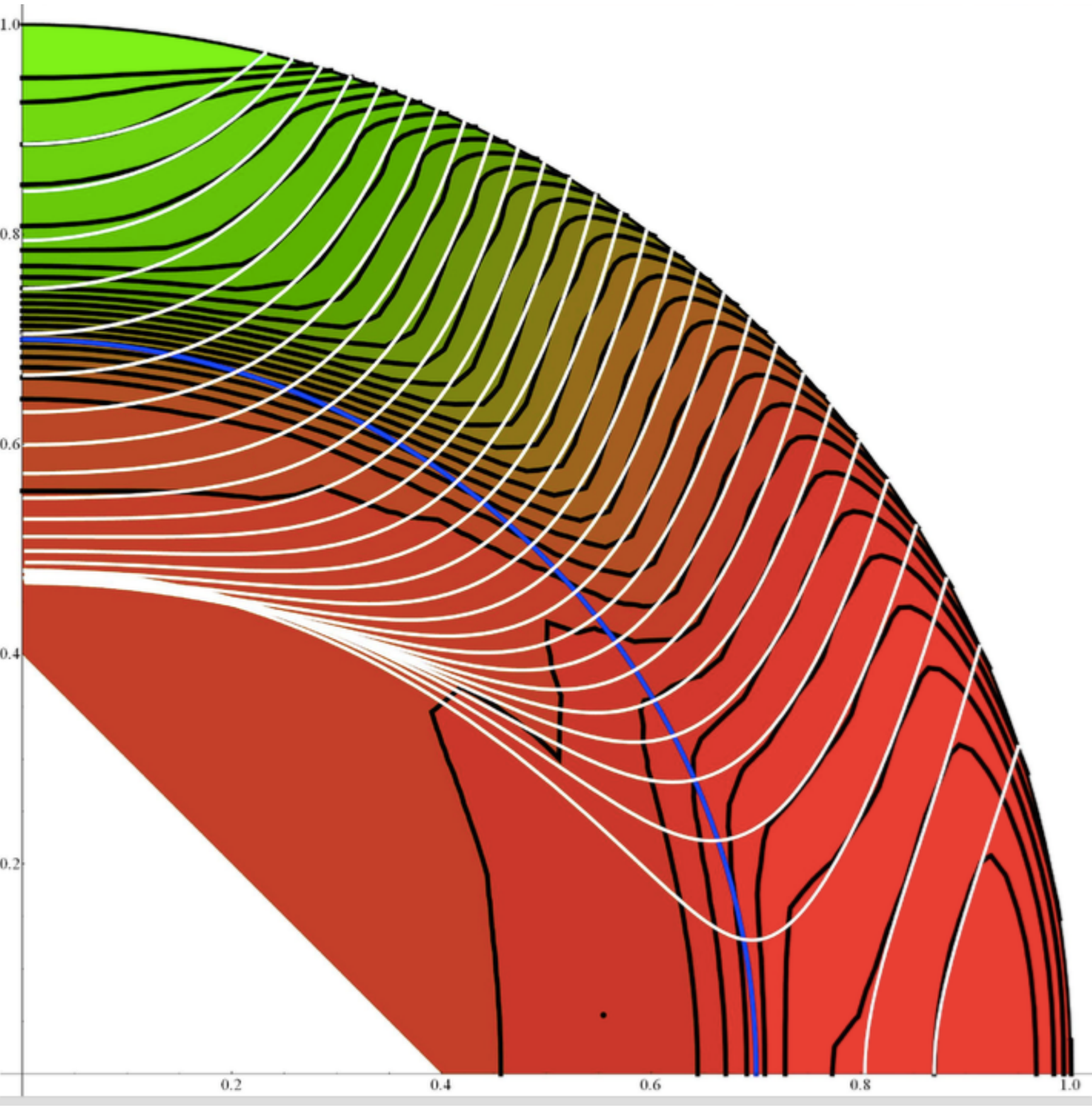} 
\caption {Contours of constant angular velocity in the interior
of the Sun. The axes are labelled in units of the solar radius with the x-y axes corresponding to a R-z cylindrical slice of the Sun respectively (where $z$ is aligned with the solar rotation axis). The region of strong differential rotation is limited
to the convective zone and the radiative zone immediately adjacent.
The bulk of the radiative interior is, to within the 
measurement acuracy, uniform rotation.  See text for detailed description.}
\label{solfit}
\end{figure}

Figure (8) shows the results of helioseismology analysis.  A meridional slice of the
Sun is depicted.   The outermost circular arc is the solar surface.  The black interior
curves are contours of constant angular velocity.  (Ignore the white curves for the moment.)  Reckoned in units of nano-Herz
($10^{-9}$ rotations per second, abbreviation nHz), the uppermost polar contour is
about 320, equatorial rotation is 460, and the intermediate contours are equally spaced
intervals.  An average rotation rate of 400 nHz corresponds to $2.5\times 10^{-6}$ s$^{-1}$,
about a one month period.   Latitudinal variations amount to some 15\%.  The question is
why do the contours look like this?   In particular, why are the rotation contours so
insensitive to depth in the bulk of the outer layers?  This will be our focus here.

The interior of the Sun is distinguished by three principal zones.  A small inner core,
comprising some 10\% of the Sun's mass is the region of nuclear energy production.
From this core out to $0.72 R_\odot$ the energy diffuses outwards at a rate proportional
to the temperature gradient.  (The radius of the Sun is denoted $R_\odot$.)  This 
region is known as the radiative zone.  The outer 28\% of the Sun, the third region,
is in a state
of turbulence in which the thermal energy is transported by convection: buoyant hot
gas rising, together with cooler, relatively heavy, gas sinking.  This may be thought
of as a sort of boiling due to the intense heating from below.   The convective motions
are typically very slow,  measured in 10s to 100s of meters per second, because of the
efficiency of this mechanical transport process.  A tiny velocity can move a great deal of 
bulk thermal energy compared with radiative diffusion.   Very near the surface, however, convection
velocities can become transonic.  

Regions of different energy transport --- diffusion, slow convection, 
rapid convection --- leave their distinctive imprint on the profile of the Sun's 
differential rotation.  At the base
of the convective zone and into the radiative zone the rotation is only weakly dependent on
$\theta$.  In the bulk of the convective zone, by contrast, where the turbulent velocity
is low, the rotation is only weakly dependent on $r$.  In the near surface layers there is a strong
dependence upon both $r$ and $\theta$.  The shear is very marked here.   

The question we pose here concerns the simplest portion of the solar rotation problem.
In the bulk of the convection zone, the velocities are very small and the stratification
is adiabatic to a remarkable accuracy: probably about 1 part in $10^5$.   This means that
the pressure should be very well described by a barotropic equation of state, 
$P=P(\rho)$.  Indeed, the near spherical symmetry by itself should be enough to ensure
a barotropic equation of state!   Why then is the rotation not stratified on cylinders?

Start with equation (\ref{vorti2}) in a dimensionless form:
\beq\label{001}
R {\dd(\ln \Omega^2)\over \dd z} =  {1\over\rho^2\Omega^2} (\del\rho\btimes \del P)\bcdot\bb{e_\phi}
\eeq
From the helioseismology data, the 
left side of this equation is a number of order 10\%.  The right side, if we simply go by 
magnitudes and do not worry about the alignment of the gradients, is a number of order $10^5$
(ratio of gravity to centrifugal forces).  The point is that if there is the slightest
misalignment of pressure and density isosurfaces---or, equivalently, of
temperature and density isosurfaces---this will be reflected in greatly magnified, easily
observed, baroclinic rotational velocity gradients.  

Barotropic rotation (which of course includes uniform rotation) distorts the
temperature and density surfaces from spherical, while still permitting the equilibrium forces to be derived from
a potential function.  This means that isobaric and isothermal surfaces coincide with one another (and with
equipotentials).     This nonspherical distortion, however, is generally inconsistent with radiative equilibrium
(e.g. Schwarzschild 1958, Clayton 1983).    

In a star rotating on cylinders, the fact that pressure and density surfaces coincide when $\Omega=\Omega(R)$
may be read off from
equation (\ref{vorti2}); that these surfaces coincide with those of the effective potential $\Phi$ follows directly
from inspection of the hydrostatic equilibrium equation
\beq\label{hsesch}
{1\over\rho}\del P = -\del{\Phi}
\eeq
where $\Phi$ is the effective potential
\beq
\Phi = \Phi_g -\int^R R'\Omega^2\, dR',
\eeq
and $\Phi_g$ the gravitational potential.   
The radiative flux $\bb{\cal F}$ may be written in the form
\beq
\bb{\cal F} = -\chi \del T= -\chi {dT\over d\Phi}\del\Phi
\eeq
where the diffusivity $\chi$ depends only on $\rho$ and $T$---and therefore only on $\Phi$.  Then,
\beq\label{247}
\del\bcdot\bb{\cal F} = -{d \over d\Phi}\left(\chi {dT\over d\Phi}\right) |\del\Phi|^2 -\chi {dT\over d\Phi}\nabla^2\Phi.
\eeq
Since 
\beq
\nabla^2\Phi=4\pi G\rho-{1\over R}{d \over dR}\left(R^2\Omega^2\right),
\eeq
when $\Omega$ is constant, everything on the right side of (\ref{247}) is constant on equipotential surfaces, except for $|\del\Phi|^2$.  This latter term can't possibly be constant, since the equipotentials are compressed at the poles and distended at the equator.  Thus the right side cannot
be identically zero.    In the case of $\Omega=\Omega(R)$, the same surfaces coincide and the functional constraints imposed by setting 
$\del\bcdot\bb{\cal F}=0$ are still too restrictive to be compatible with hydrostatic equilibrium (e.g. Strittmatter \& Roxburgh 1966).   If $\Omega=\Omega(R,z)$ the restrictions break down completely.  

Radiative equilibrium is generally thought to be maintained by tiny velocities (less than 
terrestrial plate tectonics) of meridional, or Eddington-Sweet, circulation. 
The classical argument, as presented by Schwarzschild (1958) and others,
is based on tapping the star's thermal energy gradient source to
move matter in bulk to offset radiative 
imbalance.   However, it is not energy {\it per se} that is needed, it is in particular heat---i.e., entropy.  
It is much more difficult to satisfy the radiative entropy equation.   The focus here is on
the region where the entropy gradient would vanish in a star with
barotropic rotation, so any bulk motion is locally ineffective.  What then?
One solution is that the rotation profile could become baroclinic.   Curiously, this is just what the solar
data show, in just the right location.  

Let us return to (\ref{165bis}), but using spherical coordinates on the right side:
\beq\label{166bisis165bis}
R{\dd\Omega^2\over \dd z} = {1\over \gamma \rho r} \left( {\dd\sigma\over \dd r}{\dd P\over \dd \theta} -
{\dd\sigma \over \dd \theta}{\dd P \over \dd r}\right) .
\eeq
The advantage of using the entropy variable $\sigma=\ln P\rho^{-\gamma}$
is that in the convective zone down to the radiative boundary
the spherical $r$-gradient of $\sigma$ is very small or zero (at the boundary itself).  
This is due to the great efficiency of the convective process: a gradient of order
$\sigma/r$ would produce an energy flow huge in comparison with the Sun's.  
On the right side of the above equation we need only retain the term involving 
$\dd P/\dd r$.  Switching to spherical coodinates,
\beq\label{tweq}
R {\dd \Omega^2\over \dd z} =  {g\over\gamma r}{\dd\sigma\over \dd\theta}
\eeq
where $g=-(1/\rho)(\dd P/\dd r)$ is the dominant radial gravitational field magnitude.
Equation (\ref{tweq}) is known as the {\em thermal wind equation}, widely
used in geophysics (Pedlosky 1987).   
The efficiency of convection is invoked explicitly by ignoring
the term in $\dd P/\dd\theta$, and tacitly by ignoring convective velocities in the equation
of motion, which assumes hydrostatic equilibrium.  For present purposes, we note
that an axial gradient in the angular velocity
is intimately associated with
latitudinal gradients in the entropy.  This has important
consequences for understanding radiative equilibrium at convective-radiative boundary.

Near the radiative boundary, the Sun's rotation is strongly baroclinic.
As such, there is no reason why a strict radiative equilibrium could not be enforced.
The argument against this---that the heat flux divergence cannot vanish everywhere
on a common equipotential/isothermal surface (Schwarzschild 1958)---breaks down for baroclinic flow,
because equipotentials are not isothermal surfaces. Indeed, one
could turn the argument on its head.   Why is the Sun's rotation baroclinic at all?
One answer might be that since the entropy gradient vanishes at the 
radiative/convective boundary in a barotropic model of solar rotation, it is ineffective
for circulation to offset a finite heat flux divergence.  The rotation profile
$\Omega(r, \theta)$ must then alter itself non-barotropically until the heat flux
divergences vanishes.  In this view, altered rotation, not the appearance of circulation,
is the key to the energetics, and the region remains in strict radiative balance.   

That there are plausible grounds for expecting steady non-circulating rotation patterns were put
forth long ago (e.g. Roxburgh 1964, 1966; Roxburgh \& Strittmatter 1966), but they seem
to have faded with time owing to problems of stability.    These profiles see at 
face value to be vulnerable to the 
Goldreich-Schubert-Fricke (GSF) instability (Goldrecih \& Schubert 1967, Fricke 1968), which
afflicts rotational flow not stratified on cylinders, i.e. baroclinic flow.
More recently, the question of the stability of baroclinic rotation in the upper radiative zone was
raised by Caleo, Balbus, \& Potter (2015), who calculated the explicit form of rotation
profiles using {\em static} radiative equilibrium as a requirement.  
Within the uncertainties of the helioseismology
observations, it is not difficult to reproduce the observed rotation profile
by imposing diffusive radiative equilibrium.  Moreover, their baroclinic form, 
the Caleo et al. (2015) profiles seem to be locally GSF stable in the upper radiative zone---with or without
a magnetic field (Caleo \& Balbus 2016).    The more complex question of global stability is not yet resolved.  
There is no consensus model of this radiative/convective transition zone, known as the tachocline, at the time
of this writing.   It is a very lively field of investigation.  

\subsection{Solution of the thermal wind equation}

Surprisingly, the thermal wind equation (\ref{tweq}) can be solved in a very
useful way for the bulk of the solar convection zone.  Let us begin by asking 
what convection actually does, both thermally and dynamically.

Thermally, convection redistributes entropy ($S$) from a higher $S$ to a lower $S$ region.
Doing so tends to flatten the radial entropy gradient, but does {\em not} eliminate it.
Some slightly negative (unstable) gradient is required to maintain convection, but 
this gradient is very tightly regulated.  It must allow precisely one solar luminosity
worth of thermal energy to pass through the convective zone via convective transport. 
A small change in the gradient would produce a large change in the outward heat flux.   

What effect does rotation have?   One might guess that Coriolis forces act more
adversely against motion perpendicular to the axis of rotation than motion along the axis.
In that case, the poles of the Sun should perhaps be slightly warmer than the equator.
The data suggest, in fact, that they are.

The surprise here has been previewed in section 7.3, namely there is
nothing in the {\em dynamics} of the
linear theory of convection to suggest
that this is what actually happens. For a uniformly rotating model of a star, Cowling (1951)
made a point of noting this, now almost 65 years ago! The most rapidily growing mode associated
with the dispersion relation (\ref{206}) corresponds to radial displacements
and purely azimuthal ($e^{im\phi}$) wavenumbers.  
Balbus \& Schaan (2012) showed the same state of affairs prevails even in a star
undergoing barotropic shear, $\Omega=\Omega(R)$.  (More precisely, they showed that
there was no effect of the shear to linear order in $\dd\Omega/\dd r$.)
The Coriolis force is anulled by the offsetting 
azimuthal pressure gradient in the $\phi$ equation of motion, a state of
geostrophic balance, common in planetary atmospheres and oceanography.  
But without a $\phi$ velocity, there is no Coriolis term in the radial
force equation to deflect hot parcels poleward!  So why are the poles warmer in the Sun?  
Shall we simply put it down to nonlinear complications?

As neither uniform rotation nor barotropic differential rotation affects
the behaviour of the most rapidly growing convective displacements, the last possibility
is baroclinic differential rotation.  Does a $z$ gradient cause any departures from
radial motion to leading linear order?  

Indeed it does.  Balbus \& Schaan (2012) carried out the analysis and found that the
components of the displacement vector $\bb{\xi}$ satisfy the equations
\beq
\ddot{\xi}_r  = \left(-{1\over\rho\gamma}{\dd P\over \dd r}{\dd\sigma\over\dd r} \right) \xi_r
\eeq
\beq
\ddot{\xi}_\theta = R{\dd\Omega^2\over \dd z} t \dot{\xi}_r
\eeq
so that there is a slight (northern) poleward drift of high entropy elements
when $\dd\Omega^2/\dd z<0$, as
it is throughout most of the (northern) convective zone.
This appears to show a nicely self-consistent feature, namely that the convergence of
high entropy fluid parcels at the poles could serve to raise the local temperature by
a few degrees and maintain the small latitudinal entropy gradient needed for baroclinic
rotation in the first place.   

To understand why the $z$ gradient appears whereas the $R$ gradient does not, recall
that we are dealing with embedded perturbations in a shearing medium.   Under these circumstances,
the wavenumber of a perturbation is sheared with the medium itself (see equation [\ref{167}]).
The relevant perturbations are dominated by the azimuthal $m$ wavenumber, so that the 
poloidal components $k_R$ and $k_z$ are directly proportional to $\del \Omega$.  The coupling
to the differential rotation disappears when $k_z$ does, so it is this purely kinematic
relationship between $k_z$ and $\dd\Omega/\dd z$ that brings baroclinic shear effects into 
our problem.

Let us return now to the helioseismology data of figure [8].   What is striking is
that in the bulk of the convective zone the isorotation curves seem to
resemble convective cells:
predominantly radial, but with a slight poleward bias.  In fact, Balbus \& Schaan (2012)
show both that the perturbed Eulerian azimuthal velocity $\delta v_\phi$ is very small
{\em and} that it is proportional to $\bb{\dot\xi}\bcdot\del\Omega$.   In other words,
the fluid element displacement velocities tend to lie in constant $\Omega$ surfaces
This of course means that the elements do not conserve their individual angular momenta.  
There is no mystery here; it is due to the presence of strong azimuthal pressure gradients.
This also means that the $\bb{\xi\cdot\nabla}\Omega^2$ term in equation (\ref{L196}) is
very small: the excess centrifugal term is minimised as the elements move radially.   There
are thus neither important Coriolis nor important centrifugal effects to disrupt the convective
dynamics.  

What a convective fluid parcel is trying to do is to eliminate the entropy
gradient, as we have discussed.  Both because the radial entropy gradient is tightly 
regulated and because there is a latitudinal entropy gradient, we expect
that within a (nearly radial) convective cell the radial behaviour of the entropy profile
will be given by some function,
$\sigma_r(r)$ say, with a (nearly) constant offset,
i.e.,
\beq
\sigma = \sigma_r + {\rm constant}
\eeq
The ``constant'' need not be the same constant at each latitude of course---indeed
it cannot be, if $\dd\sigma/\dd\theta$ is present.   Moreover, the convective mixing process
itself has no effect whatsoever on this entropy constant; it is the entropy gradient along
the gravitational field that is affected.    The latitudinal entropy gradient may need to be present
to drive baroclinic flow, thereby ensuring radiative equilibrium in the presence of ineffective meridional
circulation in the upper radiative zone. 
What emerges in the convective zone is a picture
in which fluid elements move in surfaces of constant $\Omega$ and at the 
same time in surfaces of constant
$\sigma-\sigma_r$  (Balbus et al. 2009).  We shall refer to this difference as ``residual entropy,'' $\delta\sigma$.  
Since only the $\theta$ gradient appears in the thermal wind equation (\ref{tweq}), we
may replace $\sigma$ with  $\delta \sigma$ on the right side.   
Our reasoning then suggests that we investigate
solutions of $(\ref{tweq})$ with its right side a function $f$ of $\Omega^2$ only:
\beq\label{tweq1}
{\dd \sigma\over \dd\theta}  \equiv
{\dd\delta \sigma\over \dd\theta} = 
{\dd f(\Omega^2)\over \dd\theta} = \left(df\over d\Omega^2\right)
{\dd\Omega^2\over \dd\theta}
\eeq
At the poles and the equator, this {\em ansatz} must be true just by the symmetry of our
problem.  The real content applies to the bulk of the convective zone, away from these symmetry
regions.  

Combining equations (\ref{tweq}) and (\ref{tweq1}) and switching to fully spherical coordinates,
\beq\label{tweq2}
{\dd\Omega^2\over \dd r} - \left(
{\tan\theta\over r} + {gf'\over \gamma r^2\sin\theta \cos\theta} \right)
{\dd\Omega^2\over \dd\theta} = 0 
\eeq
where we have written $f'$ for $df/d\Omega^2$.   Equation (\ref{tweq2}) is an equation for
precisely what we would like to know: the isorotation contours of $\Omega^2$.  In particular,
$\Omega^2$ must be constant on the characteristic curve given by
\beq\label{char1}
{d\theta\over dr} = - \left(
{\tan\theta\over r} + {gf'\over \gamma r^2\sin\theta \cos\theta} \right)
\eeq
Now $f'$ is not known, but as it depends only upon $\Omega^2$, we may be sure that along
the curve of interest, $f'$ is a constant.   So equation (\ref{char1}) is a perfectly
well-posed ordinary differential equation.  Moreover, despite is awkward appearance, it folds up
nicely:
\beq
{d(r^2\sin^2\theta)\over dr} = -{2f'\over \gamma} g = -{2f'\over \gamma} {d\Phi\over dr}
\eeq
where we have assumed that $g$, the gravitational field strength, is a function of $r$;
$\Phi(r)$ is the potential.  Our equation integrates immediately to
\beq\label{char2}
r^2\sin^2\theta \equiv R^2 = A + B\Phi(r)
\eeq
where $A$ is an integration constant and 
\beq
B = - 2f'/\gamma > 0
\eeq
is in practice a parameter to be fit from the data (in principle known from turbulence theory) and 
$B$ is positive since $f'$ is negative.

The surprise here is how simple equation (\ref{char2}) is.  After a lengthy and somewhat laboured
discursion
of the nature of convection and rotation (motivated by linear theory) we come to the conclusion
that the isorotation contours are curves on which $R^2$ is a linear function of the potential! 
For the Sun, 98\% of the mass is below the convective zone, so to
an excellent degree of approximation, $\Phi = -GM_\odot/r$, where $M_\odot$ is the mass of the 
Sun.  If $\Omega^2$ is specified as a function of $\theta$ on some surface $r=r_0$, as it
must be for a proper formulation of the solution of this class of partial differential equation, then the   
characteristic isorotation curves may be written
\beq\label{char3}
r^2\sin^2\theta \equiv R^2 = 
r_0^2\sin^2\theta_0  + \beta r_0^2 \left( 1 -{r_0\over r} \right)
\eeq
where 
\beq
\beta = - {2f'GM\over \gamma r_0^3}
\eeq
is a dimensionless number of order unity and $\theta_0$ is the value of $\theta$ at the start
of the contour.  The $\beta$ parameter must be constant along a contour, but can vary from
one contour to another with $f'(\Omega^2)$. 

Even without any detailed calculations, it is clear that equation (\ref{char3}) has the right
kind of properties to account for the general appearance of the solar isorotation contours.
As $r$ increases, the dominant balance is between $R^2$ and the constants on the right, i.e.
the flow tends toward constant on cylinders.  As $R$ becomes small near the rotation axis,
the dominant balance is between the $1/r$ and constant terms on the right, i.e. the flow
tends toward constant rotation on spherical surfaces.  

Using only one free parameter to fit to the entire Sun, $\beta=0.55$, the match is already quite
striking (e.g. Balbus, Latter, \& Weiss  2012).   If we allow ourselves the indulgence of a three parameter
polynomial fit for the same region:
\beq
\beta = A + B\sin\theta +C \sin^2\theta
\eeq
the result is nearly perfect (the white curves in figure \ref{solfit}), apart from the boundary layers.  
This level of agreement suggests that the basic assumptions of the theory---
the validity of the thermal wind equation and the sharing of angular velocity and 
residual entropy surfaces---is basically sound.   Thus, this is a potentially useful
approach for analysing the rotation profiles arising in many other problems involving convection and rotation.   The surprise here
has been that the dynamics of a complicated turbulent system has allowed itself to be encapsulated in
such a simple mathematical prescription: $\sigma' = f(\Omega^2)$ plus thermal wind balance
is enough to understand the shapes of the Sun's isorotation contours.   In displaying its isorotational contours, the Sun is acting like a great analogue computer, graphically presenting the solution characteristics of the thermal wind equation.  

\section{Concluding remarks}

The dedicated reader will be aware of the recurring lessons running through many of these problems.   Among the most significant is the remarkable dichotomy between problems in which a simple, naive treatment works wonders because of the insensitivity of the problem to anything but the dominant dynamical forces (i.e.\  the Newcomb-Parker Instability); and, in sharp contrast, problems in which the inclusion of a subtle dynamical or microphysical effect discretely and profoundly changes the behaviour of a system (i.e. weak magnetic fields). One of the reasons that, regardless of one's experience, the subject of astrophysical gasdynamics is so rich, counterintuitive and surprising, is that it is rarely obvious {\it a priori} which of these two regimes a particular problem will fall into.  Even an intuition honed by experience can be led astray by the underlying complex or delicate interactions amongst different dynamical and thermal processes, which may be on very distinct timescales.   Starting afresh with a more careful and rigorous approach is then necessary.  

The potent ability of weak magnetic fields to destabilise otherwise stable configurations is particularly striking.  That magnetic fields wield such a profound influence seems bizarre, how can it make any difference when the magnetic energy density is such a tiny fraction of the thermal energy density?   The answer lies not in strength, but in the new degrees of freedom that a field imparts to a fluid.   A magnetic field needn't compete with pressure.  Instead it propagates disturbances through the gas strictly on its own terms, in the form of an Alfv\'en wave.   These are shear modes, without any hydrodynamic counterpart to compete with.    Masses on linked springs are standard models for solids and magnetic media, both of which are in turn venues for propagating elastic waves.   Hydrodynamic fluids are not.

Another important lesson to bear in mind is the surprising ability of gravity to reverse the effects of usually straightforward interactions. Remember that the world of orbital dynamics is one in which pulling on a body in the direction of its motion slows it down and hindering its motion speeds it up.   In a gravitationally stratified medium adding heat to a body can lower its temperature and extracting heat can raise it. Forces that are attractive in a static environment become repulsive in a rotating one, and ordinary thermal conduction can enhance temperature differences in the presence of gravitational fields. These already counterintuitive systems are further complicated with added magnetic degrees of freedom.   It is, in retrospect, perhaps not surprising that so many new instabilities appear in the weak magnetic field regime.   It is all but certain that there will be more to follow, and that we don't yet fully understand the consequences of the ones we already supposedly know about.   

Beware of instabilities that appear simple but are in fact complex.    A slowly evolving process may be hypersensitve to the background equilibrium state and especially to its microphysics.    Many of the most knotty current problems of theoretical astrophysics, for example, involve the thermal behaviour of astrophysical plasmas.  On the other hand, physics is full of examples in which complexity is only
apparent.   Once established, it is usually the case that fundamental explanations are rarely unduly complicated.   On the contrary, they generally seem embarrassingly obvious.   {\it What ever was the problem in the first place?   How could so many clever people have overlooked that?   For shame!   How unfortunate that our current problems are ever so much more complicated.}   
Let naivity be a guide and maintain a playful spirit.  There are 
patterns and constraints to be discerned. Linear theory can be surprisingly helpful, so make sure it is understood before taking on a full analysis.   Unless we have been the victim of coincidences, the example of solar rotation offers hope that the days of pencil and paper analysis remain viable even in domains normally viewed as the province of large scale numerical simulation.

The current trend in our discipline has been to ever larger numerical simulations, and ever more strained 
rationalisations attempting to justify additions and alterations to the
governing equations so that the desired answers appear.   Some of this is perhaps inevitable
as our ambitions continue to mount;  the course of time will tell us whether this is a healthy trend.   In the meanwhile, one would do 
well to distinguish between a subtle explanation and a complicated one.
Genuine solutions to profound puzzles are usually the former, and only occassionally the latter. 
When they are the latter, they will, as a rule, involve the former as well.

Finally, we should never forget the wisdom of the ancients. There is still much to learn from the investigations that have attained classic status.   The challenge may simply be to recognise the same underlying physics in what may be a very different or somewhat more general setting.  

Astrophysical fluid dynamics is teeming with tractable but unsolved problems.  Regardless of our
methods, there will always be a need for powerful---and often simple--- insights.   
We may all look forward
to the surprises to come.

\section{Acknowledgements}

SAB acknowledges support from a gift from the Hinze Charitable Fund and a grant from the Royal Society in the form
of a Wolfson Research Merit award.  WJP acknowledges funding from University College, Oxford
in the form of a Junior Research Fellowship.

\end{document}